\documentclass[12pt,letterpaper]{article}
\pdfoutput=1 
\usepackage[utf8x]{inputenc} 
\usepackage[T1]{fontenc} 

\usepackage{jheppub}\hypersetup{pdfencoding=auto}
\usepackage{latexsym,amsmath,amsfonts,amssymb,bbm,bm,mathrsfs,mathtools,xfrac}
\usepackage[makeroom]{cancel}
\usepackage{graphicx}\graphicspath{{Figures/}}
\usepackage{paralist}

\DeclareUnicodeCharacter{"0393}{\Gamma}
\DeclareUnicodeCharacter{"0394}{\Delta}
\DeclareUnicodeCharacter{"0398}{\Theta}
\DeclareUnicodeCharacter{"039B}{\Lambda}
\DeclareUnicodeCharacter{"039E}{\Xi}
\DeclareUnicodeCharacter{"03A0}{\Pi}
\DeclareUnicodeCharacter{"03A3}{\Sigma}
\DeclareUnicodeCharacter{"03A5}{\Upsilon}
\DeclareUnicodeCharacter{"03A6}{\Phi}
\DeclareUnicodeCharacter{"03A8}{\Psi}
\DeclareUnicodeCharacter{"03A9}{\Omega}
\DeclareUnicodeCharacter{"03B1}{\alpha}
\DeclareUnicodeCharacter{"03B2}{\beta}
\DeclareUnicodeCharacter{"03B3}{\gamma}
\DeclareUnicodeCharacter{"03B4}{\delta}
\DeclareUnicodeCharacter{"03B5}{\epsilon}
\DeclareUnicodeCharacter{"03B6}{\zeta}
\DeclareUnicodeCharacter{"03B7}{\eta}
\DeclareUnicodeCharacter{"03B8}{\theta}
\DeclareUnicodeCharacter{"03D1}{\vartheta}
\DeclareUnicodeCharacter{"03B9}{\iota}
\DeclareUnicodeCharacter{"03BA}{\kappa}
\DeclareUnicodeCharacter{"03BB}{\lambda}
\DeclareUnicodeCharacter{"03BC}{\mu}
\DeclareUnicodeCharacter{"03BD}{\nu}
\DeclareUnicodeCharacter{"03BE}{\xi}
\DeclareUnicodeCharacter{"03C0}{\pi}
\DeclareUnicodeCharacter{"03C1}{\rho}
\DeclareUnicodeCharacter{"03C3}{\sigma} 
\DeclareUnicodeCharacter{"03C4}{\tau}
\DeclareUnicodeCharacter{"03C5}{\upsilon}
\DeclareUnicodeCharacter{"03C6}{\phi}
\DeclareUnicodeCharacter{"03D5}{\varphi}
\DeclareUnicodeCharacter{"03C7}{\chi}
\DeclareUnicodeCharacter{"03C8}{\psi}
\DeclareUnicodeCharacter{"03C9}{\omega}
\DeclareUnicodeCharacter{"21D0}{\Leftarrow}
\DeclareUnicodeCharacter{"0212B}{\AA}
\DeclareUnicodeCharacter{"00B7}{\cdot}
\DeclareUnicodeCharacter{"266A}{\eighthnote}
\DeclareUnicodeCharacter{"266B}{\twonotes}
\PrerenderUnicode{³}\PrerenderUnicode{¹}
\PrerenderUnicode{×}\PrerenderUnicode{→}
\PrerenderUnicode{Σ}\PrerenderUnicode{₆}

\newcommand{\be}{\begin{equation}} 
\newcommand{\ee}{\end{equation}}
\newcommand{\bea}{\begin{equation} \begin{aligned}}
\newcommand{\eea}{\end{aligned} \end{equation}}

\newcommand\eqs[1] {\begin{align}#1\end{align}}

\newcommand\eqss[1] {\begin{align}\begin{split}#1\end{split}\end{align}}
\newcommand\eqst[1] {\begin{multline}#1\end{multline}}

\newcommand\eqsc[1] {\begin{gather}#1\end{gather}}

\newcommand\equ[1] {\begin{equation}#1\end{equation}}

\renewcommand\i {i}
\newcommand\half {\tfrac{1}{2}}

\renewcommand\( {\left(}
\renewcommand\) {\right)}

\DeclareMathOperator{\tr}{tr}

\DeclareMathOperator{\Tr}{Tr}

\DeclareMathOperator{\sgn}{sgn}

\newcommand{\dilog}{{\text{Li}_2}}

\newcommand{\tsigma}{{\tilde{\sigma}}}
\newcommand{\tu}{{\tilde{u}}}
\newcommand{\tnu}{{\tilde{\nu}}}
\newcommand{\tgamma}{{\tilde{\gamma}}}

\newcommand\bC {{\mathbb C}}

\newcommand\bR {{\mathbb R}}
\newcommand\bS {{S}}

\newcommand{\Z}{\mathbb{Z}}
\renewcommand{\C}{\mathbb{C}}
\newcommand{\R}{\mathbb{R}}
\newcommand{\CP}{\mathbb{CP}}

\newcommand\N {{\mathcal N}} 
\renewcommand\O {{\mathcal O}}
\renewcommand\S {{\mathcal S}}

\newcommand{\cA}{\mathcal{A}}

\newcommand{\cC}{\mathcal{C}}

\newcommand{\cF}{\mathcal{F}}

\newcommand{\cH}{\mathcal{H}}

\newcommand{\cM}{\mathcal{M}}
\newcommand{\cN}{\mathcal{N}}

\newcommand{\cQ}{\mathcal{Q}}

\newcommand{\cS}{\mathcal{S}}
\newcommand{\cT}{\mathcal{T}}

\newcommand{\cV}{\mathcal{V}}
\newcommand{\cW}{\mathcal{W}}

\newcommand{\fg}{\mathfrak{g}}
\newcommand{\fm}{\mathfrak{m}}
\newcommand{\fn}{\mathfrak{n}}

\newcommand{\g}{\mathfrak{g}}
\newcommand{\m}{\mathfrak{m}}
\newcommand{\n}{\mathfrak{n}}
\newcommand{\q}{\mathfrak{q}}

\newcommand\sT {{\sf T}} 

\newcommand\bs[1] {\boldsymbol{#1}}
\newcommand{\ds}{\displaystyle}
\newcommand{\ts}{\textstyle}

\newcommand{\eg}{\textit{e.g.}}

\newcommand{\ie}{\textit{i.e.}}

\newcommand\nn {\nonumber\\}

\title{5d Partition Functions with A Twist}

\author[a]{P. Marcos Crichigno,}
\author[b]{Dharmesh Jain,}
\author[c]{and Brian Willett}

\affiliation[a]{Institute for Theoretical Physics, University of Amsterdam,\\ Science Park 904, Postbus 94485, 1090 GL, Amsterdam, The Netherlands}
\affiliation[b]{Theory Division, Saha Institute of Nuclear Physics,\\	1/AF Bidhan Nagar, Kolkata 700064, India}
\affiliation[c]{Kavli Institute for Theoretical Physics,\\ University of California, Santa Barbara, CA 93106, U.S.A.}

\emailAdd{p.m.crichigno@uva.nl}
\emailAdd{d.jain@saha.ac.in}
\emailAdd{bwillett@kitp.ucsb.edu}

\preprint{}

\abstract{We derive the partition function of 5d $\cN=1$ gauge theories on the manifold $S^3_b \times \Sigma_\g$ with a partial topological twist along the Riemann surface, $\Sigma_\g$.  This setup is a higher dimensional uplift of the two-dimensional A-twist, and the result can be expressed as a sum over solutions of Bethe-Ansatz-type equations, with the computation receiving nontrivial non-perturbative contributions.  We study this partition function in the large $N$ limit, where it is related to holographic RG flows between asymptotically locally AdS$_6$ and AdS$_4$ spacetimes,  reproducing known holographic relations between the corresponding free energies on $S^{5}$ and $S^{3}$ and predicting new ones. We also consider cases where the 5d theory admits a UV completion as a 6d SCFT, such as the maximally supersymmetric $\cN=2$ Yang-Mills theory, in which case the partition function computes the 4d index of general class $\cS$ theories, which we verify in certain simplifying limits.  Finally, we comment on the generalization to $\cM_3 \times \Sigma_\g$ with more general three-manifolds $\cM_3$ and focus in particular on $\cM_3=\Sigma_{\fg'}\times S^{1}$, in which case the partition function relates to the entropy of black holes in AdS$_6$. }

\begin{document}

\maketitle

\section{Introduction and summary}\label{sec:intro}

There has been tremendous progress in obtaining exact results for supersymmetric gauge theories in various numbers of spacetime dimensions.  These computations typically involve protected observables, which may be studied using non-renormalization theorems or, in the case of partition functions on compact manifolds, the localization technique \cite{Witten:1988ze,Pestun:2007rz}.  These results have opened new windows on quantum field theories and dualities between them.  Moreover, they have shown how theories in different dimensions are intricately interrelated.  Often, subtle properties of a theory, such as its dual descriptions or space of marginal parameters, can become obvious when we embed this theory into a higher-dimensional framework.

In this paper, we extend the list of exact observables by computing the partition function of arbitrary 5d $\cN=1$ gauge theories on  manifolds of the form $\cM_{3} \times \Sigma_\g$, \ie, a product of a three-manifold and a genus-$\g$ Riemann surface.  We will focus mainly on the case $\cM_{3}=S^{3}_{b}$, the ``squashed sphere,'' but we also discuss other cases.  This is an interesting observable for a variety of reasons.  First, five-dimensional quantum field theories are a fascinating arena with many counter-intuitive properties, and thus exact results can lead to valuable new insights.  Although 5d gauge theories are infrared trivial, in many cases they are believed to arise as a relevant deformation of nontrivial UV SCFTs, which we may probe by computing suitable quantities protected under RG flows.  In particular, there have been many such exact results for partition functions of these theories on closed manifolds; see \cite{Qiu:2016dyj,Kim:2016usy} for recent reviews and many examples and references.  These results can be used to study interesting and subtle properties of these theories, such as the appearance of enhanced global symmetries at the SCFT point \cite{Seiberg:1996bd}; see, \eg, \cite{Kim:2012gu}.  Then we expect the exact partition function on $S^3_b \times \Sigma_\g$ to lead to further probes into these 5d theories.

However, the observable $Z_{S^3_b \times \Sigma_\g}$ in particular is interesting because of its various connections to quantum field theories in other spacetime dimensions. In the remainder of this Introduction, we briefly review some of these features and summarize our main results.

\paragraph{The higher dimensional A-model and 2d TQFT.}

The first important connection is to two-dimensional QFT, and it is this connection which makes the computation of the partition function possible.  Specifically, the background we consider involves a topological twist along $\Sigma_\g$.   Higher-dimensional theories compactified on $\Sigma_\g$ with a partial topological twist have been well-studied in recent years \cite{Nekrasov:2014xaa,Benini:2015noa,Closset:2016arn,Benini:2016hjo,Closset:2017zgf,Closset:2017bse,Closset:2018ghr}.  These computations can be expressed as observables in a certain 2d topological quantum field theory (TQFT), namely the topological A-twist \cite{witten1988} of the effective theory obtained by compactification of the higher dimensional parent theory.

Here we take a similar approach, and study the effective 2d $\cN=(2,2)$ theory obtained by compactifying a 5d $\cN=1$ theory on $\cM_{3} \times \R^2$.  As described in Section \ref{sec:derivation}, the result for  $\cM_{3}=S^3_b$ takes the form of a sum over supersymmetric ``Bethe vacua'' of this 2d theory,\footnote{Here $\nu$ and $\n$ are, respectively, supersymmetric mass parameters and gauge fluxes for background symmetries through $\Sigma_\g$, as we describe in more detail in Section \ref{sec:derivation} below.}
\be \label{BEsumintro} Z_{S^3_b\times \Sigma_\g}(\nu)_{\n}   = \sum_{\hat{u} \in \cS_{\mathit{BE}}} \Pi_i(\hat{u},\nu)^{\n_i} \cH(\hat{u},\nu)^{\g-1} \,,\ee
where the objects
\be \Pi_i(u,\nu) = \exp \bigg(2 \pi i \frac{\partial \cW_{S^3_b \times \R^2}}{\partial \nu_i} \bigg)\,,\quad \cH(u,\nu) = e^{2 \pi i \Omega_{S^3_b \times \R^2}} \det_{a,b} \frac{\partial^2 \cW_{S^3_b \times \R^2}}{\partial u_a \partial u_b}\,, \ee
are refered to as the ``flux operator'' and ``handle-gluing operator,'' respectively, and are built in terms of the {\it effective twisted superpotential}, $\cW_{S^3_b \times \R^2}$, and the {\it effective dilaton}, $\Omega_{S^3_b \times\R^2}$, controlling the low energy effective theory on the 2d Coulomb branch. Finally, the set of supersymmetric Bethe vacua of the theory, $\cS_{\mathit{BE}}$, is defined as the solutions to certain Bethe-Ansatz-type equations, which may be written as
\be \label{SBEdefintro}\cS_{\mathit{BE}} = \;\; \left \{ \hat{u} \;\; \big| \;\; \Pi_a(\hat{u}) \equiv \exp \bigg( 2 \pi i \frac{\partial \cW_{S^3_b \times \R^2}}{\partial u_a}(\hat{u})\bigg) = 1, \;\;\; a=1,...,r_G\right\} /W_G\,.\ee

One novel feature of the five-dimensional computation, relative to lower-dimensional setups, is that the effective action controlling the 2d TQFT includes nontrivial non-perturbative corrections.  That is, we may write the effective twisted superpotential as
\be \cW_{S^3_b \times \R^2}(u,\nu,\gamma) =  \cW^{pert}_{S^3_b \times \R^2}(u,\nu,\gamma) +  \cW^{inst}_{S^3_b \times \R^2}(u,\nu,\gamma)\,, \ee
and similarly for the effective dilaton $\Omega_{S^3_b \times\R^2}$. In particular, these depend on the gauge coupling, $g_5$, through a parameter $\gamma = - \frac{\pi (b+b^{-1})}{g_5^2}$, with the perturbative piece dominating for small $g_{5}$.

Let us first state our result for the perturbative contribution, for definiteness.  Consider a 5d $\cN=1$ gauge theory with gauge group $G$, hypermultiplets in a representation $R = \bigoplus_i R_i$, and with a 5d CS term corresponding to the gauge-invariant functional $\text{Tr}_{CS}( \cdot)$ on the Lie algebra of $G$.  Then we may write the various operators above to perturbative accuracy as:
\eqs{\({\Pi^{pert}_a}\)^{\m_a} \({ \Pi^{pert}_i}\)^{\n_i} e^{2 \pi i (\g-1)\Omega^{pert}} =&\,  e^{-\pi i \text{Tr}_{CS}(\m u^2) + 2 \pi i \gamma \text{Tr}(\m u)} \prod_{\underset{\alpha \neq 0}{\alpha \in Ad(G)}} s_b(\alpha(u)-iQ)^{-\alpha(\m) + 1-\g} \nn
& \times \prod_i \prod_{\rho \in R_i} s_b(\rho(u) + \nu_i)^{\rho(\m)+\n_i +(r_i-1)(\g-1)}\,,
}
which are expressed in terms of the double sine function \cite{kurokawa1991,Hama:2011ea}, and we define $Q=\frac{1}{2}(b+b^{-1})$.\footnote{To be precise, this result is expressed in the variables familiar from the $S^3_b$ partition function, while the variables more natural for the A-model on $S^3_b \times \R^2$ turn out to be $\tu= i Q^{-1} u\,,  \tnu = i Q^{-1} \nu\,,$ and $\tgamma = i Q^{-1} \gamma\,$, as we explain below.} The above result is very reminiscent of the integrand of the $S^3_b$ partition function of a 3d $\cN=2$ theory \cite{Hama:2011ea}, and we comment on this relation below.  

However, as mentioned above, there are also non-perturbative contributions from instantons.  To compute these, we take a detour into equivariant localization on the space $S^3_b \times S^2_{\epsilon}$.  This is an uplift of the $S^2_{\epsilon_1} \times S^2_{\epsilon_2}$ partition function of \cite{Bawane:2014uka}, which receives contributions from point-like instantons.  This allows us to extract the non-perturbative contributions to the effective action.  We find that the full twisted superpotential and dilaton can be written in terms of the Nekrasov-Shatashvili (NS) limit of the 5d instanton partition function \cite{Nekrasov:2002qd}.  Namely, the latter has the following asymptotic behavior as we take one of the equivariant parameters to zero \cite{Nekrasov:2009rc}:
\bea \label{5dNSlimitintro} Z_{\R^2_{\q_1=e^{2 \pi i \epsilon_1}} \times \R^2_{\q_2=e^{2 \pi i \epsilon_2}} \times S^1}&\(x=e^{2 \pi i \tu}, y= e^{2 \pi i \tnu}, z= e^{2 \pi i \tgamma}\) \\
\underset{\epsilon_2 \rightarrow 0}{\longrightarrow}  \, &\exp \bigg\{ 2 \pi i \( \frac{1}{\epsilon_2} \cW_{NS}^{(5d)}(\tu,\tnu,\tgamma;\epsilon_1) - \Omega_{NS}^{(5d)}(\tu,\tnu,\tgamma;\epsilon_1) + O(\epsilon_2) \) \bigg\} \,. \eea
Then we may write the full, non-perturbative twisted superpotential and dilaton of the theory on $S^3_b \times \R^2$ as:
\eqs{ \label{wfactintro} 
\cW_{S^3_b \times \R^2}(\tu,\tnu,\tgamma) =\;&  \frac{1}{Q b}\cW^{(5d)}_{NS}(\tu,\tnu,\tgamma;-b^2) + \frac{1}{Q b^{-1}}\cW^{(5d)}_{NS}(\tu,\tnu,\tgamma;-b^{-2})\,, \\
 \Omega_{S^3_b \times \R^2}(\tu,\tnu,\tgamma) =\;& \Omega^{(5d)}_{NS}(\tu,\tnu,\tgamma;-b^2) + \Omega^{(5d)}_{NS}(\tu,\tnu,\tgamma;-b^{-2})\,.  
 }
Although in principle this gives the complete answer, in practice it is difficult to compute the non-perturbative corrections in a useful form, and so we will mainly focus on various simplifying limits where the instanton contributions can be explicitly characterized.

Finally, we comment that the form of  \eqref{wfactintro} as a sum of two contributions is closely related to the factorization of the $S^3_b$ partition function of 3d $\cN=2$ theories into two holomorphic blocks \cite{Beem:2012mb}, which are associated to the solid tori in the Heegard decomposition of $S^3$.  Our result then naturally generalizes to the case of $\cM_3 \times \Sigma_\g$ for arbitrary lens spaces, $\cM_3=L(p,q)$.   In Section \ref{sec:AdS6} we briefly comment on this generalization in the case of $\cM_3 = S^2 \times S^1$, with a topological twist on $S^{2}$,  and point out an interesting relation to the 5d prepotential.

\paragraph{Large $\bs{N}$ limit and universal RG flows to 3d.}

As we discuss in detail in Section~\ref{sec:largeN}, the large $N$ analysis of the matrix model computing the $S^{3}_{b}\times \Sigma_{\fg}$ partition function reveals an interesting structure. In particular, we will show that to leading order in $N$ the twisted superpotential $\mathcal W$ described above as well as the free energy, $F=-\text{Re}\log Z$, are specified by the $S^{5}$ partition function of the same theory. In particular, we find the large $N$ relation
\equ{
	F_{S^{3}\times \Sigma_{\fg}}= -\frac{8}{9}(\fg-1)Q^{2}\, F_{\kappa}(\fn) F_{S^{5}}\,,
}
where the function $F_{\kappa}(\fn)$ is given explicitly in \eqref{FS3S5nk} and generically depends on the theory under consideration. The case of the universal twist, however, is special \cite{Bobev:2017uzs}. This corresponds to a topological twist along the exact superconformal R-symmetry in the UV, in this case the Cartan of $SU(2)_{R}$, which amounts to setting $\fn=0$. In this case, and setting $\kappa=-1$, the relation above becomes universal:
\equ{
	F_{S^{3}\times \Sigma_{\fg}}^{\mathit{univ}}= -\frac{8}{9}(\fg-1)Q^{2} F_{S^{5}}\,.
}
These large $N$ results have an interesting interplay with holography. Indeed, the relation above was predicted in \cite{Bobev:2017uzs} (for the round sphere, $Q=1$) to hold for any 5d $\cN=1$ theory with an AdS$_{6}$ dual using properties of 6d supergravity. Our result is a field theory derivation of this relation.   

For nonzero flavor fluxes, the relation is no longer universal and one must consider specific theories. The holographic description of the twisted compactification of the Seiberg theory with flavor flux was recently constructed in \cite{Bah:2018lyv}. Specifying the above formula to this case we recover the results of this reference as well.  

Similar universal relations hold for the case $\Sigma_{\fg_{1}}\times \Sigma_{\fg_{2}}\times S^{1}$, which is described holographically by black holes in asymptotically locally AdS$_{6}$. We discuss this case in Section~\ref{sec:AdS6}.  

The interplay with holography goes both ways. While the twisted partition functions can be computed exactly by the localization methods we describe, these are not always well suited to answering  dynamical questions, such as the existence of interacting  fixed points in the IR. The explicit construction of holographic RG flows such as the ones described above then indicate the existence of interacting IR fixed points, at least at large $N$.

\paragraph{5d $\longrightarrow$ 6d $\longrightarrow$ 4d.}

One of the surprising features of QFTs in five dimensions is that there are examples of 5d effective theories whose ultraviolet completion is not itself a 5d QFT, but rather a 6d theory where one of the dimensions has been compactified on a circle, $S^1_{\beta}$, with radius $\beta$.  Here one identifies, roughly,
\be \label{r6g5} \beta \sim g_5^2\,. \ee
In particular, the KK excitations of this compactified theory, with action proportional to $\beta^{-1}$, can be naturally identified with instanton configurations in the low energy 5d gauge theory.   The prototypical example of this phenomenon is the maximally supersymmetric $\cN=2$ super Yang-Mills theory in 5d for a gauge group of ADE type, which is believed to be equivalent to the $S^1$ compactification of the $\cN=(2,0)$ SCFT associated to the corresponding Lie algebra.  

Given the above discussion, we expect that the 5d partition function on $S^3_b \times \Sigma_\g$ computes the partition function of the corresponding 6d UV theory on $S^3_b \times \Sigma_\g \times S^1_\beta$. Then, compactifying the 6d theory on $\Sigma_{\fg}$ with flavor fluxes $\n$, we may obtain a 4d theory, $\cT^{(4d)}_{\Sigma_\g,\n}$, and then we may also interpret this partition function as computing its  $S^3_b \times S^1_\beta$ partition function, better known as the ``supersymmetric index,''\footnote{More precisely, the partition function and index differ by an overall factor which can be identified with the Casimir energy of the vacuum state of the theory on $S^3_ b \times \R$.  
}   of this theory,
\be \label{indpfrel} Z_{S^3_b \times \Sigma_\g}(\nu,\gamma)_\n[\cT^{(5d)}] = Z_{S^3_b \times S^1_\beta}(p,q,\mu)[\cT^{(4d)}_{\Sigma_\g,\fn}] \,.\ee
Here the precise mapping of parameters is given in \eqref{indrel2} below.

There has been much work on understanding the compactification of 6d SCFTs on a compact Riemann surface, and the 4d $\cN=1$ and $\cN=2$ SQFTs that one obtains as a result.  This began with the work of Gaiotto on the compactification of the $A$-type $\cN=(2,0)$ theory on a punctured Riemann surface, with a twist preserving $\cN=2$ supersymmetry in 4d, leading to the celebrated theories of class $\cS$ \cite{Gaiotto:2009we}.  Subsequently this has been generalized in many directions, including compactifications of $D$ and $E$-type $\cN=(2,0)$ SCFTs, compactifications preserving only 4d $\cN=1$ SUSY \cite{Benini:2009mz,Bah:2012dg}, and, more recently, compactifications of 6d $\cN=(1,0)$ theories, leading to new classes of 4d $\cN=1$ theories and dualities \cite{Bah:2017gph,Kim:2017toz}.  Many of these 4d theories have the property that they do not have known Lagrangian descriptions, and so more indirect methods are required for studying their properties and computing observables, such as their supersymmetric index.  

However, these 5d QFTs, which are believed to be low energy descriptions of the 6d SCFTs on a circle, typically do have Lagrangian descriptions, and so we may, in principle, directly compute their $S^3_b \times \Sigma_\g$ partition function by the methods outlined above.  In practice, we will perform these computations in certain simplifying limits, namely, large $N$, and large gauge coupling.  By \eqref{r6g5}, the latter limit corresponds to large radius $\beta$.  In the 4d index, the large $\beta$ limit is dominated by the ``Casimir energy’’ of the vacuum state of the 4d theory quantized on $S^3_ b$.  In these limits the instanton contributions are under better control, and so analytic evaluation of the index is possible.  In addition, in the case of the maximal $\cN=2$ SYM theory, there is a special limit of parameters with additional supersymmetry, and where the instanton contributions are greatly simplified, and we will also be able to evaluate the index analytically in this limit.  This turns out to be closely related to the ``Schur'' \cite{Gadde:2011ik} and ``mixed Schur'' \cite{Beem:2012yn} limit of the superconformal index.

In all of these cases, we will find strong consistency checks of our computations by comparing to expected properties of the 4d index of the above theories.  We stress that our computation in principle gives a new approach to computing the index of 4d theories without a known Lagrangian description.  The form of the computation is in terms of the 2d TQFT dual to the 4d index of these theories, and so manifestly exhibits various 4d dualities, such as $S$-duality.

\paragraph{Outline.}

The paper is organized as follows. In  Section~\ref{sec:derivation}  we provide the derivation of the exact partition function by localization methods, as described above. In Section~\ref{sec:largeN} we study the large $N$ limit of the $S^{3}_{b}\times \Sigma_{\fg}$ partition function, its relation to the partition function on $S^{5}$ in this limit, and discuss universal RG flows and holography.  In Section~\ref{sec:5d6d4d} we consider theories with a 6d UV completion and compute the 4d superconformal index of theories obtained by reduction on the Riemann surface. Finally, in Section~\ref{sec:AdS6} we discuss the partition function on more general manifolds, including $S^2 \times S^1 \times \Sigma_\g$, and discuss the counting of black hole microstates in  AdS$_{6}$. 

\paragraph{Note added:} In the final stages of this work we learned about the work \cite{Hosseini:2018uzp}, which has some overlap with our discussion in Section~\ref{sec:AdS6}.


\section{\texorpdfstring{Derivation of partition function on $S^3_b \times \Sigma_\g$}{Derivation of partition function on S³b×Σg}}\label{sec:derivation}

In this section, we derive the exact partition function of 5d $\N=1$ gauge theories on $S^3_b \times \Sigma_\g$, with a partial topological twist on $\Sigma_{\fg}$. We discuss different ways of carrying out the computation, which are complementary to each other and offer different points of view on this observable.

\subsection{\texorpdfstring{5d $\cN=1$ theories on curved backgrounds}{5d N=1 theories on curved backgrounds}}\label{sec:5dN=1curved}

Let us first review some basic properties of 5d $\cN=1$ supersymmetric gauge theories in flat, Euclidean space, in preparation for studying them on curved backgrounds.  The 5d $\cN=1$ Yang-Mills action can be obtained by dimensional reduction from 6d $\cN=1$ on $\R^{5,1}$, and is given by \cite{Qiu:2016dyj} 
\be S = \frac{1}{g_{5}^2} \text{Tr} \int d^5 x \bigg( \frac{1}{2} F^{\mu \nu} F_{\mu \nu} + i \Lambda_I D \!\!\!\!/ \; \Lambda^I - D^\mu \sigma D_\mu \sigma - \Lambda_I [ \sigma,\Lambda^I] - \frac{1}{2} D^{IJ} D_{IJ}  \bigg) \,. \ee
Here $F = dA+i[A,  A]$  is the gauge field strength, $\sigma$ is a real scalar, $\Lambda_I$ is the symplectic Majorana gaugino, and $D_{IJ}$ is an auxiliary scalar field.  The indices $I,J=1,2$ correspond to the $SU(2)_R$ symmetry, and we impose the symplectic Majorana condition,
\be \big(\Lambda_\alpha^I\big)^* = \epsilon_{IJ} \omega^{\alpha \beta} \Lambda_\beta^J \,, \ee
where $\alpha=1,...,4$ is the spinor index, which corresponds to the fundamental representation of $USp(4) \cong Spin(5)$.

As will be important below, these theories are closely related to 4d $\cN=2$ theories, which are obtained upon dimensional reduction, and as for 4d $\cN=2$ theories their actions can be written in terms of a holomorphic {\it prepotential}, $\cF(\cA)$. In a notation adapted to reduction to 4d, this can be written as
\be S= \frac{1}{4 \pi} \text{Im} \bigg( \frac{1}{2} \int d^2 \theta d^5 x \frac{\partial^2 \cF}{\partial \cA^i \partial \cA^j} W^i W^j + \int d^4 \theta d^5 x \bar{\cA}^i \frac{\partial \cF}{\partial \cA^i} \bigg)\,, \ee where $W^i$ and $\cA^i$ are the  usual 4d $\cN=1$ chiral superfields making up the 4d $\cN=2$ vector multiplet, and the complex scalar in the 4d vector multiplet can be identified with
\be a = A_5 + i \sigma \,,\ee
where $A_5$ is the fifth component of the 5d gauge field.  Then the most general action in 5d can be written in terms of a cubic prepotential \cite{Seiberg:1996bd,Intriligator:1997pq},
\be \label{f5d} \cF(\cA) = \frac{1}{2 {g_5}^2} \Tr \cA^2 + \frac{1}{6} c \Tr \cA^3 \,,\ee
where $g_5$ is the 5d gauge coupling, and the cubic term defines a 5d Chern-Simons term, which may be included for certain choices of gauge group $G$.

In addition to the vector multiplet, we may include hypermultiplets.  Their field content consists of complex scalars $q_I^A$ and spinors $\psi^A$, where $I$ is the $SU(2)_R$ index, as above, and $A$ is a gauge or flavor index, for which we assume the matter is in a psueudoreal representation, $R =\bigoplus_{i} R_i$.  These are taken to satisfy reality conditions,
\be \big(q^A_I\big)^* = \Omega_{AB} \epsilon^{IJ} q_J^B, \;\;\; \big(\psi^A_\alpha\big)^* = \Omega_{AB} \omega^{\alpha \beta} \psi^B_\beta \,.\ee
These give rise to 4d $\cN=2$ hypermultiplets upon dimensional reduction.  We may also turn on background vector multiplets coupled to flavor symmetries, and including a background value, $m_i$, for the real scalar in these multiplets gives a supersymmetric real mass for hypermultiplets charged under this symmetry.  Integrating out these massive hypermultiplets gives rise to a simple correction to the prepotential \eqref{f5d} \cite{Seiberg:1996bd,Intriligator:1997pq},

\bea \label{Feffdef} \cF_{eff} &= \frac{1}{2 {g_5}^2} \Tr \cA^2 + \frac{1}{6} c \Tr \cA^3-  \frac{1}{12} \sum_i \sum_{\rho \in R_o} |\rho(\cA) + m_i|^3 + \frac{1}{12} \sum_{\alpha \in Ad(G)} |\alpha(\cA)|^3  \\
& \approx \frac{1}{2} t_{ab} \cA^a \cA^b + \frac{1}{6} c_{abc} \cA^a \cA^b \cA^c + \cdots\eea
where the latter expansion may be made for different asymptotic directions on the Coulomb branch, and defines the effective gauge coupling and Chern-Simons terms which are generated there.

\paragraph{5d $\cN=1$ supersymmetry on curved spacetimes.}

There has been much work on placing 5d supersymmetric theories on curved spacetimes and computing their partition functions by localization.  Some examples include $S^5$ \cite{Kallen:2012cs,Hosomichi:2012ek,Kim:2012qf}, $\CP^2 \times S^1$ \cite{Kim:2013nva}, $Y^{p,q}$ \cite{Qiu:2013pta}, and $S^4 \times S^1$ \cite{Kim:2012gu}; see \cite{Qiu:2016dyj,Kim:2016usy} for recent reviews, and many additional references.  In addition, maximally supersymmetric 5d Yang-Mills theory was studied on $S^3 \times \Sigma_\g$ in \cite{Kawano:2012up,Fukuda:2012jr,Kawano:2015ssa}, although our result appears to differ from theirs.\footnote{Specifically, we find additional contributions from instantons and fermionic zero modes relative to their computation.}

To write supersymmetric actions on such manifolds, we can employ the philosophy of \cite{Festuccia:2011ws} and consider a rigid limit of 5d $\cN=1$ supergravity.  Such an approach was studied in \cite{Pan:2013uoa,Imamura:2014ima,Alday:2015lta}.  In particular, in \cite{Alday:2015lta} it was shown that we may define a supersymmetric background for a 5d $\cN=1$ theory on any manifold, $\cM_5$, admitting a transversally Hermitian structure.  This means we may write a metric on $\cM_5$ of the form
\be ds_{M_5}^2 = S^2 (d \psi+ \rho)^2 + ds_4^2\,. \ee
Here $\psi$ is a coordinate generating an isometry of the metric, \ie, $K=\partial_\psi$ is a Killing vector.  The transverse directions admit a complex structure, and $ds_4^2$ is a Hermitian metric.  With this structure, we may find solutions to the Killing spinor equations, which allow us to write actions on $\cM_5$ preserving some supersymmetry.

Let us note that the above classification is very similar to that of \cite{Closset:2012ru} for coupling 3d $\cN=2$ supersymmetric theories to three-manifolds.  Namely, in that case the necessary structure is a transversally holomorphic foliation.  Then it is straightforward to check that, given a three-manifold $\cM_3$ with background gravity fields, we may define a supersymmetric background on $\cM_3 \times \R^2$ with
\be \zeta^{(5d)} = \zeta^{(3d)} \otimes \epsilon\,, \ee
for any constant $\epsilon$.  In particular, in the case where $\cM_3$ is a Seifert manifold \cite{Closset:2012ru}, we may preserve one supercharge, $\zeta^{(3d)}$, of R-charge $1$ and another, $\tilde{\zeta}^{(3d)}$, of R-charge $-1$, and so on $\cM_3 \times \R^2$ we may preserve four supercharges. 

Let us consider the algebra satisfied by the supercharges.  First, in 3d, we have \cite{Closset:2012ru}
\be \{ \delta_\zeta,\delta_\zeta \} = \{ \delta_{\tilde{\zeta}} , \delta_{\tilde{\zeta}} \} = 0\,,  \nonumber \ee
\be \{ \delta_{\zeta} , \delta_{\tilde{\zeta}} \} = -2i {\cal L}_K' + \zeta \tilde{\zeta} (z -\Delta H)\,, \ee
where $z$ is the central charge of the field being acted on, $\Delta$ is its R-charge, $H$ is a scalar in the background supergravity multiplet, and ${\cal L}_K'$ is the modified Lie derivative, which is covariant under R- and central-charge symmetry gauge transformations, acting along the Killing vector,
\be K^\mu = \zeta \gamma^\mu \tilde{\zeta} \,.\ee
Returning to the 5d superalgebra, let us define $\delta_\pm$ and $\tilde{\delta}_\pm$ as the transformations associated to $\zeta \otimes \epsilon_\pm$ and $\tilde{\zeta} \otimes \epsilon_\pm$, respectively, where we define $\epsilon_\pm$ as the $\pm 1$ eigenvectors of $\sigma_3$.  Then, for example, we may compute
\be \{ \delta_+, \delta_- \} = (\epsilon_+ \epsilon_-) ( -2i {\cal L}_K' + \zeta \tilde{\zeta} (z - \Delta H )) + (\zeta \zeta) (\epsilon_+ \sigma_i \epsilon_-) =  0 \,,\ee
\be \{ \delta_+,\tilde{\delta}_+ \} = (\zeta \tilde{\zeta}) (\epsilon_+ \sigma_i \epsilon_+) = 2i \partial_z\,, \ee
\be \{ \delta_+, \tilde{\delta}_- \} = (\epsilon_+ \epsilon_-) ( -2i {\cal L}_K' + \zeta \tilde{\zeta} (z -\Delta H) ) + (\zeta \tilde{\zeta}) (\epsilon_+ \sigma_i \epsilon_+) =-2i {\cal L}_K' + (z -\Delta H) \,.\ee
We can repeat this for the remaining supercharges.  Then, if we identify ${\cal Q}_+ = \delta_+, {\cal Q}_- = \tilde{\delta}_-, \tilde{\cal Q}_+ = \tilde{\delta}_+$, and $\tilde{\cal Q}_- = {\delta}_-$, one can check that ${\cal Q}_\alpha$ and $\tilde{\cal Q}_\alpha$ generate the $\cN=(2,2)$ superalgebra, with a central charge given by the operator
\be \label{Zdef} \tilde{Z} =   -2i {\cal L}_K' + (z -\Delta H) \,. \ee
In other words, the 5d $\cN=1$ theory compactified on $\cM_3$ gives rise to an effective 2d $\cN=(2,2)$ theory, whose central charge is determined by the operator $\tilde{Z}$ above, which depends on the KK momentum along $\cM_3$, R-charge, and real mass parameters, which contribute to the central charge, $z$.

Finally, to construct a background on the compact manifold, $\cM_3 \times \Sigma_\g$, we may perform a partial topological twist along $\Sigma_\g$.  From the 5d point of view, this amounts to turning on a flux, $\g-1$, on $\Sigma_\g$ for the background gauge field coupled to the $U(1)_R \subset SU(2)_R$ symmetry.  From the point of view of the effective 2d $\cN=(2,2)$ theory, this is simply the topological A-twist \cite{witten1988}.  We then expect the partition function to be an observable in an appropriate 2d topological quantum field theory (TQFT), which constrains the form of the answer.  We review this structure in the next subsection.

In the rest of the paper, we will mostly focus on the case $\cM_3=S^3_b$.\footnote{We will return to consider more general lens spaces in Section \ref{sec:AdS6}.} Specifically, we will take the supersymmetric background of \cite{Imamura:2011wg}, which exhibits $S^3_b$ as an $S^1$ fiber bundle over the round metric on $S^2$, \ie,
\be ds_3^2 = \tilde{\beta}^2(d \psi + a)^2 + d\Omega_{S^2}^2\,. \ee
This preserves an $SU(2) \times U(1)$ isometry, and the Killing vector  appearing in the superalgebra is $K=\partial_\psi$.  Then, in this background we have $H=\frac{i}{2}(b+b^{-1}) \equiv i Q$ \cite{Closset:2012ru}, and so the central charge in the $\cN=(2,2)$ algebra  \eqref{Zdef} is
\bea \label{ZS3} \tilde{Z}& =  -2i {\cal L}_{\partial_\psi} +\(z + A_\psi  - i \Delta Q \)\, \\
&\equiv -2i {\cal L}_{\partial_\psi} +i \sigma \,,\eea
which defines a variable, $\sigma$, valued in the complexification of the Cartan subalgebra of the flavor symmetry, which is a holomorphic combination of the background real scalar, $z$, and component, $A_\psi$, of the background gauge field along the direction $K$, as discussed in more detail in \cite{Closset:2017zgf}.

\subsection{The 2d A-twist and Bethe Ansatz equations}\label{sec:2datwist}

As observed above, the supersymmetric background on $S^3_b \times \Sigma_\g$, when considered along the $\Sigma_\g$ directions, takes the form of a 2d topological twist.  Such setups, where a $d$-dimensional gauge theory is placed on a manifold of the form $\cM_{d-2} \times \Sigma_{\fg}$, and subjected to a topological twist along the Riemann surface, have been studied in many examples recently--see, \eg, \cite{Nekrasov:2014xaa,Benini:2015noa,Closset:2016arn,Benini:2016hjo,Closset:2017zgf,Closset:2017bse,Closset:2018ghr}.  These works are related to the gauge-Bethe correspondence of \cite{Nekrasov:2009uh}, and can be described as a ``higher dimensional A-twist'' \cite{Closset:2017zgf}.  

On general grounds, we expect such an observable to be computed by an appropriate 2d TQFT, which tightly constrains the form the result may take.   In fact, by studying the effective action of the low energy theory compactified to 2d, we may express the full answer in terms of two functions, the ``effective twisted superpotential,'' $\cW(\sigma)$ and the ``effective dilaton,'' $\Omega(\sigma)$, which are functions of the twisted chiral field strength multiplet, $\Sigma$, associated to the 2d vector multiplet, $\cV$,
\be \Sigma = -i D_- \tilde{D}_+ \cV = \sigma + i \sqrt{2} (\theta^+ \tilde{\Lambda}_+ - \tilde{\theta}^- \Lambda_-) + \sqrt{2} \theta^+ \tilde{\theta}^- (D - i F_{12}) + \cdots \,. \ee
Then the effective action can be written in terms of these objects as \cite{Nekrasov:2014xaa}
\be S = \int d^2 x \sqrt{g} \bigg( - 2 F_{12}^a \frac{\partial \cW(\sigma)}{\partial \sigma_a} + \tilde{\Lambda}^a \Lambda^b \frac{\partial^2 \cW(\sigma)}{\partial \sigma_a \partial \sigma_b}  + \frac{i}{2} \Omega(\sigma) R \bigg)  + {\cal Q}( \cdots) \,.\ee
Here,  $\sigma$ may denote both dynamical and background vector multiplets associated to the gauge group, $G$, and flavor group, $G_F$, respectively, expanded in a Cartan basis.  For the rest of the paper, we set the notation
\be \sigma\;\;  \rightarrow  \;\;\;\;\;\; \{ \;\;u_a,\;\;\; a=1,...,r_G, \;\;\;\;\;\;\; \nu_i, \;\;\; i=1,...,r_{G_F} \;\; \} \,,\ee
defining the gauge and flavor symmetry parameters, $u$ and $\nu$, respectively.

Using this low energy action, we may construct the partition function of this effective theory on a Riemann surface $\Sigma_\g$ with the topological A-twist background,  where we include twisted masses $\nu$ and background magnetic fluxes $\n$ for the flavor symmetry.  We find \cite{Nekrasov:2014xaa,Closset:2017zgf}
\be \label{BEsum} Z_{\cM_{d-2} \times \Sigma_\g}(\nu)_{\n}   = \sum_{\hat{u} \in \cS_{\mathit{BE}}} \Pi_i(\hat{u},\nu)^{\n_i} \cH(\hat{u},\nu)^{\g-1} \,,\ee
where
\be \Pi_i(u,\nu) = \exp \bigg(2 \pi i \frac{\partial \cW}{\partial \nu_i} (u,\nu)\bigg), \;\;\;\; \cH(u,\nu) = e^{2 \pi i \Omega(u,\nu)} \det_{a,b} \frac{\partial^2 \cW(u,\nu)}{\partial u_a \partial u_b}\,, \ee
which we refer to as the ``flux operator'' and ``handle-gluing operator,'' respectively,
and $\cS_{\mathit{BE}}$ is the set of supersymmetric ``Bethe vacua'' of the theory, defined by
\be \label{SBEdef}\cS_{\mathit{BE}} = \;\; \left \{ \hat{u} \;\; \big| \;\; \Pi_a(\hat{u},\nu) \equiv \exp \bigg( 2 \pi i \frac{\partial \cW}{\partial u_a}(\hat{u},\nu)\bigg) = 1, \;\;\; a=1,...,r_G\right\} /W_G\,,\ee
where we quotient by the action of the Weyl group, $W_G$, of $G$ and discard any solutions on which it does not act freely.

This result may often be alternatively derived by direct UV localization.  There, one arrives at an expression of the form
\be
\label{jkformula} 
Z_{\cM_{d-2} \times \Sigma_\g}(\nu)_{\n} = \frac{1}{|W_G|} \sum_{\m \in \Lambda_G} \oint_{\cC_{JK}} du\; \Pi_i(u,\nu)^{\n_i} \Pi_a(u,\nu)^{\m_a} \cH(u,\nu)^\g e^{-2 \pi i \Omega(u,\nu)}\,, 
\ee
where we sum over gauge fluxes, $\m$, in the lattice, $\Lambda_G$, of coweights of $G$, and $\cC_{JK}$ is the so-called ``Jeffrey-Kirwan contour'' \cite{JK1995,Benini:2013xpa}.  Roughly speaking, we may define this contour by removing a small neighborhood of any singularities or boundaries at infinity in the integrand, leaving a non-singular region, $\hat{\cM}$, of the complexified Cartan of $G$, and then the contour runs over those portions of $\partial \hat{\cM}$ with
\be \sgn \; \text{Im}(\partial_a \cW(u,\nu)) =- \sgn(\eta_a)\,, \ee
where $\eta_a$ is an auxiliary parameter valued in the Cartan of the gauge group.  In addition, we discard contributions from the neighborhood of points with enhanced Weyl symmetry.  Then, if we choose $\eta_a \propto \m_a$ with a positive constant, one can check the sum over $\m_a$ is a convergent geometric series on the JK contour, and after performing it we find
\be Z_{\cM_{d-2} \times \Sigma_\g}(\nu)_{\n} = \frac{1}{|W_G|} \oint_{\partial{\hat{\cM}}} du\; \Pi_i(u,\nu)^{\n_i} \prod_a \frac{1}{1-\Pi_a(u,\nu)} \cH(u,\nu)^\g e^{-2 \pi i \Omega(u,\nu)}\,. \ee
The poles in this integrand are precisely at the solutions to \eqref{SBEdef} (appearing with multiplicity $|W_G|$, which cancels the prefactor), and taking their residues we recover the formula \eqref{BEsum}, demonstrating the equivalence of these two approaches.   We refer to \cite{Closset:2017zgf} for more details on this argument and the JK contour. 

Thus, to derive the partition function on $\cM_{d-2} \times \Sigma_\g$ for general $\Sigma_\g$, it suffices to compute the effective objects, $\cW(u,\nu)$ and $\Omega(u,\nu)$, describing the low energy theory obtained after compactifying on $\cM_{d-2}$, which in the case of present interest is $S^3_b$.  In the next subsection we attempt to carry out this procedure directly by expanding the 5d fields in KK modes on $S^3_b$.  However, the result we obtain in this way turns out to encode only the perturbative contribution to $Z_{S^3_b \times \Sigma_\g}$.  We then describe, in Section \ref{sec:5d4d}, another approach which captures the full non-perturbative twisted superpotential and dilaton.

\subsection{\texorpdfstring{Reduction on $S^3_b$ and the perturbative partition function}{Reduction on S³b and the perturbative partition function}}\label{sec:amodreduc}

Let us consider a 5d $\cN=1$ theory on $S^3_b \times \R^2$.  As discussed above, this may be described by an effective 2d $\cN=(2,2)$ theory with infinitely many fields, arising from the KK modes on $S^3_b$.  We begin by describing this in the case of a free 5d $\cN=1$ hypermultiplet.

\paragraph{Hypermultiplet.}

To write the effective twisted superpotential and dilaton generated by a hypermultiplet, we will first need to describe how its modes on $S^3_b \times \R^2$ decompose into 2d fields on $\R^2$.  Recall that the dimensional reduction of a 5d hypermultiplet to three dimensions is a 3d $\cN=4$ hypermultiplet, or equivalently, a pair of 3d $\cN=2$ chiral multiplets in conjugate representations.  Thus, if we restrict to a point, $x$, on $\R^2$, the field content of the 5d hypermultiplet on the three transverse directions is that of a 3d $\cN=4$ hypermultiplet.  We may then understand the 2d field content by expanding this 3d hypermultiplet in modes on $S^3_b$.

\begin{flushleft}
	{\it Round sphere}
\end{flushleft}

For simplicity, let us first consider the round sphere, $b=1$.  Following \cite{Drukker:2012sr}, we may decompose a chiral multiplet on $S^3$ into modes via
\be \phi = \sum_{\ell,m,n} \phi_{\ell,m,n} Y_{\ell,m,n}, \;\;  \psi = \sum_{\ell,m',n} \psi^+_{\ell,m',n} \chi^+_{\ell,m',n} + \sum_{\ell,m,n'} \psi^-_{\ell,m,n'} \chi^-_{\ell,m,n'}, \;\; F = \sum_{\ell,m,n} F_{\ell,m,n} Y_{\ell,m,n},\ee
where $Y_{\ell,m,n}$ run over the spherical harmonics on $S^3$, with $\ell \in \Z_{\geq 0}$ and $m$, $n \in \{-\frac{\ell}{2},...,\frac{\ell}{2} \}$ the angular momenta under the $SU(2)_L \times SU(2)_R \cong SO(4)$ isometry group, and similarly for the spinor spherical harmonics, $\chi^\pm$.\footnote{Specifically, there $m'$ and $n'$ take values in $\{-\frac{\ell+1}{2},...,\frac{\ell+1}{2} \}$, representing the decomposition of the spinors into representations $\bigoplus_{\ell \geq 0} (\ell,\ell+\frac{1}{2}) \oplus (\ell+\frac{1}{2},\ell)$ of $SU(2) \times SU(2)$.} Then these modes organize into the following  short and long 0d $\cN=1$ multiplets of the action of the supercharges $\delta$ and $\tilde{\delta}$:
\begin{align}
\text{short:}&           & \qquad &\{ \phi_{\ell,\frac{\ell}{2},n}, \;\; \psi^+_{\ell,\frac{\ell+1}{2},n} \}\,,  \qquad  \{ \psi^{-}_{\ell,-\frac{\ell+1}{2},n}, \;\; F_{\ell,-\frac{\ell}{2},n} \}\,, \\
\text{long:}&        &   &\{ \phi_{\ell,m,n}, \;\; \psi^+_{\ell,m+\frac{1}{2},n}, \;\; \psi^-_{\ell,m+\frac{1}{2},n}, \;\; F_{\ell,m+1,n} \}    \,, \qquad -\frac{\ell}{2} \leq m <  \frac{\ell}{2}\,.
\end{align}
We expect contribution from long multiplets cancels out of protected observables, and so we will henceforth focus on the short multiplets.

For a 3d $\cN=4$ hypermultiplet, which consists of two 3d $\cN=2$ multiplets, $\Phi$ and $\tilde{\Phi}$ in conjugate gauge and flavor representations, we may organize the short $\cN=1$ multiplets into $\cN=2$ chiral multiplets, for $\ell \in \Z_{\geq  0}$ and $n \in \{-\frac{\ell}{2},...,\frac{\ell}{2} \}$,
\eqs{ \Phi^+_{\ell,n} \equiv&\, \{ \phi_{\ell,\frac{\ell}{2},n}, \;\; \psi^+_{\ell,\frac{\ell+1}{2},n}, \tilde{\psi}^{-}_{\ell,-\frac{\ell+1}{2},n}, \;\; 
	\tilde{F}_{\ell,-\frac{\ell}{2},n} \}  \,,\\ 
	\Phi^-_{\ell,n} \equiv&\,\{ \tilde{\phi}_{\ell,\frac{\ell}{2},n}, \;\; \tilde{\psi}^+_{\ell,\frac{\ell+1}{2},n}, {\psi}^{-}_{\ell,-\frac{\ell+1}{2},n}, \;\; {F}_{\ell,-\frac{\ell}{2},n} \} \,.}
Then one can check that, for a hypermultiplet with R-charge $1$ coupled to a background gauge field with scalar $\tsigma$, one has
\be \label{comm} [\delta, \tilde{\delta}] \Phi^\pm_{\ell,n} = \(\ell+1 \pm \tsigma \) \Phi^\pm_{\ell,n}\,. \ee

The above analysis was carried out at a single point, $x \in \R^2$. In general, we obtain 2d superfields $\Phi^\pm_{\ell,n}(x)$ which, as discussed above, form multiplets of the 2d $\cN=(2,2)$ superalgebra. Specifically, it is easy to see that each mode gives rise to an independent 2d chiral multiplet, and from \eqref{comm} and \eqref{ZS3}, we can see that these chiral multiplets have twisted masses
\be \label{Phipmb1} \tilde{m}^\pm_{\ell,n} = \ell+1 \pm \tsigma \,. \ee
In addition, we have modes which contribute long multiplets in 2d, which have unprotected masses, and which do not contribute to protected observables.

We may now compute the effective twisted superpotential contributed by integrating out this compactified 5d hypermultiplet.  Recall that the contribution of a single 2d chiral multiplet of twisted mass $\tilde{m}$ is \cite{Witten:1993yc}\footnote{More precisely we should introduce a dynamical scale, $\mu$, and write the logarithm as $\log\big(\frac{\tilde{m}}{\mu}\big)$, but this would drop out of the computation below.}
\be \cW_{\R^2}(\tilde{m}) =- \frac{1}{2 \pi i} \tilde{m} (\log \tilde{m}-1)\,. \ee
The full twisted superpotential from summing over all the short multiplets appearing in \eqref{Phipmb1} is
\be 
\cW^{hyp,U(1)}_{S^3_{b=1} \times \R^2}(\tsigma) =  \sum_{\ell=0}^\infty \sum_{\pm} (\ell+1) \cW^\Phi_{\R^2}\( (\ell+1) \pm \tsigma \) \equiv g_{b=1}(\tsigma) \,.
\ee
After suitable regularization (see \eqref{defgb} below), this infinite sum may be explicitly evaluated to give
\be g_{b=1}(\tsigma) = -2 \frac{1}{(2 \pi i)^3} \text{Li}_3(e^{2 \pi i \tsigma}) + \tsigma \frac{1}{(2 \pi i)^2} \dilog(e^{2 \pi i \tsigma}) + \frac{1}{24}\tsigma(2\tsigma^2-1)\,. \ee 

\begin{flushleft}
	{ \it Squashed sphere}
\end{flushleft}

The above computation is  generalized in a straightforward manner to $S^3_b$ for general $b$ by expanding the fields in spherical harmonics, following \cite{Imamura:2011wg}.\footnote{Here we find it convenient to use a slightly non-standard normalization of the scalar $\tsigma$.  Our normalization is related to the usual scalar, $\sigma$, on $S^3_b$ by $\sigma= -iQ \tsigma$, where $Q$ is as in \eqref{Qdef}.  We will comment more on this relation below.}  One finds,
\be \label{comm2} [\delta, \tilde{\delta}] \Phi^\pm_{\ell,n} = \left\{ \ell+1 + \frac{2(b-b^{-1})}{b+b^{-1}} n \pm \tsigma\right\} \Phi^\pm_{\ell,n} \,. \ee
A more general R-charge, $\Delta$, can be obtained by shifting, in our convention, $\tsigma \rightarrow \tsigma+ 1-\Delta$. 

As above, each of these modes corresponds to a 2d chiral multiplet, with twisted masses
\be \label{Phipmb} \tilde{m}^\pm_{\ell,n} = \ell+1+ \frac{2(b-b^{-1})}{b+b^{-1}}n \pm \tsigma \,. \ee
Then, the twisted superpotential obtained by integrating out these chirals reads
\be 
\cW^{hyp,U(1)}_{S^3_b \times \R^2}(\tsigma) =  \sum_{\ell=0}^\infty  \sum_{n=-\frac{\ell-1}{2}}^{\frac{\ell-1}{2}} \sum_{\pm} \cW^\Phi_{\R^2}\(  \ell+1+ \frac{2(b-b^{-1})}{b+b^{-1}}n \pm \tsigma \) \equiv g_b(\tsigma) \,,
\ee
which, after suitable regularization, defines a function $g_b(\tsigma)$.  We regularize this sum as follows.  Note that, using $e^{2 \pi i \partial_\tsigma \cW_{\R^2}(\tsigma)} = \tsigma^{-1}$, we may formally write
\eqs{\exp\bigg( 2\pi i \frac{\partial g_b(\tsigma)}{\partial \tsigma} \bigg) & = \prod_{\ell=0}^\infty  \prod_{n=-\frac{\ell-1}{2}}^{\frac{\ell-1}{2}} \frac{ \ell+1+ \frac{2(b-b^{-1})}{b+b^{-1}}n - \tsigma}{ \ell+1+ \frac{2(b-b^{-1})}{b+b^{-1}}n + \tsigma} \nn
	&=\prod_{j,k \geq 0} \frac{(j+\frac{1}{2}) b + (k+\frac{1}{2})b^{-1} -Q \tsigma}{(j+\frac{1}{2}) b + (k+\frac{1}{2})b^{-1} + Q \tsigma} \equiv s_b(-i Q \tsigma)\,,
	\label{dsdef}}
where we have defined
\be  \label{Qdef} Q = \tfrac{1}{2}(b+b^{-1}) \,, \ee
and $s_b(x)$ is the double sine function \cite{kurokawa1991,Hama:2011ea}, which can be rigorously defined as a meromorphic function of $\sigma$. 
We may then set
\be \label{defgb}
g_b(\tsigma) = \frac{1}{2 \pi i}\int d\tsigma \log s_b(-iQ \tsigma) \,.
\ee

Next consider the effective dilaton.  This depends on the choice of R-symmetry used to twist the theory on the Riemann surface, $\Sigma_\g$, and the contribution of a hypermultiplet will depend on its charge, $r$, under this R-symmetry.  We stress that $r$, which is in general integer-quantized due to the nontrivial flux, $\g-1$, through $\Sigma_\g$ for the R-symmetry background gauge field, is distinct from the R-charge appearing in the coupling to $S^3_b$, which we denote $\Delta$, and which in general is a real number.  Then the contribution of a single 2d chiral multiplet to the effective dilaton is:
\be \Omega_{\R^2}(\tilde{m})  = -(r-1) \log \tilde{m},\ee
and proceeding as above, we find:
\be \Omega^{hyp,U(1)}_{S^3_b \times \R^2}(\tsigma) = (r-1) \ell_b(\tsigma)\,,\ee
where we have defined
\be \label{elldef} \ell_b(\tsigma) \equiv \frac{1}{2 \pi i} \log s_b(-iQ \tsigma)\,. \ee
Below it will often be natural to take $r=1$, as this is the R-charge obtained by twisting by the $U(1)_R \subset SU(2)_R$ symmetry appearing in the $\cN=1$ superalgebra, and in this case the hypermultiplet has vanishing contribution to the effective dilaton.

We now have all the elements we need to compactify the $\R^2$ factor to $\Sigma_\g$ using the 2d A-twist.  Namely, from \eqref{BEsum} above we have, for a flux $\m$ on $\Sigma_\g$, 

\be \label{u1hyp} Z^{hyp,U(1)}_{S^3_b \times \Sigma_\g}(\tsigma)_\m = \Pi(\tsigma)^\m e^{2 \pi i (\g-1) \Omega(\tsigma)} = s_b(-iQ \tsigma)^{\m+(r-1)(\g-1)}\,. \ee

\paragraph{General gauge theory.}

It is straightforward to generalize this result to a set of hypermultiplets living in a representation $R = \oplus_i R_i$ of a gauge group $G$.  As above we take $\tsigma \rightarrow \tu$ and $\tnu$ for the gauge and flavor symmetry (real mass) parameters, respectively.  Then we have
\eqs{\cW^{hyp}_{S^3_b \times \R^2}(\tu,\tnu) =&\, \sum_{i} \sum_{\rho \in R_i} g_b(\rho(\tu) + \tnu_i)\,, \\
	\Omega^{hyp}_{S^3_b \times \R^2}(\tu,\tnu) =&\, \sum_{i} \sum_{\rho \in R_i}(r_i-1) \ell_b(\rho(\tu) + \tnu_i)\,, 
}
where we have introduced masses $\tnu_i$ and R-charges $r_i$ for the hypermultiplets.  In addition to the hypermultiplets, we expect a contribution to the 2d A-model from the vector multiplets and the classical action.  For the former, we write
\eqs{ \cW^{vec}_{S^3_b \times \R^2}(\tu) =&\, -\sum_{\alpha \in \mathit{Ad(G)'}} g_b( \alpha(\tu) + 1 )\,, \\ \nonumber
	\Omega^{vec}_{S^3_b \times \R^2}(\tu) =&\,- \sum_{\alpha \in \mathit{Ad(G)'}} \ell_b(\alpha(\tu)+1) \\ 
	=&\,- \sum_{\alpha>0} \frac{1}{2 \pi i} \log\left[ 2 \sin\(\pi b Q \alpha(\tu)\) 2 \sin\(\pi b^{-1} Q \alpha(\tu) \)\right] \,.
}
where the sums are over the set, $Ad(G)$, of roots of $G$, and the primes denote that we include only the non-zero roots in the sum.  This can be motivated as follows.  Given a $\cN=1$ 5d vector multiplet and an adjoint hypermultiplet of R-charge zero, we may give the latter an expectation value,  without breaking R-symmetry, and completely Higgsing the gauge group.  At low energies this leaves no degrees of freedom for either the vector or hypermultiplet  and thus their contributions should cancel out in the partition function.  This leads to the expression above.  We will also derive this by an alternate method in Section \ref{sec:5d4d} below.

Finally, the classical contribution to $\cW(u)$ and $\Omega(u)$ can in principle be computed directly, but we can also take a more indirect approach and compute them on the background $S^3_b \times \Sigma_\g$.  The contribution of the Yang-Mills term was computed in \cite{Kawano:2015ssa}, and we may similarly evaluate the CS contribution and find:
\be \label{s3bclass} Z_{S^3_b \times \Sigma_\g} = e^{\pi i Q^2 \text{Tr}_{\mathit CS} (\m \tu^2) -2 \pi i Q^2 \tgamma \m \tu}\,,\ee
where we defined\footnote{The 5d Yang-Mills coupling naturally appears in a complex combination with the 5d theta angle, $\theta_{5d}$, and we may naturally extend the definition of $\tgamma$ as $\tgamma = \frac{2 \pi i}{{g_5}^2} + \frac{\theta_{5d}}{2 \pi}$.  However, we will mostly restrict our attention to $\theta_{5d}=0$ in this paper, and do not write the theta angle explicitly below.}
\be \label{tgamdef} \tgamma = - \frac{2 \pi i}{{g_5}^2}\,. \ee 
From this and the general form of \eqref{BEsum}, we can read off
\be \label{wclass} \cW^{classical}_{S^3_b \times \R^2}(\tu) =- \frac{1}{2} Q^2 \tgamma \tu^2 - \frac{1}{6} Q^2\text{Tr}_{\mathit CS} (\tu^3), \;\;\;\;\;\; \Omega^{\mathit classical}_{S^3_b \times \R^2} = 0\,.  \ee

Putting these pieces together, we arrive at the following result for the twisted superpotential of a 5d $\cN=1$ theory
\be \label{Wpert} \cW^{\mathit pert}_{S^3_b \times \R^2}(\tu,\tnu) =  \cW^{\mathit hyp}_{S^3_b \times \R^2}(\tu,\tnu)  +  \cW^{\mathit vec}_{S^3_b \times \R^2}(\tu)  +  \cW^{\mathit classical}_{S^3_b \times \R^2}(\tu)\,,  \ee
and similarly for $\Omega^{\mathit pert}_{S^3_b \times \R^2}$, where, with foresight, we have labeled these as the perturbative contributions, for reasons that will be clear below.  

We may nevertheless proceed as in the discussion above in Section \ref{sec:2datwist} and write the answer for the (perturbative) partition function on $S^3_b \times \Sigma_\g$.  Before doing this, it will be convenient to define rescaled variables, which are normalized in a way which will look more natural in the context of $S^3_b$ partition function, namely,
\be u = - i Q \tu\,, \qquad  \nu = - iQ \tnu\,, \qquad \gamma = -i Q \tgamma \,. \ee
Then we can write the perturbative contribution to the $S^3_b \times \Sigma_\g$ partition function as
\be \label{ZpertSec2}
Z^{\mathit pert}_{S^3_b \times \Sigma_\g}(\nu)_\n = \sum_{\hat{u} \in \cS_{\mathit{BE}}} \Pi^{pert}_i(\hat{u},\nu)^{\n_i} \cH^{pert}(\hat{u},\nu)^{\g-1}\,,
\ee
where
\be \label{ZpertSec2Parts}
\Pi^{pert}_i(u,\nu) \equiv \prod_{\rho \in R_i} s_b(\rho(u) + \nu_i), \;\;\; \Pi^{pert}_a(\hat{u},\nu) \equiv \prod_i \prod_{\rho \in R_i } s_b(\rho(\hat{u}) + \nu_i)^{\rho_a} \prod_{\alpha \in \mathit{Ad(G)'}} s_b(\alpha(\hat{u}) -iQ)^{-\alpha_a} \,,  \nonumber \ee
\be \cH^{pert}(u) \equiv \prod_i \prod_{\rho \in R_i} (s_b(\rho(u) + \nu_i))^{r_i-1} \prod_{\alpha>0} \left[2 \sin\(\pi b \alpha(u)\) 2 \sin\(\pi b^{-1} \alpha(u) \)\right]^{-1} \det_{a,b}\frac{-Q}{2 \pi} \frac{\partial \log \Pi^{pert}_a}{\partial u_b} \,,\ee
and
\be \cS^{pert}_{\mathit{BE}} = \; \{ \; \hat{u} \;\; \big| \;\; \Pi_a(\hat{u},\nu)  = 1 , \;\;\; a=1,...,r_G \;\}/W_G \,. \ee
Alternatively, we may write this in terms of an integral over the JK contour, as in \eqref{jkformula}, 
\be \label{s3bpertjkform} Z^{\mathit pert}_{S^3_b \times \Sigma_\g}(\nu)_{\n} = (-iQ)^{-r_G} \frac{1}{|W_G|} \sum_{\m \in \Lambda_G} \oint du\; \Pi^{pert}_i(u,\nu)^{\n_i} \Pi^{pert}_a(u,\nu)^{\m_a} \cH^{pert}(u,\nu)^\g e^{-2 \pi i \Omega^{pert}(u,\nu)} \,.\ee

\paragraph{Alternative perspective: A direct sum of 3d theories.}

Before moving on to evaluate the validity of this formula more closely, let us mention an alternative perspective on the above result.  Let us again return to the case of a single hypermultiplet on $S^3_b \times \Sigma_\g$.  Then, rather than expanding in modes on $S^3_b$, we may expand in modes on $\Sigma_\g$.  After the topological twist we expect many of the modes to cancel, with the number that remain determined by the total flux felt by the chiral multiplet.  

The 2d index theorem states that for a Dirac fermion, $\chi$, the difference in the number of right- and left-moving fermionic zero modes on $\Sigma_{\fg}$ is given by
\equ{\label{indexth}
	n_{R}^{\chi}-n_{L}^{\chi}= -t_{\chi}\,\eta_{\Sigma}\,, \qquad  \eta_{\Sigma}=\begin{cases} 
		2 |\fg-1| & \text{for $\fg \neq  1$} \\
		1& \text{for $\fg=1$} 
	\end{cases}\,,
}
where $t_{\chi}$ is the charge of the corresponding  fermion in the background $F=T\text{dvol}(\Sigma_{\fg})$,  $T=\frac{\kappa}{2} \(T_{R}+  \m_{I} \,  T_{I}\)$, where $T_{R,I}$ are the R-symmetry and other possible background symmetry generators, respectively, and $\kappa=\{1,0,-1\}$ for $\fg=0, \;1,$ and $\fg>1$, respectively.  Applying this to a fermion with R-charge $(r-1)$ and gauge charge $1$, and denoting $\hat \m\equiv \m+(r-1)(\g-1)$,  with $\fm$ an integer-quantized gauge flux, gives the reduction rule
\be \label{2dIndexThm}\text{5d hyper of charge $1$} \rightarrow\left\{ \begin{array}{ccc} |\hat \m|  \; &\text{3d chirals of charge $1$} &\;\;\;
	\text{for $ \hat \m>0$} \\
	|\hat \m |  \; &\text{3d chirals of charge $-1$} &\;\;\; \text{for $\hat \m <0$} \end{array} \right.\,. \ee
Thus, since each chiral multiplet contributes a factor of $s_b(u)$ to the $S^3_b$ partition function, the latter may be written as (using also $s_b(-u) = s_b(u)^{-1}$)
\be Z_{S^3_b \times \Sigma_\g}^{hyp,U(1)}(u) = s_b(u)^{\m + (r-1)(\g-1)}\,, \ee
which  is precisely what we found in \eqref{u1hyp}.

More generally, given a gauge theory, we may consider the effective 3d field content obtained for each choice of flux, $\m \in \Lambda_G$, on the Riemann surface, $\Sigma_\g$.  This matter content describes an effective 3d $\cN=2$ theory, which we may denote $\cT^{(3d)}_{\Sigma_\g,\m}$.  Then we expect that, schematically, the full 5d theory compactified on $\Sigma_\g \times \R^3$ can be written as a direct sum of 3d theories
\be \cT^{(5d)}_{\Sigma_\g} = \bigoplus_{\m \in \Lambda_G} \cT^{(3d)}_{\Sigma_\g,\m} \,.\ee
If we take this relation literally, then we may also compute the $S^3_b \times \Sigma_\g$ partition function as
\be Z_{S^3_b \times \Sigma_\g} = \sum_{\m \in \Lambda_G} Z_{S^3_b}[\cT^{(3d)}_{\Sigma_\g,\m}] \,.\ee
We claim this is precisely the interpretation of the formula \eqref{s3bpertjkform}.  Namely, each summand in that formula takes the form of the integral of a product of double sine functions, and such an integral describes the $S^3_b$ partition function of a particular 3d $\cN=2$ theory, which we claim is precisely $\cT^{(3d)}_{\Sigma_\g,\m}$.

\subsection{Reduction to 4d and the instanton partition function}\label{sec:5d4d}

While the above two perspectives nicely complement each other, they also suffer from an important shortcoming.  In both approaches we considered the compactification of the manifold from $d=5$ to $d'=2$ (in the first case) or $d'=3$ (in the second).  This compactification can be viewed as studying the theory in a limit where the compactified manifold is very small, and deriving the effective dynamics on the directions that remain large.  While this is valid for any finite relative size of the $S^3_b$ and $\Sigma_\g$ factors, owing to the topological invariance along $\Sigma_\g$, the strict limit relevant to the compactification may not commute with the inclusion of non-perturbative effects, since in 5d these are associated to codimension-$4$ field configurations, namely, instantons.  As we will see below, instantons do indeed modify the dynamics of the compactified theory, and are important for describing the full effective twisted superpotential and dilaton which compute the $S^3_b \times \Sigma_\g$ partition function.  In this section, we describe an alternative method, based on reduction to 4d, which captures these non-perturbative contributions.  We obtain a perturbative piece which matches the result above but which is supplemented by an instanton contribution, which we write below.

To start, let us consider a 5d $\cN=1$ theory on $\R^4 \times S^1$.  This gives an effective 4d $\cN=2$ theory on $\R^4$, with towers of KK modes corresponding to the Fourier modes of $S^1$.  For example, a 5d $\cN=1$ hypermultiplet, $\Phi$, with real mass, $m$, gives rise to a tower of 4d hypermultiplets, $\Phi_n$, with complex masses
\be m_n = i m + A_5 + \frac{n}{r} \,,\ee
where $r$ is the radius of $S^1$, and $A_5$ is the holonomy of the background gauge field along the $S^1$.  Below we will mostly work in units with $r=1$, but it will sometimes be important to keep it as a free parameter.

For the purpose of computing the $S^3_b \times \Sigma_\g$ partition function, we may realize the $S^3_b$ factor as an $S^1$ fibration over $S^2$.  Then the partition function on this five-manifold is equivalent to a partition function on $S^2 \times \Sigma_\g$, where we insert one unit of flux on $S^2$ for the $U(1)_{\mathit{KK}}$ symmetry, corresponding to translations along the $S^1$ fiber.

For the rest of this subsection, let us restrict our attention to the case of $\Sigma_{\g=0} = S^2$, so that the underlying four-manifold is $S^2 \times S^2$.  The partition function of 4d $\cN=2$ gauge theories on this space was considered in \cite{Bawane:2014uka}, and so we may apply their results to gain another perspective on our computation.

\paragraph{The 4d computation.}

Let us briefly review the computation of the partition function on $S^2 \times S^2$, following \cite{Bawane:2014uka}.  Consider the following vector field on $S^2 \times S^2$, which generates an infinitesimal isometry
\be v = 2 \epsilon_1  (i z_1 \partial_{z_1} - i \bar{z}_1 \partial_{\bar{z}_1})+2 \epsilon_2  (i z_2 \partial_{z_2} - i \bar{z}_2 \partial_{\bar{z}_2})\,, \ee
where $z_i$ are the complex coordinates on each $S^2$ factor.  Then we may consider the equivariant deformation of the Donaldson-Witten topological twist on $S^2 \times S^2$ \cite{Witten:1988ze}.  The partition function can then be computed by equivariant localization, and reduces to a contribution from the four fixed points of the $U(1)_1 \times U(1)_2$ isometry above, namely, at the products of the poles of the two $S^2$ factors.  More precisely, one finds that the path integral localizes to the locus
\be [\Phi,\bar{\Phi}] =  [F,\Phi] = [F,\bar{\Phi}]=0, \;\;\;\;  \iota_v F = i d \Phi , \;\;\;  \iota_v D \bar{\Phi} = 0\,,\ee
where $\Phi$ and $F$ are the complex scalar and field strength, respectively, in the $\cN=2$ vector multiplet.  The first conditions imply we may take $\Phi, \bar{\Phi}$, and $F$ in the same Cartan subalgebra of the gauge group, $G$, with rank $r_{G}$.  The second and third then imply that we may write, in this Cartan basis,
\be \frac{F_a}{2 \pi i} = \m_{1,a} \omega_1 +  \m_{2,a} \omega_2 , \;\;\;\; \Phi_a = \tu_a + \frac{1}{2}\m_{1,a} \epsilon_1 h_1 + \frac{1}{2}\m_{2,a} \epsilon_2 h_2 , \;\;\; a=1,...,r_G \,,\ee
where $\tu_a \in \C$ parameterizes the allowed profiles of $\Phi$, $\m_{i,a} \in \Z$ label the GNO fluxes of the gauge field on the two $S^2$ factors, and $\omega_i$ and $h_i$ are the volume forms and height functions, respectively, on the two $S^2$ factors, \ie, $\omega_i = \frac{1}{4 \pi} \sin \theta_i d\theta_i d\phi_i$ and $h_i = \cos \theta_i$ in the usual coordinates on $S^2$.  In addition, the path integral receives contributions from point-like instantons localized at each of the fixed points.

Let us consider a 4d $\cN=2$ theory with gauge group $G$ and hypermultiplets in the representation $R = \bigoplus_{i=1}^s R_i$.  Then the partition function is given by an integral over the BPS locus above

\be Z_{S^2_{\epsilon_1} \times S^2_{\epsilon_2}}(\tnu,\tau)_{\n_1,\n_2} = \sum_{\m_1, \m_2} \oint d\tu\; \prod_{\ell} Z_{\R^2_{\epsilon_1^{(\ell)}} \times \R^2_{\epsilon_2^{(\ell)}}}(\tu^{(\ell)},\tnu^{(\ell)},\tau) \ee
Here $\tau$ is the complexified gauge coupling, $\tu=(\tu_a)$ and $\tnu=(\tnu_i)$ are complex scalar parameters associated to the gauge and flavor symmetry, respectively, $\m_{1,2}$ and $\n_{1,2}$ are the corresponding fluxes through the two $S^2$ factors, the index $\ell$ runs over the four fixed points, $\ell \in \{ nn, ns, sn, ss\}$, sitting at the north ($n$) and south ($s$) poles of the two spheres, and we have defined
\bea \label{ueldef} 
& \epsilon_1^{(\ell)} = \left\{ \begin{array}{cc} \epsilon_1\,, & \ell=nn \; \text{or} \; ns \\  -\epsilon_1\,, & \ell=sn \; \text{or} \; ss \end{array}\right. \qquad &\epsilon_2^{(\ell)} =
\left\{ \begin{array}{cc} \epsilon_2\,, & \ell=nn \; \text{or} \; sn \\  -\epsilon_2\,, & \ell=ns \; \text{or} \; ss \end{array}\right.  \\
& {\tu^{(\ell)}}_a = \tu_a + \frac{1}{2}(\m_{1,a} \epsilon_1^{(\ell)}+ \m_{2,a} \epsilon_2^{(\ell)}) \qquad\qquad  &{\tnu^{(\ell)}}_i = \tnu_i + \frac{1}{2}(\n_{1,i} \epsilon_1^{(\ell)}+ \n_{2,i} \epsilon_2^{(\ell)}). \eea
Finally, $Z_{\R^2_{\epsilon_1}\times\R^2_{\epsilon_2}}$ is the partition function on $\R^4$ in the $\Omega$-background \cite{Nekrasov:2002qd}, which describes the contribution in the local neighborhood of each fixed point.  It can be decomposed as
\be Z_{\R^2_{\epsilon_1}\times\R^2_{\epsilon_2}}(\tu,\tnu,\tau) = Z^{classical}_{\R^2_{\epsilon_1}\times\R^2_{\epsilon_2}}(\tu,\tau)
Z^{1-loop}_{\R^2_{\epsilon_1}\times\R^2_{\epsilon_2}}(\tu,\tnu) Z^{inst}_{\R^2_{\epsilon_1}\times\R^2_{\epsilon_2}}(\tu,\tnu,\tau)\,. \ee

Let us analyze each of these pieces in turn.  First, the classical contribution comes from the Yang-Mills term, and is given by
\be Z_{\R^2_{\epsilon_1}\times\R^2_{\epsilon_2}}^{classical}(\tu,\tau)= \exp \bigg(-\pi i \tau \text{Tr} \frac{\tu^2}{\epsilon_1 \epsilon_2}  \bigg) \,,\ee
where we define $\text{Tr}(\tu^2) = K^{ab} \tu_a \tu_b$, with $K^{ab}$  the Killing form.  Taking the contribution from the four fixed points using \eqref{ueldef}, we find
\be Z^{classical}_{S^2_{\epsilon_1} \times S^2_{\epsilon_2}}(\tu,\tau)_{\m_1,\m_2} \equiv \prod_{\ell} Z_{\R^2_{\epsilon^{(\ell)}_1} \times\R^2_{\epsilon^{(\ell)}_2}}^{classical}(\tu^{(\ell)}) = e^{-8 \pi i \tau K^{ab} \m_{1,a} \m_{2,b} }\,. \ee

For the perturbative and instanton part, it is useful to first write the contribution to the corresponding equivariant indices \cite{Gomis:2011pf,Kim:2012qf}.  The appropriate index for the vector multiplet is that for the so-called self-dual complex, and for the hypermultiplet it is the Dirac complex, where both are twisted by the vector bundle, $V$, in which the fields take values.  These can both be related to the index of the Dolbeault complex for $V$, $\text{ind}(\bar{\partial}_V)$, namely,
\be \label{indvechyp} \text{ind}_{self-dual} = \frac{1+e^{i (\epsilon_1+\epsilon_2)}}{2} \text{ind}(\bar{\partial}_V) , \;\;\;\; \text{ind}_{Dirac}  =  -e^{i \frac{\epsilon_1+\epsilon_2}{2}} \frac{e^{i \tnu}+e^{-i \tnu}}{2} \text{ind}(\bar{\partial}_V) \,. \ee
Thus, we first consider the Dolbeault index, $\text{ind}(\bar{\partial}_V)$.  This may be decomposed into a perturbative piece and an instanton contribution,
\be \label{inddv} \text{ind}(\bar{\partial}_V) \equiv  \text{ind}_{pert}(\bar{\partial}_V) + \text{ind}_{\mathit{inst}}(\bar{\partial}_V) \,. \ee
The former can be written for an arbitrary representation $R$ of an arbitrary gauge group $G$ as
\be \text{ind}_{pert}(\bar{\partial}_V) = \frac{\tr_R e^{i \tu}}{(e^{i \epsilon_1}-1)(e^{i \epsilon_2}-1)} \,.\ee
The form of the instanton contribution is more subtle, and depends in a detailed way on the representation and gauge group, with explicit expressions typically known only for the classical gauge groups.  For example, for the adjoint representation of $G=U(N)$, and working in the $k$ instanton sector, we have \cite{Nekrasov:2004vw,Kim:2012qf}
\be \label{indinst} \text{ind}_{\mathit{inst}}(\bar{\partial}_V) =-e^{-i\frac{\epsilon_1+\epsilon_2}{2}}\big(\tr_N (e^{i \tu})  \tr_{\bar{k}} (e^{i \phi} )+ c.c.\big)+(1-e^{-i\epsilon_1})(1-e^{-i \epsilon_2}) \tr_{adj_k}(e^{i \phi})\,, \ee
where $\phi$ are the equivariant parameters for the $U(k)$ symmetry acting on the $k$-instanton moduli space.

The $1$-loop determinant in a given instanton background is given by making the formal replacement,
\be \label{indrep} \text{ind} = \sum_\lambda w_\lambda e^{\lambda} \;\;\; \rightarrow \;\;\; \prod_\lambda \lambda^{w_\lambda}\,. \ee

Let us focus on the perturbative contribution first.  Let us choose $\delta_i \in \{ \pm 1 \}$ so that $|e^{i \epsilon_i \delta_i}|<1$ (we always assume $\epsilon_i \in \C \backslash \R$).  Then the natural expansion of the index of the Dolbeault complex is (taking a charge $1$ representation of $U(1)$ for simplicity)
\be \text{ind}_{pert}^{U(1)}(\bar{\partial}_V) =\delta_1 \delta_2 \sum_{k_1,k_2 \geq 0} e^{i \tu + i \delta_1 \epsilon_1\( k_1 + \tfrac{1-\delta_1}{2}\) + i \delta_2 \epsilon_2\( k_2 + \tfrac{1-\delta_2}{2}\) } \,,\ee
and so the corresponding $1$-loop determinant is, using \eqref{indrep},
\be Z^{pert,U(1),\bar{\partial}_V}_{\R^2_{\epsilon_1} \times \R^2_{\epsilon_2}} = \prod_{k_1,k_2 \geq 0}\( \tu + \delta_1 \epsilon_1\( k_1 + \tfrac{1-\delta_1}{2}\) +  \delta_2 \epsilon_2\( k_2 + \tfrac{1-\delta_2}{2}\) \)^{\delta_1 \delta_2} \,.\ee
If we consider instead the Dirac complex, corresponding to a hypermultiplet, we find, using \eqref{indvechyp}
\eqst{\label{u1hypr4}
	Z^{pert,U(1),hyp}_{\R^2_{\epsilon_1} \times \R^2_{\epsilon_2}} = \prod_{k_1,k_2 \geq 0} \Big[ \(\tu +  \delta_1 \epsilon_1\( k_1 + \tfrac{1}{2}\) + \delta_2 \epsilon_2\( k_2 + \tfrac{1}{2}\) \) \\
	\times \(-\tu + \delta_1 \epsilon_1\( k_1 + \tfrac{1}{2}\) + \delta_2 \epsilon_2\( k_2 + \tfrac{1}{2}\) \)\Big]^{-\delta_1 \delta_2/2}
}

Returning to $S^2_{\epsilon_1} \times S^2_{\epsilon_2}$, we may use the above result to compute the contributions from the four fixed points, identifying parameters as in \eqref{ueldef}.  Let us assume now for concreteness that $|e^{i \epsilon_i}|<1$.  Then one finds many of terms in the infinite products above cancel, and we are left with
\be Z^{pert,U(1),hyp}_{S^2_{\epsilon_1} \times S^2_{\epsilon_2}}(\tu)_{\m_1,\m_2} = \Big([\tu;\epsilon_1,\epsilon_2]_{\m_1,\m_2}[-\tu;\epsilon_1,\epsilon_2]_{-\m_1,-\m_2}  \Big)^{-1/2}\ee
where we defined
\be [\tu;\epsilon_1,\epsilon_2]_{\m_1,\m_2} \equiv \prod_{k_1=-\frac{1}{2}(|\m_1|-1)}^{\frac{1}{2}(|\m_1|-1)}\prod_{k_2=-\frac{1}{2}(|\m_2|-1)}^{\frac{1}{2}(|\m_2|-1)} (
\tu+\epsilon_1 k_1 + \epsilon_2 k_2)^{ \sgn(\m_1) \sgn(\m_2)} \,.\ee
We can simplify this using
\be [\tu;\epsilon_1,\epsilon_2]_{\m_1,\m_2} =  [\tu;\epsilon_1,\epsilon_2]_{-\m_1,-\m_2}=(-1)^{\m_1 \m_2} [-\tu;\epsilon_1,\epsilon_2]_{\m_1,\m_2}\,. \ee
Then we find\footnote{Here we have chosen a convenient overall phase, which was ambiguous in the above infinite products.  We similarly fix several such phases below.  It would be desirable to fix these phases more rigorously from first principles, perhaps by careful consideration of the 4d 't Hooft anomalies and the 5d parity anomaly, as discussed in \cite{Chang:2017cdx,Closset:2018TA}.}
\be Z^{pert,U(1),hyp}_{S^2_{\epsilon_1} \times S^2_{\epsilon_2}}(\tu)_{\m_1,\m_2} = [\tu;\epsilon_1,\epsilon_2]_{\m_1,\m_2}^{-1} \,.\ee

We note for later convenience that in the limit of vanishing equivariant parameters, $\epsilon_{1,2} \rightarrow 0$, we have
\be \label{epszero}  [\tu;\epsilon_1,\epsilon_2]_{\m_1,\m_2}\;\; \underset{\epsilon_2 \rightarrow 0}{\longrightarrow}  \;\; \prod_{k_1=-(|\m_1|-1)/2}^{(|\m_1|-1)/2}(
\tu+\epsilon_1 k_1)^{ \sgn(\m_1)\m_2} \;\; \underset{\epsilon_1 \rightarrow 0}{\longrightarrow}  \;\; \tu^{\m_1 \m_2}\,.\ee
The perturbative contribution for a more general hypermultiplet is then computed by
\be Z^{1-loop,hyp}_{S^2_{\epsilon_1} \times S^2_{\epsilon_2}}(\tu,\tnu)_{\m_1,\n_1;\m_2,\n_2} = \prod_i \prod_{\rho \in R_i}  [\rho(\tu) + \tnu_i;\epsilon_1,\epsilon_2]_{ \rho(\m_1)+{\n_1}_i,\rho(\m_2)+{\n_2}_i}^{-1} \ee
For the vector multiplet, a similar argument, using \eqref{indvechyp}, gives
\eqs{Z^{1-loop,vec}_{S^2_{\epsilon_1} \times S^2_{\epsilon_2}}(\tu)_{\m_1,\m_2} &= \prod_{\alpha \in Ad(G)'} \Big([\alpha(\tu);\epsilon_1,\epsilon_2]_{\alpha(\m_1)-1,\alpha(\m_2)-1}[\alpha(\tu);\epsilon_1,\epsilon_2]_{\alpha(\m_1)+1,\alpha(\m_2)+1} \Big)^{1/2} \nn
	&=\prod_{\alpha>0} [\alpha(\tu);\epsilon_1,\epsilon_2]_{\alpha(\m_1)-1,\alpha(\m_2)-1}[\alpha(\tu);\epsilon_1,\epsilon_2]_{\alpha(\m_1)+1,\alpha(\m_2)+1}
	\label{4dvecpert}}
Then the full perturbative contribution is
\be \label{4dperte1e2}
Z^{pert}_{S^2_{\epsilon_1} \times S^2_{\epsilon_2}}(\tu,\tnu,\tau)_{\m_1,\n_1;\m_2,\n_2}= Z^{classical}_{S^2_{\epsilon_1} \times S^2_{\epsilon_2}}(\tu,\tau)_{\m_1,\m_2} Z^{1-loop,vec}_{S^2_{\epsilon_1} \times S^2_{\epsilon_2}}(\tu)_{\m_1,\m_2}  Z^{1-loop,hyp}_{S^2_{\epsilon_1} \times S^2_{\epsilon_2}}(\tu,\tnu)_{\m_1,\n_1;\m_2,\n_2} \ee

The full nonperturbative expression is obtained by including the contribution from the remaining terms in \eqref{indinst}, which correspond to the instanton contribution.  The detailed form of the instanton contribution will depend on the gauge group and matter representation, but has the general form
\be Z^{inst}_{\R^2_{\epsilon_1} \times \R^2_{\epsilon_2}}(\tu,\tnu,\tau) = \sum_{k=0}^\infty z^k Z^{(k)}_{\R^2_{\epsilon_1} \times \R^2_{\epsilon_2}}(\tu,\tnu)\,, \ee
where we defined $z=e^{2 \pi i \tau}$, the classical contribution from the instanton action, and $Z^{(k)}_{\R^2_{\epsilon_1} \times \R^2_{\epsilon_2}}$ is the contribution to the $1$-loop determinant in the $k$-instanton background.  For example, in the $U(N)$ case, it is given by integration over the $U(k)$ equivariant parameters appearing in \eqref{indinst}, and can be expressed in some cases as a sum over $N$-colored Young diagrams \cite{Nekrasov:2002qd}.  We will discuss an explicit example when we consider the 5d uplift below.  

Putting these ingredients together, we may write
\be Z_{S^2_{\epsilon_1} \times S^2_{\epsilon_2}}(\tnu,\tau)_{\n_1,\n_2} = \frac{1}{|W_G|} \sum_{\m_1, \m_2} \oint d\tu\, Z^{pert}_{S^2_{\epsilon_1} \times S^2_{\epsilon_2}}(\tu,\tnu,\tau)_{\m_1,\n_1;\m_2,\n_2} \prod_{\ell} Z^{inst}_{\R^2_{\epsilon^{(\ell)}_1} \times \R^2_{\epsilon^{(\ell)}_2}}(\tu^{(\ell)},\tnu^{(\ell)},\tau)  \,.\ee

\paragraph{Uplifting to 5d.}

We may now uplift this result to our desired background, $S^3_b \times S^2$, by applying this computation to the effective 4d $\cN=2$ theory, $\tilde{\cT}^{(4d)}$, obtained by dimensional reduction of our chosen 5d theory, $\cT^{(5d)}$, as outlined above.

More precisely, our strategy will be to exhibit $S^3_b$ as an $S^1$ fibration over the topologically twisted $S^2$.  Let us first describe the case $b=1$.  Then it was shown in \cite{Closset:2017zgf} that the usual supersymmetric background on the round sphere is an $S^1$ fibration over the topological A-twist background on the $2$-sphere.  Here we must include one unit of flux for the connection fibering the $S^1$.  Then this background may be equivalently obtained by considering  the effective 2d theory obtained by dimensional reduction, which has a $U(1)_{\mathit{KK}}$ global symmetry corresponding to translations along the $S^1$ direction, and inserting a unit flux on $S^2$ for this global symmetry.  We will employ the same strategy here, this time with an additional $S^2$ factor in the geometry.  In other words, we consider the effective 4d $\cN=2$ theory obtained by dimensional reduction along the $S^1$ fiber.  Here we perform an ordinary topological twist along each $S^2$ factor, which corresponds to the $\epsilon_1,\epsilon_2 \rightarrow 0$ limit of the equivariant background discussed above.  Finally, we must turn on a unit flux for the $U(1)_{\mathit{KK}}$ symmetry along one of the $S^2$ factors.

In the case of a non-round sphere, $b \neq 1$, we will argue below that this can also be exhibited as an $S^1$ fibration over $S^2$, but now with a nonzero Omega-background on the $S^2$ \cite{Closset:2015rna}.  This will again fit into the framework described above, but now rather than taking both $\epsilon_1$ and $\epsilon_2$ to zero, we will keep a nonzero $\epsilon_1$, specifically, we find that the appropriate value is
\be \label{epsilonbdef} \epsilon_1 = \epsilon_b \equiv -\frac{2}{r}\;\frac{b-b^{-1}}{b+b^{-1}}\,, \ee
where $r$ is the $S^1$ radius, which we will often set to $1$ below.  Note this vanishes when we set $b=1$.  Thus, we arrive at the following schematic relation
\be \label{s3sgaslim} Z_{S^3_b \times S^2}^{\cT^{(5d)}}(\tnu)_{\n_1,\n_2} \cong \lim_{\epsilon \rightarrow 0} Z_{S^2_{\epsilon_b}\times S^2_{\epsilon}}^{\tilde{\cT}^{(4d)}}\(\tnu,\tnu_{\mathit{KK}}=\tfrac{1}{r}\)_{\n_1,\n_{1,\mathit{KK}}=1;\n_2,\n_{2,\mathit{KK}}=0}\,, \ee
where $\tilde{\cT}^{(4d)}$ is the effective 4d theory.  Here the arguments of the partition functions correspond to the mass and fluxes for flavor symmetries.  On the RHS, there is an additional flavor symmetry, the $U(1)_{\mathit{KK}}$ symmetry corresponding to translations along $S^1$, which we have assigned a mass $\frac{1}{r}$ and a unit flux on the first $S^2$ factor, which gives rise to the Hopf fibration of $S^3$ described above.  We will also see in a moment that the flux $\n_1$ can be absorbed into a shift of $\tnu$, as expected since there is no nontrivial $2$-cycle on $S^3$ on which to support a flux.  We will also demonstrate that the partition function can be written in terms of the 5d Nekrasov partition function, or more precisely, of the Nekrasov-Shatashvili limit of this partition function.

Let us now make these statements more precise.  As above, the integrand of the partition function factorizes into contributions from the perturbative (classical and $1$-loop) piece, and the instanton contribution.  First, the classical contributions on $S^3_b \times \Sigma_\g$ were already presented above in \eqref{s3bclass}, and we may use that result in the present case, $\g=0$, as well.

Next we have the perturbative and instanton contributions.  To compute these, let us return to the contribution to the Dolbeault index in 4d from a single fixed point, given by \eqref{inddv},
\be \label{inddv2} \text{ind}(\bar{\partial}_V) \equiv  \text{ind}_{pert}(\bar{\partial}_V) + \text{ind}_{\mathit{inst}}(\bar{\partial}_V) \,. \ee
Then when we add a tower of KK modes with masses $\frac{n}{r}$, $n \in \Z$, the index is modified to
\be \label{inddv5d} \text{ind}^{(5d)}(\bar{\partial}_V) \equiv   \sum_{n \in \Z} e^{\frac{i n}{r} }  \text{ind}(\bar{\partial}_V) \,,\ee
and similarly for the vector and hypermultiplet indices.  Then we expect that when we construct the $1$-loop determinants on $\R^4 \times S^1$, these will be related to those in 4d by, schematically,
\be Z^{1-loop}_{\R^2_{\epsilon_1} \times \R^2_{\epsilon_2}} = \prod_{\lambda} \lambda \;\;\; \Rightarrow \;\;\; Z^{1-loop}_{\R^2_{\epsilon_1} \times \R^2_{\epsilon_2} \times S^1} = \prod_{n \in \Z} \prod_{\lambda} \(\lambda + \frac{n}{r}\) \;\; \propto \; \; \prod_{\lambda} 2 \sin \pi r \lambda\,. \ee
This holds for both the perturbative contribution and the $1$-loop contribution in an instanton background, as discussed in \cite{Kim:2012qf} in the context of the $S^5$ partition function.

To see how this works in more detail, let us consider the perturbative contribution for a single hypermultiplet. Recall that the perturbative contribution of the hypermultiplet on $\R^2_{\epsilon_1} \times \R^2_{\epsilon_2}$ is given by \eqref{u1hypr4}, namely,
\eqs{Z^{pert,U(1),hyp}_{\R^2_{\epsilon_1} \times \R^2_{\epsilon_2}} &= \prod_{k_1,k_2 \geq 0} \Big[\!\(\tu +  \delta_1 \epsilon_1( k_1 + \tfrac{1}{2}) +  \delta_2 \epsilon_2( k_2 + \tfrac{1}{2}) \) \nn
	&\qquad\qquad ×\(-\tu + \delta_1 \epsilon_1( k_1 + \tfrac{1}{2}) +  \delta_2 \epsilon_2( k_2 + \tfrac{1}{2}) \)\!\Big]^{-\delta_1 \delta_2/2}\,.
	\label{u1hypr4b}}
The uplift to $\R^2_{\epsilon_1} \times \R^2_{\epsilon_2} \times S^1$ is then given by\footnote{Here we fix the overall normalization arising from regularizing the product over $n$ for later convenience.  As mentioned above, it would be interesting to fix this from first principles.}
\eqs{Z^{pert,U(1),hyp}_{\R^2_{\epsilon_1}  \times \R^2_{\epsilon_2} \times S^1}(\tu) &= \prod_{n \in \Z} \prod_{k_1,k_2 \geq 0} \Big[\!\(\tu + \tfrac{n}{r} +  \delta_1 \epsilon_1( k_1 + \tfrac{1}{2}) +  \delta_2 \epsilon_2( k_2 + \tfrac{1}{2}) \) \nn
	&\qquad\qquad ×\(-\tu - \tfrac{n}{r}+ \delta_1 \epsilon_1( k_1 + \tfrac{1}{2}) +  \delta_2 \epsilon_2( k_2 + \tfrac{1}{2}) \)\!\Big]^{-\delta_1 \delta_2/2} \nn
	&= \prod_{k_1,k_2 \geq 0} \Big[ 2i \sin\!\left[ \pi r\(\tu +  \delta_1 \epsilon_1( k_1 + \tfrac{1}{2}) +  \delta_2 \epsilon_2( k_2 + \tfrac{1}{2}) \)\right] \nn
	&\qquad\qquad ×2i \sin\!\left[ \pi r\(-\tu + \delta_1 \epsilon_1( k_1 + \tfrac{1}{2}) +  \delta_2 \epsilon_2( k_2 + \tfrac{1}{2}) \)\right]\!\Big]^{-\delta_1 \delta_2/2} \,.
	\label{u1hypr4s1}}
Note this is naturally a function of the parameters
\be \label{xqidef} x=e^{2 \pi i r \tu}, \qquad  \q_i = e^{2 \pi i r \epsilon_i}\,. \ee
We will sometimes emphasize this by writing the 5d partition function as
\be Z_{\R^2_{\q_1}  \times \R^2_{\q_2}\times S^1}(x) \,,\ee
where the meaning of the subscripts and arguments should be clear from context.  Now we may glue four copies of this function with parameters identified as in \eqref{ueldef}, and one finds that the infinite product simplifies to
\be \label{zs2s2s1pert} Z^{pert,U(1),hyp}_{S^2_{\epsilon_1} \times S^2_{\epsilon_2} \times S^1}(\tu)_{\m_1,\m_2} = \(x;\q_1,\q_2\)_{\m_1,\m_2}^{-1}\,, \ee
where the parameters are as in \eqref{xqidef}, and we have defined
\be
(x;\q_1,\q_2)_{\m_1,\m_2} = \prod_{\substack{k_1=\\ -\frac{1}{2}(|\m_1|-1)}}^{\frac{1}{2}(|\m_1|-1)} \prod_{\substack{k_2=\\ -\frac{1}{2}(|\m_2|-1)}}^{\frac{1}{2}(|\m_2|-1)} \(x^{1/2} {\q_1}^{k_1/2} {\q_2}^{k_2/2}-x^{-1/2} {\q_1}^{-k_1/2} {\q_2}^{-k_2/2}\)^{\sgn(\m_1) \sgn(\m_2)}.
\ee
This result may also be obtained by directly uplifting the perturbative contribution,
\eqs{ \nonumber Z^{pert,U(1),hyp}_{S^2_{\epsilon_1} \times S^2_{\epsilon_2} \times S^1}(\tu)_{\m_1,\m_2} =&\, \prod_{n \in \Z}  Z^{pert,U(1),hyp}_{S^2_{\epsilon_1} \times S^2_{\epsilon_2}}\(\tu+ \tfrac{n}{r}\)_{\m_1,\m_2} =\prod_{n \in \Z} \left[\tu+\tfrac{n}{r};\epsilon_1,\epsilon_2\right]_{\m_1,\m_2}^{-1} \\ \label{s2s2s1pertdir}
	=&\, \(x;\q_1,\q_2\)_{\m_1,\m_2}^{-1} \,.
}

The above computation represents the $1$-loop determinant on the background $S^2_{\epsilon_1} \times S^2_{\epsilon_2} \times S^1$, where the $S^1$ appears as as trivial product.  We will return to this case in Section \ref{sec:AdS6} below, but our present interest is in the partition function on $S^3_b \times S^2$.  Then we must include a magnetic flux for fields charged under the $U(1)_{\mathit{KK}}$ symmetry.  Recall that the magnetic flux enters the fixed point contributions on $S^2_{\epsilon_1} \times S^2_{\epsilon_2}$ by shifting the eigenvalues by $\frac{\m_i \epsilon_i}{2}$.  Thus, including a unit $U(1)_{\mathit{KK}}$ flux, the perturbative contribution of the hypermultiplet in \eqref{u1hypr4s1} is modified to
\eqs{Z^{pert,U(1),hyp}_{\R^2_{\epsilon_1} \times \R^2_{\epsilon_2} \times S^1}(\tu) &= \prod_{n \in \Z} \prod_{k_1,k_2 \geq 0} \Big[\!\(\tu + \tfrac{n}{r} + \tfrac{n \epsilon_1}{2}+  \delta_1 \epsilon_1( k_1 + \tfrac{1}{2}) +  \delta_2 \epsilon_2( k_2 + \tfrac{1}{2})\) \nn
	&\qquad\qquad \times \(- \tu - \tfrac{n}{r} -\tfrac{n \epsilon_1}{2}+ \delta_1 \epsilon_1( k_1 + \tfrac{1}{2}) +  \delta_2 \epsilon_2( k_2 + \tfrac{1}{2}) \)\!\Big]^{-\delta_1 \delta_2/2} \nn
	&=\prod_{k_1,k_2 \geq 0} \Big[2 i \sin\!\left[\pi \tilde{r}\(\tu +  \delta_1 \epsilon_1( k_1 + \tfrac{1}{2}) +  \delta_2 \epsilon_2( k_2 + \tfrac{1}{2}) \)\right] \nn
	&\qquad\qquad \times 2i \sin\!\left[\pi \tilde{r} \(-\tu + \delta_1 \epsilon_1( k_1 + \tfrac{1}{2}) +  \delta_2 \epsilon_2( k_2 + \tfrac{1}{2}) \)\right]\!\Big]^{-\delta_1 \delta_2/2},
	\label{u1hypr4s1fib}}
where we defined
\be \tilde{r} = \frac{r}{1+\frac{r \epsilon_1}{2}} \,.\ee
This is the same function we obtained in \eqref{u1hypr4s1} above, but now evaluated at the arguments
\be \label{xqideffib} x'=e^{2 \pi i \tilde{r} \tu}\,, \qquad \q_i' = e^{2 \pi i \tilde{r} \epsilon_i} \,.\ee

Now to construct the full $S^3_b \times S^2$ partition function, we must include the contribution from the four fixed points, that is (writing this now as a function of $x$ and the $\q_i$),
\be Z^{pert,U(1),hyp}_{S^3_b \times S^2}(x)_{\m_1,\m_2} = \prod_{\ell} Z^{pert,U(1),hyp}_{\R^2_{\q_1^{(\ell)}}  \times \R^2_{\q_2^{(\ell)}} \times S^1}(x^{(\ell)}) \,,\ee
where the parameters at the four fixed points can be read off from \eqref{ueldef} and \eqref{xqideffib}, and we find, for vanishing fluxes, and renaming $\epsilon_2 \rightarrow \epsilon$ as in \eqref{s3sgaslim},
\bea \label{ueldef5d2} 	 & x^{(\ell)} = \left\{ \begin{array}{cc} e^{2 \pi i \frac{r \tu}{1+\frac{1}{2}r \epsilon_b} }  = e^{2 \pi i b Q r \tu}  \equiv x \,, &\quad \ell=nn \; \text{or} \; ns \\ e^{2 \pi i \frac{r \tu}{1-\frac{1}{2}r \epsilon_b} }  = e^{2 \pi i b^{-1} Q r\tu}  \equiv \bar{x}\,, & \quad\ell=sn \; \text{or} \; ss \end{array}\right.&\\
&\q_1^{(\ell)} = \left\{ \begin{array}{cc} e^{2 \pi i \frac{r \epsilon_b}{1+\frac{1}{2}r \epsilon_b} }  = e^{-2 \pi i b^2 }  \equiv \q \,, &\quad \ell=nn \; \text{or} \; ns \\ e^{-2 \pi i \frac{r \epsilon_b}{1-\frac{1}{2}r \epsilon_b} }  = e^{-2 \pi i b^{-2}}  \equiv \bar{\q} \,,&\quad \ell=sn \; \text{or} \; ss \end{array}\right.\;\;\;\; \\
&\q_2^{(\ell)} =
\left\{ \begin{array}{cc} 
	e^{2 \pi i \frac{r\epsilon}{1+\frac{1}{2} r \epsilon_b} } = e^{2 \pi i b Q r \epsilon} \,,& \quad \ell=nn  \\
	e^{-2 \pi i \frac{r\epsilon}{1+\frac{1}{2} r \epsilon_b} } = e^{-2 \pi i b Q r \epsilon} \,,& \quad\ell=ns  \\
	e^{2 \pi i \frac{r\epsilon}{1-\frac{1}{2} r \epsilon_b} } = e^{2 \pi i b^{-1} Q r \epsilon}\,, &\quad \ell=sn  \\
	e^{-2 \pi i \frac{r\epsilon}{1-\frac{1}{2} r \epsilon_b} } = e^{-2 \pi i b^{-1} Q r \epsilon}\,, &\quad \ell=ss  \\
\end{array}\right. 
\eea
where we introduced the parameters
\be \label{qxdef} \q = e^{-2 \pi i b^2}\,, \quad \bar{\q} = e^{-2 \pi i b^{-2}}\,, \quad  x =e^{2 \pi \i b Q r \tu}\,, \quad \bar{x} = e^{2 \pi i b^{-1} Q r \tu}\,,  \ee 
and $Q=\frac{1}{2}(b+b^{-1})$ as usual.  We can then introduce fluxes, as above, by shifting $u^{(\ell)} \rightarrow u^{(\ell)}+ \frac{1}{2}(\m_1 \epsilon^{(\ell)}_1+ \m_2 \epsilon^{(\ell)}_2)$.  We will describe the dependence on fluxes in more detail below.

\paragraph{Perturbative contribution and holomorphic blocks.}

We now evaluate this perturbative contribution for the hypermultiplet explicitly using two methods.  First, we may directly uplift the perturbative contribution on $S^2_{\epsilon_1} \times S^2_{\epsilon_2}$, as in \eqref{s2s2s1pertdir}, but now including the $U(1)_{\mathit{KK}}$ flux.  We find
\eqs{Z^{pert,U(1),hyp}_{S^3_b \times S^2}(\tu)_{\m_1,\m_2} &= \prod_{n \in \Z}  Z^{pert,U(1),hyp}_{S^2_{\epsilon_b} \times S^2_{\epsilon}}\(\tu+ \frac{n}{r}\)_{\m_1+n,\m_2} = \prod_{n \in \Z} \left[\tu+\frac{n}{r};\epsilon_b,\epsilon\right]_{\m_1+n,\m_2}^{-1} \nn
	&\underset{\epsilon \rightarrow 0}{\longrightarrow} \;\; \prod_{n \in \Z} \prod_{\substack{k_1=\\-\frac{1}{2}(|\m_1+n|-1)}}^{\frac{1}{2}(|\m_1+n|-1)} \(\tu + \frac{n}{r} + k_1 \epsilon_b\)^{-\sgn(\m_1+n)\m_2},
}
where we used \eqref{epszero} to take the $\epsilon \rightarrow 0$ limit.  Defining $\ell=|\m_1 +n|$, this may be further rewritten as
\be Z^{pert,U(1),hyp}_{S^3_b \times S^2}(\tu)_{\m_1,\m_2} = \prod_{\ell=1}^{\infty} \prod_{k_1=-\frac{1}{2}(\ell-1)}^{\frac{1}{2}(\ell-1)} \bigg(\frac{\tu + \frac{-\ell-\m_1}{r} + k_1 \epsilon_b}{\tu + \frac{\ell-\m_1}{r} + k_1 \epsilon_b} \bigg)^{-\m_2}\,.  \ee
Finally, after substituting the definition of $\epsilon_b$ and defining $j=\ell-1+2k_1, k=\ell-1-2k_1$, which take values in $\Z_{\geq 0}$, and comparing to \eqref{dsdef}, we see that
\be Z^{pert,U(1),hyp}_{S^3_b \times S^2}(u)_{\m_1,\m_2} = s_b\(-i Q(r\tu- \m_1) \)^{\m_2} \,.\ee

We may take another approach to this computation which will give another useful perspective on this partition function.  Let us consider the contribution from the fixed points $\ell=nn$ and $\ell=ns$.  We also take $|\q| = |e^{-2 \pi i b^2}|<1$ for concreteness.  Then one finds the infinite products over $k_2$ in \eqref{u1hypr4s1fib} reduce to a finite product, and we can take a finite $\epsilon \rightarrow 0$ limit,
\eqs{Z^{pert,U(1),hyp}_{\R^2_{\epsilon_b}  \times \R^2_{\epsilon} \times S^1} Z^{pert,U(1),hyp}_{\R^2_{\epsilon_b}  \times \R^2_{-\epsilon} \times S^1} &=\prod_{k_1 \geq 0} \prod_{\mathclap{\substack{k_2= \\ -\frac{1}{2}(|\m_2|-1)}}}^{\frac{1}{2}(|\m_2|-1)}\!\! \(\frac{ \sin\!\left[ \pi \tilde{r}(\tu + \epsilon_b( k_1 + \frac{\m_1}{2}+ \frac{1}{2}) + \epsilon k_2)\right]}{\sin\!\left[ \pi \tilde{r}(-\tu + \epsilon_b( k_1 - \frac{\m_1}{2}+ \frac{1}{2}) + \epsilon k_2 )\right]} \)^{-\sgn(\m_2)/2} \nn
	& \underset{\epsilon \rightarrow 0}{\longrightarrow} \;\;\prod_{k_1 \geq 0} \bigg(\frac{\sin\!\left[\pi \tilde{r}\(\tu + \epsilon_b( k_1 + \frac{\m_1}{2}+ \frac{1}{2}) \)\right]}{ \sin\!\left[\pi \tilde{r}\(-\tu + \epsilon_b( k_1 - \frac{\m_1}{2}+ \frac{1}{2}) \)\right]} \bigg)^{-\m_2/2} \nn
	& \propto \(\frac{((-\q^{\frac{1}{2}})^{1+\m_1} x;\q)}{((-\q^{\frac{1}{2}})^{1-\m_1} x^{-1};\q)}\)^{-\m_2/2},
}
where $\q$ and $x$ are as in \eqref{qxdef}, and we introduced the $\q$-Pochhammer symbol
\be (x;\q) = \prod_{k \geq 0} (1-x \q^k)\,. \ee
A similar argument for the contribution from the fixed points $\ell=sn$ and $\ell=ss$ gives
\bea Z^{pert,U(1),hyp}_{\R^2_{\epsilon_b}  \times \R^2_{\epsilon} \times S^1} Z^{pert,U(1),hyp}_{\R^2_{\epsilon_b}  \times \R^2_{-\epsilon} \times S^1}  \rightarrow \(\frac{((-\bar{\q}^{\frac{1}{2}})^{-1-\m_1} \bar{x}^{-1};\bar{\q}^{-1})}{((-\bar{\q}^{\frac{1}{2}})^{-1+\m_1} \bar{x};\bar{\q}^{-1})}\)^{-\m_2/2} 	\,.\eea
Then, using the relation \cite{Beem:2012mb,Closset:2018ghr}

\be s_b(i r Q \tu) = e^{\frac{\pi i}{2} r^2 Q^2 \tu^2 - \frac{\pi i}{24}(b^2 + b^{-2})} \frac{\(-\q^{1/2} x;\q\)}{\(-\bar{\q}^{-1/2} \bar{x};\bar{\q}^{-1}\)}  = s_b(-i r Q\tu)^{-1}\,,\ee
where the parameters are as in \eqref{qxdef}, we again arrive at
\be Z^{pert,U(1),hyp}_{S^3_b \times S^2} = s_b(-i Q(r\tu-\m_1))^{\m_2} \,.\ee
This method demonstrates that the partition function of the hypermultiplet naturally factorizes into a contribution from the fixed points at the north and south poles of the $S^2$ base of $S^3_b$.

Note that the dependence on the flux $\m_1$ can be absorbed into a shift of $u$, as expected since the $S^3_b$ factor cannot support a topologically nontrivial flux.  Thus we will henceforth set $\m_1=0$, and rename the flux on $S^2$ by $\m_2 \rightarrow \m$.  
Then we see that this agrees with the result \eqref{u1hyp} derived above by KK reduction, where in the present case we are implicitly taking R-charge one for the hypermultiplet.  We may also define, as in Section \ref{sec:amodreduc}, the rescaled parameters
\be u = -i Q r \tu\,, \qquad \nu = - i Q r \tnu\,. \ee
We will sometimes work in terms of these parameters below.

It is straightforward to extend the argument above to a hypermultiplet in a general representation, giving
\be \label{zperthyp} Z^{pert,hyp}_{S^3_b \times S^2}(u,\nu)_{\m,\n} = \prod_i \prod_{\rho \in R_i} s_b\(\rho(u) + \nu_i\)^{\rho(\m) + \n_i}\,. \ee
Similarly, for the vector multiplet we find
\bea \label{zpertvec} Z^{pert,vec}_{S^3_b \times S^2}(u)_\m & = \prod_{\alpha >0} s_b\(-i Q( r\alpha(\tu)-1)\)^{-\alpha(\m)-1} s_b\(-i Q( r\alpha(\tu)+1)\)^{-\alpha(\m)+1} \\ 
& = \prod_{\alpha \in Ad(G)'} s_b\(\alpha(u) -i Q\)^{1-\alpha(\m)}\,.\eea
We note this perturbative piece precisely agrees with what we found by the na\"ive KK reduction of Section \ref{sec:amodreduc}.  This confirms our claim that the previous computation only reproduced the perturbative contribution to the $S^3_b \times \Sigma_\g$ partition function, and missed the instanton corrections.  The latter can be seen in the present approach, and we will discuss them below.

The factorization of the perturbative contribution noted above also holds for the general hypermultiplet and vector multiplet in \eqref{zperthyp} and \eqref{zpertvec}.  This is closely related to the factorization of the $S^3_b$ partition function of 3d $\cN=2$ theories into {\it holomorphic blocks} \cite{Beem:2012mb}.  It is therefore natural to expect that a similar approach as the one here applies to the spaces $L(p,q)_b \times S^2$ for arbitrary (squashed) lens space $L(p,q)$, as such lens spaces can be constructed by gluing two such holomorphic blocks.  Indeed, we have seen this already in the case of $S^2 \times S^2 \times S^1$, and will return to this example in Section \ref{sec:AdS6} below.  We also note that 3d holomorphic blocks have also appeared in 5d partition functions in other contexts -- see, \eg, \cite{Nieri:2013yra,Pasquetti:2016dyl} -- and it would be interesting to understand the relation to their appearance here.  As we will see in the next subsection, this factorization continues to hold at the non-perturbative level.

\paragraph{Instanton contribution and the Nekrasov-Shatashvili limit.}

The argument above extends to the $1$-loop determinant of a general hypermultiplet or vector multiplet in a general instanton background.  Specifically, we may similarly uplift the contribution in the background of point-like instantons on $\R^2_{\epsilon_1} \times \R^2_{\epsilon_2}$, which are now configurations supported on loops wrapping the $S^1$ factor.  Thus, we may write
\be Z_{\R^2_{\q_1} \times \R^2_{\q_2} \times S^1}\(x,y,z\) = Z^{pert}_{\R^2_{\q_1} \times \R^2_{\q_2} \times S^1}\(x,y,z\)\sum_{k=0}^\infty z^k Z^{(k)}_{\R^2_{\q_1} \times \R^2_{\q_2} \times S^1}\(x,y\) \,,\ee
where $Z^{pert}_{\R^2_{\q_1} \times \R^2_{\q_2} \times S^1}$ is the perturbative contribution from vector and hypermultiplets, and $Z^{(k)}_{\R^2_{\q_1} \times \R^2_{\q_2} \times S^1}$ is the $1$-loop determinant in the $k$-instanton background.  We have also defined the classical contribution to the instanton action, which is now integrated over $S^1$,
\be z = e^{2 \pi i r \tgamma} \,,  \qquad \tgamma =- \frac{2 \pi i}{{g_5}^2} \,.\ee
This 5d instanton partition function was originally defined in \cite{Nekrasov:2002qd}, and plays an important role in many of the 5d partition functions mentioned at the beginning of this section.  We see the same object controls the $S^3_b \times S^2$ partition function.

To form the integrand of the $S^3_b \times S^2$ partition function, we take the product of the contributions from the four fixed points,
\be \label{s3s2inst} \text{lim}_{\epsilon \rightarrow 0} \prod_\ell Z_{\R^2_{\q^{(\ell)}_1} \times \R^2_{\q^{(\ell)}_2} \times S^1}(x^{(\ell)},y^{(\ell)},z^{(\ell)}) \,,\ee
where the arguments of the instanton partition functions at the four fixed points are $x^{(\ell)}$ and $\q_i^{(\ell)}$, defined as in \eqref{ueldef5d2}, and $y^{(\ell)}$ and $z^{(\ell)}$, defined analagously. Explicitly, 
\be \label{yzdef} y^{(\ell)} = \left\{ \begin{array}{cc}  e^{2 \pi i b Q r \tnu}  \equiv y\,, &\quad \ell=nn \; \text{or} \; ns \\  e^{2 \pi i b^{-1} Q r \tnu}  \equiv \bar{y} \,,& \quad\ell=sn \; \text{or} \; ss \end{array}\right. \;\;\;\;\; z^{(\ell)} = \left\{ \begin{array}{cc}  e^{2 \pi i b Q r \tgamma}  \equiv z \,,&\quad \ell=nn \; \text{or} \; ns \\  e^{2 \pi  i b^{-1} Q r \tgamma}  \equiv \bar{z} \,,& \quad\ell=sn \; \text{or} \; ss \end{array}\right. \,.\ee

The expression \eqref{s3s2inst} involves taking the limit of the instanton partition function where one of the equivariant parameters is sent to zero.  This limit has been widely studied, beginning with \cite{Nekrasov:2009rc}, and is known as the {\it Nekrasov-Shatashvili limit} of the instanton partition function.  As shown in \cite{Nekrasov:2009rc}, the leading behavior of the instanton partition function in this limit can be expanded as (considering the 4d case first)
\be Z_{\R^2_{\epsilon_1}\times\R^2_{\epsilon_2}}(\tu) \underset{\epsilon_2 \rightarrow 0}{\longrightarrow} \exp \bigg\{2 \pi i \( \frac{1}{\epsilon_2} \cW_{NS}(\tu,\epsilon_1) -\frac{1}{2} \Omega_{NS}(\tu,\epsilon_1) + \mathcal O(\epsilon_2) \) \bigg\} \,,\ee
which implicitly defines functions, $\cW_{NS}(\tu,\epsilon_1)$ and $\Omega_{NS}(\tu,\epsilon_2)$, which depend on the theory under consideration.\footnote{The notation here should not be confused with that of the fixed  points at the north and south poles of the $S^2$ factors, which we denote with lower case letters.}  We may similarly define, for the 5d instanton partition function, adapted to the notation above (and setting $r=1$ from now on),
\bea \label{5dNSlimit} Z_{\R^2_{\q_1=e^{2 \pi i \epsilon_1}} \times \R^2_{\q_2=e^{2 \pi i \epsilon_2}} \times S^1}&\(x=e^{2 \pi i \tu}, y= e^{2 \pi i \tnu}, z= e^{2 \pi i \tgamma}\) \\
\underset{\epsilon_2 \rightarrow 0}{\longrightarrow}  \, &\exp \bigg\{ 2 \pi i \( \frac{1}{\epsilon_2} \cW_{NS}^{(5d)}(\tu,\tnu,\tgamma;\epsilon_1) - \Omega_{NS}^{(5d)}(\tu,\tnu,\tgamma;\epsilon_1) + O(\epsilon_2) \) \bigg\} \,. \eea
The notation we have chosen anticipates the role these functions will play as the twisted superpotential and effective dilaton below.

Let us consider again the expression \eqref{s3s2inst} in light of this expansion.  If we again consider the product of the $\ell=nn$ and $\ell=ns$ terms, we find (suppressing the dependence on $\tnu$ and $\tgamma$ from the notation)
\begin{align} \nonumber
&  Z_{\R^2_{\q=e^{-2 \pi i b^2}} \times \R^2_{e^{2 \pi i Q b \epsilon}} \times S^1}\; \( e^{2\pi i b Q (\tu + \frac{1}{2} \m \epsilon)}\)Z_{\R^2_{\q=e^{-2 \pi i b^2}} \times \R^2_{e^{-2 \pi i Q b \epsilon}} \times S^1}\( e^{2\pi i b Q (\tu + \frac{1}{2} \m \epsilon)}\) \;  \\ \nonumber
&\qquad \underset{\epsilon \rightarrow 0}{\longrightarrow} \exp\bigg\{2\pi i \bigg( \tfrac{1}{Q b \epsilon} \cW_{NS}^{(5d)}\(Qb (\tu+ \tfrac{1}{2} \m \epsilon),-b^2\) -\frac{1}{2} \Omega_{NS}^{(5d)}\(Qb (\tu+ \tfrac{1}{2} \m \epsilon),-b^2\) \\ \nonumber
&\qquad \qquad - \tfrac{1}{Q b \epsilon} \cW_{NS}^{(5d)}\(Qb (\tu- \tfrac{1}{2} \m \epsilon),-b^2\) - \frac{1}{2} \Omega_{NS}^{(5d)}\(Qb (\tu- \tfrac{1}{2} \m \epsilon),-b^2\) + O(\epsilon) \bigg) \bigg\} \\ 
& \qquad \longrightarrow \, \exp\bigg\{2\pi i \( \m \partial_{\tu} \frac{1}{Q b} \cW^{(5d)}_{NS}(Q b \tu,-b^2) - \Omega_{NS}^{(5d)}(Qb \tu,-b^2) \)  \bigg\}\,.
\end{align}
Similarly, the other two fixed points contribute
\begin{align} \nonumber
&Z_{\R^2_{\q=e^{-2 \pi i b^{-2}}} \times \R^2_{e^{2 \pi i Q b^{-1} \epsilon}} \times S^1}\; \( e^{2\pi i b^{-1} Q (\tu + \frac{1}{2} \m \epsilon)}\)Z_{\R^2_{\q=e^{-2 \pi i b^{-2}}} \times \R^2_{e^{-2 \pi i Q b^{-1} \epsilon}} \times S^1}\( e^{2\pi i b^{-1} Q (\tu + \frac{1}{2} \m \epsilon)}\) \;  \\
& \qquad \underset{\epsilon \rightarrow 0}{\longrightarrow} \, \exp\bigg\{2\pi i \( \m \partial_{\tu} \frac{1}{Q b^{-1} }\cW^{(5d)}_{NS}(Q b^{-1} \tu,-b^{-2}) - \Omega_{NS}^{(5d)}(Qb^{-1} \tu,-b^{-2}) \)  \bigg\}\,.
\end{align}
Putting this together, we see these objects behave in precisely the same way as we expect the twisted superpotential and effective dilaton to behave in the context of the $S^2 \times S^3_b$ partition function.  Thus it is natural to define
\be \label{wfact} \cW_{S^3_b \times \R^2}(\tu,\tnu,\tgamma) =  \frac{1}{Q b}\cW^{(5d)}_{NS}(\tu,\tnu,\tgamma;-b^2) + \frac{1}{Q b^{-1}}\cW^{(5d)}_{NS}(\tu,\tnu,\tgamma;-b^{-2})\,, \ee
\be \Omega_{S^3_b \times \R^2}(\tu,\tnu,\tgamma) = \Omega^{(5d)}_{NS}(\tu,\tnu,\tgamma;-b^2) + \Omega^{(5d)}_{NS}(\tu,\tnu,\tgamma;-b^{-2})\,. \ee
Thus, we see the factorization of the perturbative contribution continues to hold at the non-perturbative level.  We can now write, \eg,
\be \cW_{S^3_b \times \R^2}(\tu,\tnu,\tgamma) = \cW^{pert}_{S^3_b \times \R^2}(\tu,\tnu,\tgamma) + \cW^{inst}_{S^3_b \times \R^2}(\tu,\tnu,\tgamma) \,.\ee
Then we have seen the perturbative piece is given by \eqref{Wpert}, and can in principle be further factorized into $\cW^{(5d),pert}_{NS}(\tu,\tnu,\tgamma;-b^{2})$ and $\cW^{(5d),pert}_{NS}(\tu,\tnu,\tgamma;-b^{-2})$, although we do not write those expressions here.  The instanton contribution is thus determined implicitly by studying the Nekrasov-Shatashvili limit of the instanton partition function.

\paragraph{Integration contour and Bethe sum.}

Having finally written an expression for the integrand of the $S^3_b \times S^2$ partition function, we can now write the partition function itself as an integral over a suitable contour and sum over gauge fluxes on $S^2$, \ie,
\be Z_{S^3_b \times S^2}(\tnu,\tgamma)_{\n} = \frac{1}{|W_G|} \sum_{\m_a \in \Lambda_G} \oint_{\cC_{JK}} d\tu\; \Pi_a(\tu,\tnu,\tgamma)^{\m_a} \Pi_i(\tu,\tnu,\tgamma)^{\n_i} e^{-2 \pi i \Omega_{S^3_b \times \R^2}(\tu,\tnu,\tgamma)} \,,\ee
where
\be \Pi_a(\tu,\tnu,\tgamma) = e^{2 \pi i \partial_{\tu_a} \cW_{S^3_b \times \R^2}(\tu,\tnu,\tgamma)}\,, \qquad \Pi_i(\tu,\tnu,\tgamma) = e^{2 \pi i \partial_{\tnu_i} \cW_{S^3_b \times \R^2}(\tu,\tnu,\tgamma)}\,.\ee
We have so far not been careful to specify the precise contour of integration in the above integral.  However, it is natural to conjecture that this contour is such that, by a similar manipulation as described above and in \cite{Closset:2017zgf}, we may evaluate this contour integral and perform the resulting geometric series over $\m_a$, and obtain a ``Bethe sum'' formula,
\be  Z_{S^3_b \times S^2}(\tnu,\tgamma)_{\n}  = \sum_{\hat{\tu} \in \cS_{BE}} \Pi_i(\hat{\tu},\tnu,\tgamma)^{\n_i} \cH(\hat{\tu},\tnu,\tgamma)^{-1}\,, \ee
where  $\cS_{BE}$ is defined as in \eqref{SBEdef}, and we have defined
\be \cH(\tu,\tnu,\tgamma) = e^{2 \pi i \Omega_{S^3_b \times \R^2}(\tu,\tnu,\tgamma) } \det_{a,b} \frac{\partial^2 \cW_{S^3_b \times \R^2}(\tu,\tnu,\tgamma)}{\partial \tu_a \partial \tu_b} \,.\ee

Finally, although the above formula has only been motivated for genus zero, it is very natural to conjecture the generalization to arbitrary genus, $\g$, using the general A-twist formalism described above, namely,
\be \label{znonpertsec2} Z_{S^3_b \times \Sigma_\g}(\tnu,\tgamma)_{\fn}  = \sum_{\hat{\tu} \in \cS_{BE}} \Pi_i(\hat{\tu},\tnu,\tgamma)^{\n_i} \cH(\hat{\tu},\tnu,\tgamma)^{\g-1}\,. \ee
In fact, this form is essentially determined by topological invariance on $\Sigma_\g$, which means this observable must be computed by a 2d TQFT.  This represents our result for the full, non-perturbative partition function on $S^3_b \times \Sigma_\g$.

\paragraph{Summary.}

We have  computed the $S^3_b \times \Sigma_\g$ partition function by two methods.  First, we performed a na\"ive reduction to 2d, leading to the perturbative result for the twisted superpotential in \eqref{Wpert}.  However, this missed the contribution of instantons, and we then recovered the full non-perturbative result by a reduction to 4d, where we conjectured the answer is expressed in terms of the Nekrasov-Shatashvili limit of the instanton partition function, as in \eqref{wfact}.  From the twisted superpotential (and effective dilaton), we may then construct the full $S^3_b \times \Sigma_\g$ partition function as in the general discussion of Section \ref{sec:2datwist}, leading to the formulae \eqref{ZpertSec2} for the perturbative partition function, and \eqref{znonpertsec2} for the non-perturbative result.

While the latter method in principle gives the complete answer, in practice the instanton contributions may be difficult or impossible to compute analytically.  For the remainder of this paper, we will therefore consider various simplifying limits in which their contribution is suppressed, or otherwise under control.  Specifically, it is well-known that their contribution is subleading in $N$ when we take a large $N$ limit, and in the next section we consider this limit in several examples, utilizing only the perturbative contribution to the partition function.  In the following section, we will specialize to theories which we expect to have a 6d UV completion.  For such theories we will argue that in the limit of large gauge coupling, corresponding to large radius of the emergent $S^1$ direction, the instantons contribute a simple factor to  the partition function, which can be interpreted as computing the Casimir energy in this limit.  Finally, in the special case of the maximal 5d $\cN=2$ SYM theory, we will see there exists a limit with enhanced supersymmetry, where the instanton contribution is very simple.


\section{\texorpdfstring{Large $\bs{N}$ limit and holography}{Large N limit and holography}}\label{sec:largeN}

In this section, we study the large $N$ limit of the $S^{3}_{b}\times \Sigma_{\fg}$ partition function for a large class of 5d quiver gauge theories. We distinguish two classes of quivers; those leading to an $N^{5/2}$ scaling of the free energy and those leading to an $N^{3}$ scaling. The latter class of theories (which are expected to have UV completions as 6d theories on a circle) are discussed in detail at finite $N$ in Section~\ref{sec:5d6d4d}. For the former class, we find that the matrix model determining the partition function exhibits an interesting structure at large $N$, becoming closely related to the matrix model determining the $S^{5}$ partition function  \cite{Kallen:2012cs,Hosomichi:2012ek,Kallen:2012va,Kim:2012ava,Lockhart:2012vp}.  Precisely, we will establish the large $N$ relations 
\eqss{\label{FS3V}
	F_{S^{3}_{b}\times \Sigma_{\fg}}=&\,-6\pi(\fg-1)\(\frac{∑_{α∈G}c_α -{∑_{ρ∈R}}c_ρ{\hat \fn}_{ρ}\tnu_{ρ}}{{∑_{ρ∈R}}c_ρ\(1-\tnu_{ρ}^2\)}\) \cW_{S^{3}_{b}\times \mathbb R^{2}}\,,
	\\ 
	\cW_{S^{3}_{b}\times \mathbb R^{2}}=&\,\frac{4Q^2}{27\pi} \(\frac{{∑_{ρ∈R}}c_ρ(1-\tnu_{ρ}^2)}{{∑_{ρ∈R}}c_ρ}\)^{3/2}F_{S^5}\,,
}
where the sum ${∑_α}$ is over the vector multiplets and $∑_ρ$ is over various hypermultiplets and $F$ is the free energy of the theory on the corresponding manifold.\footnote{See Sections~\ref{sec:Seiberg theory and its orbifolds} and \ref{sec:General quivers} for notation and details.} Thus,  although the matrix models computing the partition functions on $S^{5}$ and on $S^{3}_{b}\times \Sigma_{\fg}$ are distinct at finite $N$, they are closely related at large $N$.

As we discuss in detail below, the universal twist $\hat \fn_{\rho}=\tnu_{\rho}=0$ is special \cite{Bobev:2017uzs}. This corresponds to a topological twist purely along $U(1)_{R}\subset SU(2)_{R}$. Combining the expressions above it follows that\footnote{As shown in Section~\ref{sec:General quivers}, for this class of quivers $\sum_{\alpha}c_{\alpha}=\sum_{\rho}c_{\rho}$ and the universal relations follow.  } 
\equ{\label{univrels} 
	\cW^{\mathit{univ}}_{S^{3}_{b}\times \mathbb R^{2}}= \frac{4Q^{2}}{27\pi} \, F_{S^{5}}\,,\qquad 
	F_{S^{3}_{b}\times \Sigma_{\fg}}^{\mathit{univ}}=-\frac{8}{9}(\fg-1)Q^{2}\,F_{S^{5}}\,,
}
for any 5d $\cN=1$ theory with a universal twist on $\Sigma_{\fg}$ at large $N$. The second relation above for $Q=1$ is in  agreement with the  supergravity prediction of \cite{Bobev:2017uzs}, valid for the compactification of any 5d $\cN=1$ theory with a universal twist. We will discuss holography in Section~\ref{sec:Holography}. 

\begin{flushleft}
	{\it Useful formulas}
\end{flushleft}

Before we proceed to the computation, let us collect the relevant results of Section~\ref{sec:derivation} for ease of reference. For a general theory with gauge group $G$ and hypermultiplets in gauge representations $R_{I}$, and collecting the terms from \eqref{ZpertSec2}, \eqref{ZpertSec2Parts}, the perturbative partition function is given by 
\eqss{\label{ZS3bpert}
	Z^{\mathit{pert}}_{S^{3}_{b}\times \Sigma_{\fg}} = \sum_{ \hat u \in \mathcal S_{\mathit{BE}}}\,  H^{\fg-1} &\prod_{\alpha\in \mathit{Ad}(G)'} s_{b}\(\alpha(\hat{u})-iQ\)^{1-\fg} \,\prod_{I}\prod_{\rho\in  R_{I}}s_{b}\(\rho(\hat{u})+\nu_{I}\)^{(\fg-1) \hat \n_{I}}\,,
}
where  $H$ denotes the Hessian contribution to the handle-gluing operator in the second line of \eqref{ZpertSec2Parts} and we have written the flavor flux as $\n_{I}=\hat \fn_{I}(\fg-1)$, with $\hat \fn_{I}$ integer-quantized. The Bethe equations are given by 
\be \label{SBESec3}\cS_{\mathit{BE}} = \;\; \left \{ \hat{u} \;\; \big| \;\; \Pi_a(\hat{u}) \equiv \exp \( 2 \pi i \frac{\partial \cW^{\mathit{pert}}_{S^3_b \times \R^2}}{\partial u_a}(\hat{u})\) = 1, \;\;\; a=1,...,r_G\right\} /W_G\,,\ee
with the twisted superpotential \eqref{Wpert}, given by
\eqs{\label{Bethepert}
	\mathcal W^{\mathit{pert}}_{S^{3}_{b}\times \mathbb R^{2}}(\tu) =\mathcal W^{\mathit{classical}}_{S^{3}_{b}\times \mathbb R^{2}}(\tu,\tnu) +\sum_{I}∑_{\rho\in R_{I}}g_b\big(\rho(\tu)+\tnu_{I}\big)-∑_{α}g_b\big( α(\tu)+1\big) \,,
}
where the classical contribution is given in \eqref{wclass}, and for the theories considered in this section the Hessian contribution, $H$, given in \eqref{ZS3bpert} is subleading in the large $N$ limit. Since instanton corrections are suppressed at large $N$, and to avoid clutter, in the remainder of this section we shall always omit the label ``{pert}.'' 

To study the large $N$ limit, we will need the following asymptotic behaviors: 
\eqs{
	g_{b}(\tu+\tnu) &\rightarrow \pm\(-\frac{1}{12}Q^{2}\tu^{3}+\frac14Q^{2}\tnu\tu^{2}-\frac14\left[\frac{1+4Q^{2}}{6}-Q^{2}(1-\tnu^{2})\right]\tu\)\,,\label{expansiongb}\\
	\ell_b(\tu+\tnu) & \rightarrow \pm \frac{Q^{2}}{2}\(\frac12 \tu^{2}+\tnu\tu \)\,, \qquad\qquad\qquad\text{for $\text{Im}(\tu)\to\pm\infty$}\,,\label{expansionlb}
}
where we have assumed that $\tnu\in \mathbb R$ and $|\tnu|\leq 1$, and took the principal branch. We note also the expansion 
\equ{ \label{sbexp}
	s_{b}(x)\rightarrow e^{\pm \frac{i\pi}{2}\(x^{2}+\frac13Q^{2}-\frac16\)}\,,\qquad\qquad \text{for $\text{Re}(x)\to\pm\infty$} \,. 
}

\subsection{Seiberg theory and its orbifolds}\label{sec:Seiberg theory and its orbifolds}

\begin{figure}
	\centering
	\includegraphics[scale=0.5]{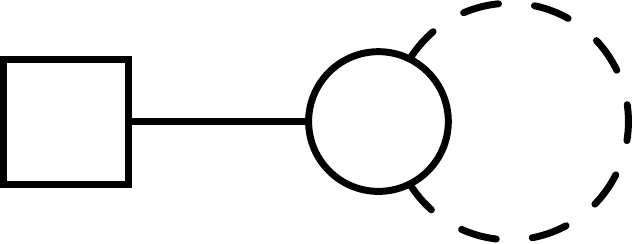}\put(-89,14){$N_f$}
	\caption{The 5d Seiberg theory. The node represents a $USp(2N)$ gauge group, the solid line $N_{f}$ hypermultiplets in the fundamental representation and the dashed line an antisymmetric hypermultiplet.}
	\label{fig:quiver1}
\end{figure}

The Seiberg theory \cite{Seiberg:1996bd} (see also \cite{Intriligator:1997pq}) consists of a single 5d $\cN=1$ vector multiplet in the adjoint of the gauge group $Sp(N)\simeq USp(2N)$\footnote{Since the terminology regarding symplectic groups differs in the literature, let us clarify the one followed here. The group of real $2N\times 2N$ matrices $U$ such that $U^{\sT}\Omega U=\Omega$, with $\Omega^{2}=-1$, is denoted $Sp(2N,\bR)$. The group of such complex matrices is denoted $Sp(2N,\bC)$. If one adds to the later the condition that $U$ is unitary, one obtains the group $USp(2N)\equiv Sp(2N,\bC)\cap U(2N)$, also denoted $Sp(N)$. It is useful to know that $SU(N)\times U(1)\subset USp(2N)\subset SU(2N)$.}, $N_{f}<8$  hypermultiplets in the fundamental representation of the gauge group, and one hypermultiplet in the antisymmetric representation (see Figure~\ref{fig:quiver1}).  The global symmetry of the theory is $SO(5)\times SU(2)_{R}\times SU(2)_{M}\times SO(2N_{f})\times U(1)_{\mathit{top}}$.\footnote{The $U(1)_{\mathit{top}}$  is a topological symmetry which together with the  $SO(2N_{f})$ is expected to combine into an enhanced $E_{N_{f}+1}$ at the conformal fixed point \cite{Seiberg:1996bd}.}  The antisymmetric hypermultiplet transforms as a doublet of $SU(2)_{M}$ and as a singlet of $SO(2N_{f})$. We turn general flavor fluxes along the Cartan of $SU(2)_{M}$ and $SO(2N_{f})$, denoted  $\hat \n  _{M}$ and $\hat \n_{I}$, $I=1,\dots,N_{f}$, respectively. 

The fundamental representation of $USp(2N)$ has weights $\pm e_{i}$, where $e_{i}$ are unit vectors of $\mathbb R^{N}$. The antisymmetric representation  has weights $\pm(e_{i} - e_{j})$  and $\pm(e_{i} + e_{j})$  with $i<j$. The adjoint has the same weights as the antisymmetric and also $\pm 2e_{i}$. Then, the twisted  superpotential \eqref{Bethepert} reads 
\eqs{
	\mathcal W_{S^{3}_{b}\times \mathbb R^{2}} =& -\sum_{\pm}∑_{i< j}\left[g_b\big(1 \pm (\tu_{i}-\tu_{j})\big)+g_b\big(1\pm (\tu_{i}+\tu_{j})\big)\right]-\sum_{i} g_b\big(1 \pm2 \tu_{i}\big) \nn
	&+\sum_{\pm}\,∑_{i< j}\left[g_b\big(\tnu_{\mathit{AS}} \pm (\tu_{i}-\tu_{j})\big)+g_b\big(\tnu_{\mathit{AS}} \pm (\tu_{i}+\tu_{j})\big)\right] \nn
	&+\sum_{I=1}^{N_{f}}\sum_{\pm}\,∑_{i}g_b\big(\tnu_{I} \pm \tu_{i}\big)\,,
}
where we have omitted the classical piece $\mathcal W^{\mathit{classical}}_{S^{3}_{b}\times \mathbb R^{2}}$ since one can check that it is subleading in $N$. To analyze the large $N$ behavior of this quantity we follow the approach introduced in \cite{Herzog:2010hf}, replacing
\equ{\label{continuum}
	\tu_{i}\to iN^{\alpha}x\,,\qquad \sum_{i}\to N \int dx \,\rho(x)\,,
}
where $\alpha>0$,  $x$ is a continuous variable of order $\mathcal O(N^{0})$, and $\rho(x)$  is the eigenvalue density, normalized as $\int dx\, \rho(x)=1$. Using the large $\tu$ expansion \eqref{expansiongb} and then going to the continuum variables, we have\footnote{Due to the structure of roots and weights for $USp(2N)$ and $SU(2N)$, only the real part of $g_b(z)$ contributes from the expansion \eqref{expansiongb} and $\mathcal W_{S^{3}_{b}\times \mathbb R^{2}}$ is real. The same holds for $\ell_b(z)$ appearing in the free energy.}
\eqst{
	\mathcal W_{S^{3}_{b}\times \mathbb R^{2}}\approx\, \frac{(8-N_{f})}{6}Q^{2} \, N^{1+3\alpha}\int dx \,\rho(x) \,|x|^{3}  \\
	-\frac{1}{2} Q^{2}\(1-\tnu_{\mathit{AS}}^2\)N^{2+\alpha}\int_{y<x} dx \,dy \, \rho(x)\rho(y)\( |x+y|+|x-y|\)\,.
}
For this function to have a nontrivial saddle at large $N$ requires both terms to be of the same order in $N$ and hence $\alpha=\frac{1}{2}$, which we set in what follows. We note that, to this order in $N$, only the mixing parameter for the antisymmetric hypermultiplet, $\tnu_{\mathit{AS}}$, is visible while the parameters for the fundamental fields, $\tnu_{I}$,  are not.  We also note there has been a cancellation of the nonlocal cubic terms $|x\pm y|^{3}$ among the vector multiplet and the antisymmetric hypermultiplet. The same cancellation occurs for the free energy of this theory on $S^{5}$  \cite{Jafferis:2012iv}.  In fact, by a simple rescaling of the coordinates
\equ{
	x\to \frac23 \sqrt{1-\tnu_{\mathit{AS}}^2}\,\, x \,,
}
and a corresponding inverse rescaling of $\rho$ to preserve its normalization, we have\footnote{We have used the symmetry of the integrand to replace $\int_{y<x}dxdy\to \tfrac12 \int dxdy$, where now both integration variables $x,y$ are integrated over their full domain.} 
\eqst{
	\mathcal W_{S^{3}_{b}\times \mathbb R^{2}}\approx\frac{4Q^{2}}{27 \pi} \(1-\tnu_{\mathit{AS}}^2\)^{3/2}\bigg[\frac{\pi(8-N_{f})}{3}\,N^{5/2}\int dx \,\rho(x) \,|x|^{3}  \\
	-\frac{9 \pi}{8} N^{5/2}\int dx \,dy \, \rho(x)\rho(y)\( |x+y|+|x-y|\)\bigg]\,.
}
We note the quantity inside the brackets is precisely the free energy functional of this theory on $S^{5}$  (see Eq. (3.4) in \cite{Jafferis:2012iv}). Thus, we find the simple relation
\equ{ \label{VFS5seiberg}
	\mathcal W_{S^{3}_{b}\times \mathbb R^{2}}=\frac{4Q^2}{27\pi} \(1-\tnu_{\mathit{AS}}^2\)^{3/2}F_{S^5}\,.
}
Since this is a functional relation in the eigenvalue density $\rho(x)$, extremization of the Bethe potential  is equivalent to the extremization of  $F_{S^{5}}$. As shown in \cite{Jafferis:2012iv} the saddle configuration is given by
\equ{\label{density}
	\rho(x)=\frac{2|x|}{x_{\ast}^{2}}\,, \qquad x_{\ast}=\frac{3}{\sqrt{2(8-N_{f})}}\,,
}
with $x\in[0,x_{\ast}]$ and at the extremum \equ{F_{S^{5}}=-\frac{9\sqrt 2 \pi N^{5/2}}{5\sqrt{8-N_{f}}}\,.} 

Now let us evaluate the free energy on ${S^{3}_{b}\times \Sigma_{\fg}}$ in this Bethe vacuum. Taking the logarithm  of \eqref{ZS3bpert} for this theory it is easy to see that the Hessian is subleading in this vacuum and only the flux operators contribute, which using the expansion \eqref{expansionlb} and vacuum \eqref{density} becomes  
\eqs{\label{FS3FS5seiberg}
	F_{S^{3}_{b}\times \Sigma_{\fg}} =-\frac{8}{9}(\fg-1)Q^2(1 -\hat \n  _{M}\tnu_{\mathit{AS}})\,\sqrt{1-\tnu_{\mathit{AS}}^{2}}\,F_{S^{5}}\,.
}
As we will show in Section~\ref{sec:General quivers} the relations \eqref{VFS5seiberg} and \eqref{FS3FS5seiberg} are not particular to the Seiberg theory but can be generalized to a large class of quiver gauge theories. This is rather nontrivial as it requires a number of special cancellations in the Bethe potential. We note that relations analogous to  these hold for 3d  $\cN=2$ theories on $S^1×Σ_{\fg}$; in that case the relations are among  the topological twisted index, corresponding Bethe potential, and the free energy on the round $S^{3}$  \cite{Hosseini:2016tor,Hosseini:2016ume}. 

Before discussing the general case we prove these relations for orbifolds of the Seiberg theory.

\begin{figure}
	\centering
	{\includegraphics[scale=0.4]{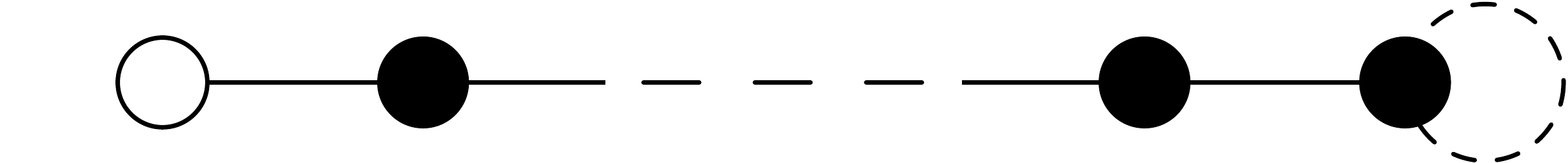}\put(-325,11){$(A)$}\put(-255,-2){$\scriptstyle 0$}\put(-208,-2){$\scriptstyle 1$}\put(-84,-2){$\scriptstyle k-1$}\put(-33,-2){$\scriptstyle k$}} \\ \vspace*{4mm}
	{\includegraphics[scale=0.4]{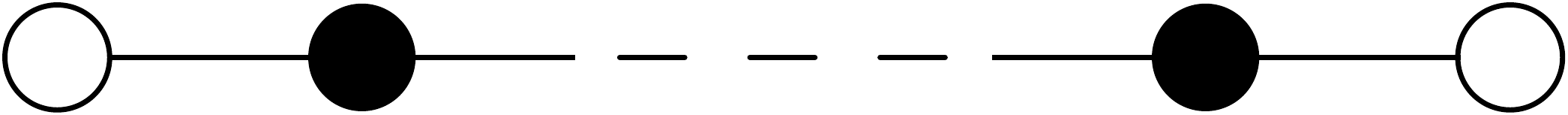}\put(-304.5,4){$(B)$}\put(-235,-7){$\scriptstyle 0$}\put(-188,-7){$\scriptstyle 1$}\put(-64,-7){$\scriptstyle k-1$}\put(-12,-8){$\scriptstyle k$}} \\ \vspace*{4mm}
	{\includegraphics[scale=0.4]{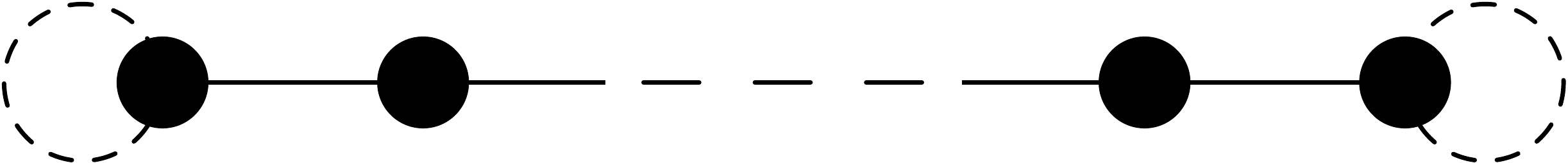}\put(-325,11){$(C)$}\put(-255,-2){$\scriptstyle 1$}\put(-208,-2){$\scriptstyle 2$}\put(-84,-2){$\scriptstyle k-1$}\put(-33,-2){$\scriptstyle k$}}
	\caption{The 5d quiver gauge theories of discussed in \cite{Bergman:2012kr}. Black nodes represent $SU(2N)$ gauge groups and white ones $USp(2N)$ gauge groups. Solid lines denote bifundamental hypermultiplets and dashed lines  hypermultiplets in the antisymmetric representation. In addition, at any given node, $a$, one may have $N_{f}^{a}$ number of fundamental hypermultiplets, which we have not depicted.}
	\label{fig:quivers}
\end{figure}
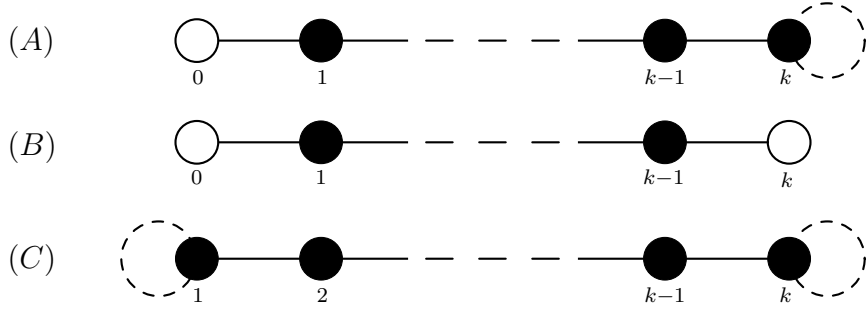

\paragraph{Orbifolds.}
These theories are obtained as $\mathbb Z_{n}$ orbifolds of the Seiberg theory discussed above and consist of linear quivers with $SU(2N)$ nodes, in addition to $USp(2N)$ nodes \cite{Bergman:2012kr}. There are three classes of quivers: class $(A)$ for $n=2k+1$ odd and; $(B),(C)$ when $n=2k$ is even, shown in Figure~\ref{fig:quivers}. The Seiberg theory corresponds to class $(A)$ with $k=0$.

Analyzing each class separately one finds that the Bethe equations for each quiver can be obtained from the following Bethe potentials:\footnote{To avoid clutter we denote $\cW_{S^{3}_{b}\times \mathbb R^{2}}= \cW$ in what follows.}
{\allowdisplaybreaks
\eqs{\label{potVS3Sigma}
	\cW^{(A)}=&\,\cW^{\circ}_{W_0} +\sum_{a=1}^{k}\cW_{W_{a}}^{\bullet}+\sum_{a=0}^{k-1}\cW_{BF_{(a,a+1)}}+\cW_{\mathit{AS}_{k}}+\sum_{a=0}^{k}N_{f}^{a}\,\cW_{F_{a}}\,,\\
	\cW^{(B)}=&\,\cW^{\circ}_{W_0} +\sum_{a=1}^{k-1}\cW_{W_{a}}^{\bullet}+\cW_{W_{k}}^{\circ}+\sum_{a=0}^{k-1}\cW_{BF_{(a,a+1)}}+\sum_{a=0}^{k}N_{f}^{a}\,\cW_{F_{a}}\,,\\
	\cW^{(C)}=&\,\sum_{a=1}^{k}\mathcal V_{W_{a}}^{\bullet}+\sum_{a=1}^{k-1}\cW_{BF_{(a,a+1)}}+\cW_{\mathit{AS}_{1}}+\cW_{\mathit{AS}_{k}}+\sum_{a=1}^{k}N_{f}^{a}\,\cW_{F_{a}}\,,
}}

\noindent where each term is given by
\eqs{\cW^{\circ}_{W} &= -∑_{±,i,j<i}\left[g_{b}(1±(\tu^i-\tu^j)) +g_{b}(1±(\tu^i+\tu^j))\right] -∑_{±,i}g_{b}(1±2 \tu^i) \,,\nn
	\cW^{\bullet}_{W} &=-2∑_{\mathclap{±,i,j<i}}\left[g_{b}(1±(\tu^i-\tu^j)) +g_{b}(1±(\tu^i+\tu^j))\right] -∑_{±,i}g_{b}(1±2 \tu^i) \,,\nn
	\cW_{BF_{(a,b)}} &=2∑_{\mathclap{±,i,j<i}}\left[g_{b}(\tnu_{(a,b)}±(\tu^i-\tu^j)) +g_{b}(\tnu_{(a,b)}±(\tu^i+\tu^j))\right] +∑_{±,i}g_{b}(\tnu_{(a,b)}±2\tu^i)\,, \nn
	\cW_{AS} &=∑_{\mathclap{±,i,j<i}}\left[g_{b}(\tnu_{AS}±(\tu^i-\tu^j)) +g_{b}(\tnu_{AS}±(\tu^i+\tu^j))\right] \,,\nn
	\cW_{F} &=∑_{±,i}g_{b}(\tnu_F± \tu^i)\,.
}
To write the above expressions, we start with fundamental weights in the $\bR^{2N}$ basis subject to the constraint $∑_{A=1}^{2N}\tu^A=0$ for $SU(2N)$ group and $\tu^i=-\tu^{N+i}$ with $i=1,⋯,N$ for $USp(2N)$ group, as explained in \cite{Jafferis:2012iv}. To extremize the Bethe potentials, we chose the same Ansatz for $\tu$'s corresponding to different nodes, \ie, $\tu^i_a≡\tu^i$. The final result in terms of $\tu^i$ then follows for all representations straightforwardly. The continuum limit now follows as before with $\tu^i → \i N^{1/2}x$ and using the large $\tu$ expansion \eqref{expansiongb}. Let us work out the coefficients of different terms at various orders of $N$ in $\mathcal{W}^{(A)}$ (ignoring overall signs and factors):
\eqs{&\O(N^{7/2}) \text{ with }|x±y|^3 : 1 +2k -2k -1 +0 = 0\,, \nn
	&\O(N^{5/2}) \text{ with }|x|^3 : 8 +8k -8k +0 -{\textstyle ∑_a}N_f^a = 8-N_f\,, \nn
	&\O(N^{5/2}) \text{ with }|x±y| : \(\tfrac{1+4Q^2}{6}\) +2k\(\tfrac{1+4Q^2}{6}\) -2{\ts ∑_{a=0}^{k-1}}\(\tfrac{1+4Q^2}{6}-Q^{2}(1-\tnu_{(a,a+1)}^{2})\) \nn
	&\; -\(\tfrac{1+4Q^2}{6}-Q^{2}(1-\tnu_{AS_k}^{2})\) +0 =Q^2\({\ts ∑_{a=0}^{k-1}}\(2(1-\tnu_{(a,a+1)}^{2})\) +(1-\tnu_{AS_k}^{2})\),
	\label{cubNLccl}}
where we defined  $N_{f}=\sum_{a}N_{f}^{a}$. Note that the field content is such that the terms at $\O(N^{7/2})$ cancel leaving the expected terms at $\O(N^{5/2})$ to be dominant. This happens for classes $(B)$ and $(C)$ as well as can be easily checked. In fact, for all the three classes, the cubic term has the same coefficient as the Seiberg theory and only the coefficient of the nonlocal linear term changes depending on the precise matter field content. Repeating the same extremization procedure done for the Seiberg theory for each class, we obtain the following Bethe potentials:
\eqs{ 
	\cW^{(A)} &=\frac{4Q^2}{27\pi}\(\frac{∑_{a=0}^{k-1}\(2(1-\tnu_{(a,a+1)}^{2})\) +(1-\tnu_{AS_k}^{2})}{2k+1}\)^{3/2}F^{(A)}_{S^5}, \\
	\cW^{(B)} &=\frac{4Q^2}{27\pi}\(\frac{∑_{a=0}^{k-1}\(2(1-\tnu_{(a,a+1)}^{2})\)}{2k}\)^{3/2}F^{(B)}_{S^5}, \\
	\cW^{(C)} &=\frac{4Q^2}{27\pi}\(\frac{∑_{a=1}^{k-1}\(2(1-\tnu_{(a,a+1)}^{2})\) +(1-\tnu_{AS_1}^{2})+(1-\tnu_{AS_k}^{2})}{2k}\)^{3/2}F^{(C)}_{S^5},
}
where $F^{(A,B,C)}_{S^5}=n^{3/2}F^{\mathit{Seiberg}}_{S^5}$ \cite{Jafferis:2012iv}.
The free energy calculation is similar to what was done for the Seiberg theory. We turn on flavor fluxes for all the $U(1)_{(a,a+1)}$ and $U(1)_{AS}$ global symmetries of the quiver theories. This leads to the following free energies
\eqs{F_{S^{3}_{b}\times \Sigma_{\fg}}^{(A)} &=-6\pi(\fg-1)\frac{2k+1 -∑_{a=0}^{k-1}2\hat \n  _{(a,a+1)}\tnu_{(a,a+1)} -\hat \n  _{AS_k}\tnu_{AS_k}}{\(∑_{a=0}^{k-1}\(2(1-\tnu_{(a,a+1)}^{2})\) +(1-\tnu_{AS_k}^{2})\)}\,\cW^{(A)}\,, \\
	F_{S^{3}_{b}\times \Sigma_{\fg}}^{(B)} &=-6\pi(\fg-1)\frac{2k -∑_{a=0}^{k-1}2\hat \n  _{(a,a+1)}\tnu_{(a,a+1)}}{∑_{a=0}^{k-1}\(2(1-\tnu_{(a,a+1)}^{2})\)}\,\cW^{(B)}\,, \\
	F_{S^{3}_{b}\times \Sigma_{\fg}}^{(C)} &=-6\pi(\fg-1)\frac{2k -∑_{a=1}^{k-1}2\hat \n  _{(a,a+1)}\tnu_{(a,a+1)} -\hat \n  _{AS_1}\tnu_{AS_1} -\hat \n  _{AS_k}\tnu_{AS_k}}{\(∑_{a=1}^{k-1}\(2(1-\tnu_{(a,a+1)}^{2})\) +(1-\tnu_{AS_1}^{2})+(1-\tnu_{AS_k}^{2})\)}\,\cW^{(C)}\,.
}
We note that for the universal twist, $\hat \n  =\tnu=0$, the formulas above reduce to $\cW=\frac{4Q^{2}}{27\pi} F_{S^5}$ and $F_{S^{3}_{b}\times \Sigma_{\fg}}=-6\pi(\fg-1)\cW$ for all classes of orbifolds. We will see this observation generalizes to a large class of quivers with an $N^{5/2}$ scaling in Section~\ref{sec:General quivers}.

\subsection{Extremization}

We recall that when $Z_{S^{3}_{b}\times \Sigma_{\fg}}$ is viewed as direct sum of 3d theories, the fugacity parameters   $\tnu$ are associated to the 3d R-charge, $Δ=1-\tnu$, of the 3d $\cN=2$ chiral fields obtained by reduction of the hypermultiplets on $\Sigma_{\fg}$. Since the theory is topological on $\Sigma_{\fg}$ we can shrink it to zero size and $F_{S^{3}_{b}\times \Sigma_{\fg}}=F_{S^{3}_{b}}$ coincides with the free energy of the 3d effective theory on $S^{3}_{b}$. In view of $F$-maximization \cite{Jafferis:2010un,Closset:2012vg} it is thus natural to propose that in order to obtain information about the 3d IR fixed point (assuming such fixed point exists), the fugacities must be set to those values extremizing the free energy:\footnote{We are assuming there are no accidental flavor symmetries in the IR.}
\equ{\label{ext}
	\frac{\partial F_{S^{3}_{b}\times \Sigma_{\fg}}}{\partial \tnu_{ρ}}=0\,.
}
Carrying this out for each class of quivers leads to a coupled set of quadratic equations with two sets of solutions:
\equ{\label{sole3Q}
	\tnu_{I}^{(A,B,C)}=\frac{\hat \n  _I^{(A,B,C)}}{4\hat \n  ^2_{(A,B,C)}}\(1±\sqrt{1+8\hat \n  ^2_{(A,B,C)}}\),
}
where we defined
\eqs{
	\hat \n  ^{2}_{(A)} &= \frac{1}{2k+1}\(∑_{a=0}^{k-1}2\hat \n  _{(a,a+1)}^2+\hat \n  _{AS_k}^2\) \,,\\
	\hat \n  ^{2}_{(B)} &= \frac{1}{2k}\(∑_{a=0}^{k-1}2\hat \n  _{(a,a+1)}^2\) \,,\\
	\hat \n  ^{2}_{(C)} &= \frac{1}{2k}\(∑_{a=1}^{k-1}2\hat \n  _{(a,a+1)}^2+\hat \n_{AS_1}^2+\hat \n_{AS_k}^2\).
}
We emphasize  the index $I$ in \eqref{sole3Q} runs over all antisymmetric and bifundamental hypermultiplets in the corresponding quiver, and not fundamental ones.\footnote{Just as in the Seiberg theory,  the fugacities for all fundamental fields are not visible at this order in $N$ and thus are not fixed by extremization. } Plugging the solutions back into $F_{S^{3}_{b}\times \Sigma_{\fg}}^{(A,B,C)}$  gives the common formula
\equ{\label{FExtQvrs}
	F_{S^{3}_{b}\times \Sigma_{\fg}}^{(A,B,C)}=-\frac{8}{9}(\fg-1)Q^2\(±\frac{\sqrt{2}\left|\hat \n  ^{2}_{(A,B,C)}-1\right|^{3/2}\(\sqrt{1+8\hat \n  ^{2}_{(A,B,C)}}±1\)}{\left|4\hat \n  ^{2}_{(A,B,C)} -1 ∓\sqrt{1+8\hat \n  ^{2}_{(A,B,C)}}\right|^{3/2}}\) F_{S^{5}}^{(A,B,C)}\,.
}
Depending on the values of the flavor fluxes, one of the roots in \eqref{sole3Q} may be discarded by the requirement $|\tnu_I|≤1$ (\ie,  $0\leq\Delta_{I}\leq2$), which we have assumed. 

Note that for the universal twist $\hat \n=\tnu=0$ this reduces to \eqref{univrels} for all classes. For non-universal twists the relation among free energies depends on the theory under consideration. Specifying \eqref{FExtQvrs} for the case of the Seiberg theory with a nonzero flux for its $SU(2)_{M}$ flavor symmetry matches the supergravity result recently found in \cite{Bah:2018lyv} where the relevant supergravity solution was constructed.

\subsection{General quivers}\label{sec:General quivers}

Consider a 5d quiver gauge theory with a number $N^{\circ}_{V}$ and $N^{\bullet}_{V}$ of symplectic and unitary gauge groups, respectively, and matter fields in the fundamental, bifundamental, antisymmetric, or adjoint representations of the gauge groups. The perturbative Bethe potential \eqref{Bethepert} receives contributions from all these fields, but the scaling with $N$ of each contribution depends on the particular weights of the representations. We recall that for $USp(2N)$ gauge group, the fundamental representation has weights $\pm e_{i}$, where $e_{i}$ are unit vectors of $\mathbb R^{N}$. The antisymmetric representation has weights $\pm(e_{i} - e_{j})$  and $\pm(e_{i} + e_{j})$  with $i<j$. The adjoint has the same weights as the antisymmetric and also $\pm 2e_{i}$.  For $U(N)$ gauge group the fundamental has weights $e_{i}$, the antisymmetric $±(e^i+e^j)$ and the adjoint $\pm(e_{i}-e_{j})$ with $i<j$. We refer to weights with two nonzero entries as ``nonlocal weights'' as in the continuum limit they lead to nonlocal terms in the Bethe potential. Since a given representation may have both kinds of weights we introduce the continuum notation $\rho(u)\to \bs{\rho}(x,y)$ for the nonlocal terms and $\rho(u)\to \bs{\rho}(x)$ for the local ones,\footnote{We denote the weights by boldface in order not to be confused with the eigenvalue density $\rho$. We will switch to normal font in expressions we get after the integration is done.} and similarly for the roots $\alpha(u)$. With this notation, using the expansion \eqref{expansiongb} and rearranging the various terms that appear, the Bethe potential reads 
\eqst{\label{fullBetheLargeN} 
	\cW = - \frac12 Q^{2}\tilde\gamma\, N^{1+2α}∑_{\bs{α}}N_{\bs{α}}∫dx ρ(x)x^2 \\
	+\frac12∑_{\bs{α}}\bigg(N^2∫dxdyρ(x)ρ(y)\left[\frac{Q^{2}}{3}N^{3α}|\bs{α}(x,y)|^3 -\frac{1+4Q^2}{6}N^{α}|\bs{α}(x,y)|\right] \\ +N∫dxρ(x)\left[\frac{Q^{2}}{3}N^{3α}|\bs{α}(x)|^3-\frac{1+4Q^2}{6}N^{α}|\bs{α}(x)|\right]\!\bigg) \\
	-\frac12∑_{\bs{\rho}}\bigg(N^2∫dxdyρ(x)ρ(y)\left[\frac{Q^{2}}{3}N^{3α}|\bs{\rho}(x,y)|^3 \!-\!\(\!\frac{1+4Q^2}{6}-Q^2(1-\tnu_{\bs{\rho}}^2)\!\)\! N^{α}|\bs{\rho}(x,y)|\right] \\
	+N∫dxρ(x)\left[\frac{Q^{2}}{3}N^{3α}|\bs{\rho}(x)|^3\!-\!\(\!\frac{1+4Q^2}{6}-Q^2(1-\tnu_{\bs{\rho}}^2)\!\)\!N^{α}|\bs{\rho}(x)|\right]\!\bigg).
}
We see that the various terms have different scalings in $N$: $1+2α, 2+3α, 2+α, 1+3α, 1+α$. In order to have a nontrivial saddle at large $N$ both a quadratic and linear term in the eigenvalue density must appear at a given order in $N$. Although in principle there seem to be six cases, demanding $\alpha>0$ there are in fact only two scalings that provide a nontrivial saddle point:
\begin{enumerate}[(I)]
	\item $2+α=1+3α ⇒ α=\half ⇒ N^{5/2} \;\, \text{Higher order to cancel: }N^{2+3α=7/2}$ 
	\item $2+α=1+2α ⇒ α=1 ⇒ N^3 \quad \text{Higher orders to cancel: }N^{2+3α=5} \;\text{and}\; N^{1+3α=4}$
\end{enumerate}
In each case the cancellation of higher orders is required to avoid a trivial saddle point, which would otherwise render the large $N$ method used here inapplicable. We consider each case in turn.

\subsubsection*{(I) Theories with $N^{5/2}$ scaling}

We assume the generic quivers have $USp(2N)$ and $SU(2N)$ gauge groups,\footnote{It should be possible to study quivers with different ranks but we do not consider this here. } with $N_V^{\circ}, N_V^{\bullet}$ number of vector multiplets corresponding to respective gauge groups and $N_{\mathit{Ad}}$, $N_{\mathit{AS}}$, $N_{\mathit{BF}}$, $N_f$ number of hypermultiplets in the adjoint, antisymmetric, bifundamental and fundamental representations of given gauge groups, respectively. Setting $\alpha=\frac{1}{2}$ in \eqref{fullBetheLargeN} the cancellation of nonlocal cubic terms at $\O(N^{7/2})$ requires
\eqsc{
	\bigg(∑_αc_α-∑_ρc_ρ\bigg)∫dxdyρ(x)ρ(y)\(|x+y|^3+|x-y|^3\)=0 \nn
	⇒∑_αc_α -∑_ρc_ρ =0\,, \label{consNF}
}
where $c_α$ and $c_ρ$ are numerical constants for each vector and hypermultiplet that appear when collecting all the contributions to the integral shown above. For instance, $c_V^{\circ}=1$, $c_V^{\bullet}=2$ for each vector multiplet  in the quiver and $c_{Ad}=1$, $c_{AS}=1$, $c_{BF}=2$, $c_{f}=0$ for each  hypermultiplet. Thus, a constraint is imposed on the number of various fields present in the theory. We have already seen an example of such a relation for type $(A)$ quiver theory in \eqref{cubNLccl}, where $∑_ρc_ρ=1+2k=N_{AS}+2N_{BF}$, which is also equal to $∑_αc_α=N^{\circ}_{V}+2N^{\bullet}_{V}$ for this quiver. Thus, the Bethe potential for these theories with an $N^{5/2}$ scaling is given by\footnote{We note that due to \eqref{consNF}, the $-\frac{1}{6}(1+4Q^2)$ part in the terms of order $N^{2+\alpha}$ in \eqref{fullBetheLargeN} also cancels.}
\eqs{\cW &=N^{5/2}Q^{2}\bigg[\frac{1}{6}∫dxρ(x)\(∑_{\bs{α}}|\bs{α}(x)|^3 -∑_{\bs{\rho}}|\bs{\rho}(x)|^3\) \nn
	&\qquad\qquad -\frac12 ∫dxdy\,ρ(x)ρ(y)∑_{\bs{\rho}}\(1-\tnu_{\bs{\rho}}^2\)|\bs{\rho}(x,y)|\bigg] \nn
	&=N^{5/2}Q^{2}\bigg[\frac{1}{6}\(∑_αc'_ α -∑_ρc'_ρ\)∫dxρ(x)|x|^3 \nn
	&\qquad\qquad -\frac12 ∑_{\rho}c_ρ\(1-\tnu_{\rho}^2\)∫dxdy\,ρ(x)ρ(y)\(|x+y|+|x-y|\)\bigg].
	\label{genBethePotN52}}
Here $c'_α$ and $c'_ρ$ are another set of numerical constants appearing when collecting all the contributions to the local $|x|^3$ term. These are in general different from the $c$'s, for example, ${c'}_V^{\circ}={c'}_V^{\bullet}=8$, $c'_{Ad}=8$, $c'_{AS}=0$, $c'_{BF}=8$, $c'_f=1$, assuming the same set of gauge groups as before. After the rescaling $x→\frac{2}{3}\sqrt{∑_{ρ}c_ρ\(1-\tnu_{ρ}^2\)}\,x$ (and corresponding inverse rescaling of $\rho$), we have 
\eqst{\cW=\frac{4Q^2}{27\pi}\bigg(∑_{ρ}c_ρ\(1-\tnu_{ρ}^2\)\bigg)^{3/2}N^{5/2}\bigg[\frac{π(8N_v-N_f)}{3}∫_{0}^{x_*}dx\,ρ(x)|x|^3 \\
	-\frac{9π}{8}∫_{0}^{x_*}dxdyρ(x)ρ(y)\(|x+y| +|x-y|\)\bigg],
	\label{VN52exp}}
where we defined $N_v=N^{\circ}_{V}+N^{\bullet}_{V}-N_{Ad}-N_{BF}$. By comparison with the free energy functional on $S^{5}$ for the same theory, it follows that
\equ{\label{VFS51}
	\cW=\frac{4Q^{2}}{27\pi}\,\(\frac{∑_{\rho}c_ρ\(1-\tnu_{\rho}^2\)}{∑_{\rho}c_ρ}\)^{3/2}\, F_{S^{5}}\,,
}
with $F_{\bS^5}=\big(∑_ρc_ρ\big)^{3/2}F^{Seiberg}_{\bS^5}$, which is extremized for
\equ{\label{rho52}
	\rho(x)=\frac{2|x|}{x_{\ast}^{2}}\,, \qquad x_{\ast}=\frac{3}{\sqrt{2(8N_v-N_f)}}\,·
}

We now evaluate the on-shell free energy. Taking the logarithm of \eqref{ZS3bpert} one can see that the Hessian part of the handle-gluing operator does not contribute to leading order in $N$ and one finds\footnote{\label{footnoteHessian}Although in principle one may worry that exponentially subleading/diverging terms in $H$ of the form $\exp{(∓N^{\alpha}x)}$  may contribute, to leading order in $N$ these are of the form $H \approx \exp{\(∑_i∂_{u^i}\mathcal{W}\)}$ and thus $H \approx 1$ when evaluated in the Bethe vacua of the theory.} 
\eqs{
	F_{S^{3}_{b}\times \Sigma_{\fg}}   =-6\pi(\fg-1)\(\frac{∑_{α}c_α -∑_{\rho}c_ρ{\hat \fn}_{\rho}\tnu_{\rho}}{∑_{\rho}c_ρ\(1-\tnu_{\rho}^2\)}\)\cW\,,
	\label{FN52}}
which using \eqref{VFS51} can also be written as 
\eqs{\label{genFS3fS5}
	F_{S^{3}_{b}\times \Sigma_{\fg}}=\,-\frac{8}{9}Q^2(\fg-1)\bigg(∑_{α}c_α -∑_{\rho}c_ρ{\hat \fn}_{\rho}\tnu_{\rho}\bigg)\bigg(∑_{\rho}c_ρ(1-\tnu_{\rho}^2)\bigg)^{1/2}\bigg(∑_{\rho}c_ρ\bigg)^{-3/2}F_{S^5}\,.
} 
Finally, extremizing with respect to the fugacities gives
\equ{\label{sole}
	\tnu_{\rho}^{\pm}=\frac{\hat \fn_ρ}{4\hat\fn^2}\(1±\sqrt{1+8\hat\fn^2}\),
}
where $\hat\fn^{2}\equiv\big(∑_ρc_ρ\big)^{-1}\big(∑_ρc_ρ\hat\fn_ρ^2\big)$ and evaluating at the extremum, 
\equ{\label{FS3S5n}
	F_{S^{3}_{b}\times \Sigma_{\fg}}=-\frac{8}{9}(\fg-1)Q^2\(±\frac{\sqrt{2}\left|\hat\fn^{2}-1\right|^{3/2}\(\sqrt{1+8\hat\fn^{2}} ±1\)}{\left|4\hat\fn^{2} -1 ∓\sqrt{1+8 \hat\fn^{2}}\right|^{3/2}}\) F_{S^{5}}\,.
}
Since $F_{\bS^5}$ is negative and $F_{\bS^3}$ should be positive, we see that for $\fg>1$, the solution with $+$ sign should be chosen above and for $\fg=0$, the solution with $-$ sign. This sign then suggests the introduction of $κ$, the normalized curvature of $Σ_{\fg}$ defined below \eqref{indexth}, by rescaling $\hat\fn_ρ→\frac{\hat\fn_ρ}{κ}$, giving
\equ{\label{FS3S5nk}
	F_{S^{3}_{b}\times \Sigma_{\fg}}=\frac{8}{9}(\fg-1)Q^2\(\frac{\sqrt{2}\left|\hat\fn^{2}-κ^2\right|^{3/2}\(\sqrt{κ^2+8\hat\fn^{2}} -κ\)}{κ\left|4\hat\fn^{2} -κ^2 +κ\sqrt{κ^2+8 \hat\fn^{2}}\right|^{3/2}}\) F_{S^{5}}\,.
}
One can also make sense of the above expression for $κ=0$ (as in \cite{Bah:2018lyv} for the Seiberg theory) by considering the limit $\frac{\fg-1}{κ}→-4π$, $\hat\fn_ρ→\frac{\hat\fn_ρ}{2π}$,  resulting in the relation $F_{S_b^3×Σ_{\fg=1}}=-\frac{8}{9}Q^2\sqrt{\hat\fn^2}F_{S^5}$.

\subsubsection*{(II) Theories with $N^{3}$ scaling}

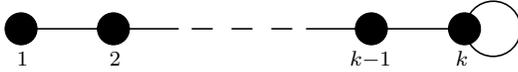
\begin{figure}
	\centering
	\includegraphics[scale=0.3]{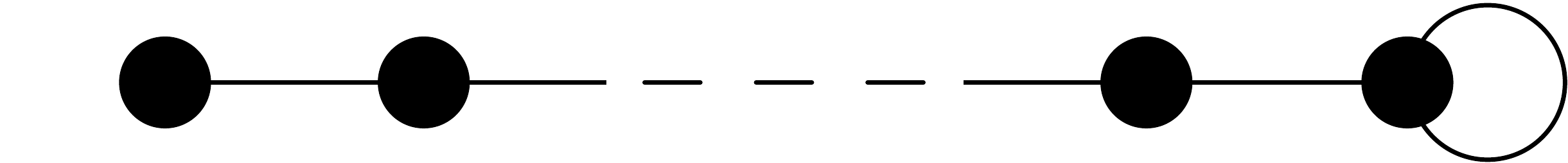}\put(-191,-3){$\scriptstyle 1$}\put(-156,-3){$\scriptstyle 2$}\put(-65,-3){$\scriptstyle k-1$}\put(-25,-3){$\scriptstyle k$} \\ \vspace*{4mm}
	\caption{A simple class of 5d quiver gauge theories with free energy scaling as $N^3$. Black nodes represent $SU(2N)$ gauge groups, a line connecting two nodes denotes a bifundamental hypermultiplet, and the line ending on the same node denotes an adjoint hypermultiplet. No fundamental hypermultiplets are allowed in this case.}
	\label{fig:N3quivers}
\end{figure}

Setting $\alpha=1$ in \eqref{fullBetheLargeN} and requiring the cancellation of nonlocal cubic terms at $\O(N^{5})$ gives the same constraint as before, \ie,  \eqref{consNF}. The cancellation of local cubic terms at $\O(N^4)$ leads to the additional condition
\equ{∑_αc'_α-∑_ρc'_ρ=0\,.
	\label{consNf}}
The constraints \eqref{consNF} and \eqref{consNf} together mean no cubic terms (either nonlocal and local) survive in the expansion of the function $g_b$. For pure $U(N)$ or $SU(N)$ theories, this constraint simply becomes $N_f=0$. The maximal 5d $\N=2$ $SU(N)$ SYM is such an example. Another known example is class $\cS_k$; circular quivers consisting of $k$ $SU(N)$ gauge groups and $k$ bifundamental hypermultiplets. The maximal theory is obtained for $k=1$ when the bifundamental multiplet turns into an adjoint one.   We also note a linear quiver with $k$ $SU(N)$ gauge groups and $k-1$ bifundamentals, and an additional adjoint matter on one node (see Figure~\ref{fig:N3quivers}) satisfies  \eqref{consNF} and \eqref{consNf}. An example with $USp(2N)$ group is the Seiberg theory with $N_f=8$ or E-string theory. In addition, all the three classes of orbifold theories studied above with $N_f=8$ also give rise to $N^3$ scaling of the free energy.  

\subsection{Holography}\label{sec:Holography}

On general grounds, we expect the gravity dual of a 5d SCFT on $S^{3}\times \Sigma_{\fg}$  with a partial topological twist on $\Sigma_{\fg}$ to be given by a supersymmetric solution interpolating between asymptotically locally AdS$_{6}$  at infinity (with an $S^{3}\times \Sigma_{\fg}$ boundary) and an AdS$_{4}\times \Sigma_{\fg}$ geometry for small values of the radial coordinate.\footnote{The holographic solutions described here correspond to the case of a round 3-sphere $S^{3}_{b=1}$.} In the case of a universal topological twist ($\fn=0$) it was argued in \cite{Bobev:2017uzs} that such a solution, originally found in \cite{Nunez:2001pt,Naka:2002jz}, is given by an extremal 2-brane solution in minimal 6d $F(4)$ gauged supergravity, with metric of the form
\eqss{\label{sol6dsugra}
	ds^{2}&=e^{2f(r)}(-dt^{2}+dz_{1}^{2}+dz^{2}_{2}+dr^{2})+e^{2g(r)}ds^{2}_{\Sigma_{\fg}}\,,
}
which is supported by a nontrivial magnetic flux for a $U(1)\subset SU(2)$ graviphoton on $\Sigma_{\fg}$ and a nontrivial scalar $\varphi(r)$.\footnote{This supergravity theory \cite{Romans:1985tw} has 16 real supercharges and bosonic field content: the graviton $g_{\mu\nu}$, an $SU(2)$ gauge potential $A_{\mu}^{I}$, an Abelian one-form potential $A_{\mu}$, a massive two-index tensor gauge field $B_{\mu\nu}$, and a scalar field $\varphi$.} The interpolating solution, which exists only for $\fg>1$, preserves four real supercharges and can be found numerically. As the radial coordinate goes to $r\to \infty$ the metric is locally asymptotic to AdS$_{6}$ with unit radius and as $r\to0$ it is of the form AdS$_{4}\times \Sigma_{\fg}$, with sizes $e^{2f(r)}=\frac{1}{r^{2}}2^{3/2}3^{-3/2}$ and $e^{2g(r)}=2^{1/2}3^{-3/2}$, and the free energy given by \cite{Bobev:2017uzs}
\equ{\label{FS3FS5sugra}
	F_{S^{3}\times \Sigma_{\fg}}^{\mathit{sugra}}=-\frac{8}{9}\,(\fg-1)\, F_{S^{5}}\,,
}
where $F_{S^{5}}$ is the free energy of the 5d field theory on $S^{5}$. The crucial point, emphasized in \cite{Bobev:2017uzs}, is that upon uplift to massive IIA on topologically $S^{4}$, the solution \eqref{sol6dsugra} describes the twisted compactification of {\it any} 5d $\cN=1$ theory with a gravity dual and thus \eqref{FS3FS5sugra}  holds for any such compactification.\footnote{See also \cite{Benini:2015bwz} and \cite{Azzurli:2017kxo}  for similar discussions in other dimensions. } Indeed, this is corroborated by our field theory results; setting $Q=1$ for the round $S^{3}$, $\kappa=-1$,  and $\fn_{\rho}=0$ in \eqref{FS3S5nk} we see this exactly matches \eqref{FS3FS5sugra}. The existence of this holographic flow is thus strong evidence for the existence of a large class of 3d SCFTs arising from compactification of 5d $\cN=1$ theories, at least at large $N$, for the universal  twist. We emphasize that the  free energy of these 3d theories scales as $N^{5/2}$ rather than the more standard  $N^{3/2}$ or $N^{5/3}$ scaling of 3d theories with gravity duals. 

The  general  result \eqref{FS3S5nk} suggests that a larger class of such holographic flows should exist, whose endpoint is described by a discrete family  of 3d SCFTs labeled by the integers $\fn_{\rho}$ for each  5d parent theory. The simplest example is the Seiberg theory with a nonzero flux for the Cartan of $SU(2)_{M}$. The explicit holographic RG flow in this case was constructed for $Q=1$ in \cite{Bah:2018lyv}. Indeed, specifying \eqref{FS3S5nk} for this case exactly matches the supergravity result (see Eq. (1.1) in \cite{Bah:2018lyv}). It would be interesting to construct the analogous holographic RG flows for the orbifold theories considered in detail in Section~\ref{sec:Seiberg theory and its orbifolds}, which now have a number of flavor symmetries visible at large $N$  and whose free energy is given by \eqref{FExtQvrs}. In general, one should be able to do this for arbitrary values of $Q$, which would be interesting.

These holographic checks give strong support for the existence a novel class of 3d SCFTs. The results of this paper provide a method for computing the exact partition function of such theories, in principle at finite $N$. 


\section{5d theories, 6d  SCFTs, and the 4d index}\label{sec:5d6d4d}

As reviewed in the Introduction, some 5d gauge theories  are believed to be  low energy descriptions of certain 6d SCFTs compactified on a circle, $S^{1}_{\beta}$. The radius of the 6d circle  is related to the 5d gauge coupling constant by
\be \label{betag5gen} \beta =  \frac{g_5^2}{2 \pi \lambda} \,, \ee
where $\lambda$ is a numerical factor which depends on the specific theory under consideration.  The prototypical example of this phenomenon, which we will discuss in detail in Section~\ref{sec:N=2SYM}, is that of the maximally supersymmetric Yang-Mills theory in 5d with an ADE gauge group, which is expected to have a UV completion as the circle compactification of the corresponding $\cN=(2,0)$ SCFT.  There are also other examples of this phenomenon with $\cN=(1,0)$ supersymmetry, including theories obtained by orbifolding the maximal theory, and the E-string, which uplifts to the 6d E-string theory.

When one computes protected observables in these 5d theories on a compact manifold, $\cM_5$, one generally expects that these may be interpreted as observables in the ``parent'' 6d theory on $\cM_5 \times S^1_\beta$.  Indeed, this philosophy was applied in the case of $\cM_5$ being the squashed five-sphere to study the superconformal index, or $S^5 \times S^1_\beta$ partition function \cite{Kim:2012qf,Kim:2013nva}, as well as other examples; see \cite{Kim:2016usy} for a review. In the case of the $S^3_b \times \Sigma_\g$ partition function, we  expect this to compute the partition function of the 6d SCFT on 
\be S^3_b \times \Sigma_\g \times S^1_\beta\,. \ee

\begin{figure}
	\centering
	\includegraphics[scale=1.1]{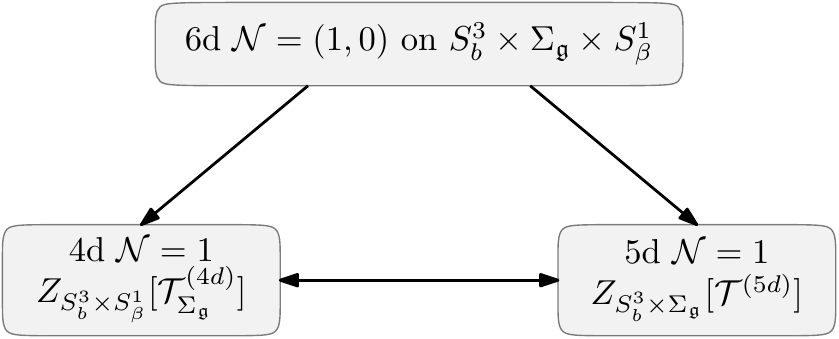}
	\caption{The partition function of 4d $\cN=1$ class $\cS$ theories on  $S^{3}_{b}\times S^{1}_{\beta}$  from the partition function of $\cN=1$ gauge theories on $S^{3}_{b}\times \Sigma_{\fg}$. The precise relation is given in \eqref{rel4d5d} with the mapping of parameters  in \eqref{indrel2}.}
	\label{fig456}
\end{figure}

This leads to another perspective on this object, as follows.  By compactifying the parent 6d $\cN=(1,0)$  theory  on the Riemann surface instead leads to a class of 4d offspring  theories labeled by the compactification manifold, $\Sigma_{\fg}$,\footnote{In the general case the Riemann surface may contain nontrivial punctures, which we do not consider in this work. These should appear as local defect operators in the TQFT on $\Sigma_\g$.}  and flavor fluxes, $\fn$, which we denote here by $\cT^{(4d)}_{\Sigma_\g,\n}$.  Then our computation can be interpreted as giving the partition function of these 4d theories on $S^3_b \times S^1_\beta$ (see Figure~\ref{fig456}), \ie, 
\equ{\label{rel4d5d}
	Z_{S^3_b \times \Sigma_\g}(\nu,\gamma)_\n[\cT^{(5d)}] = Z_{S^3_{b} \times S^1_ β}(p,q,\mu)[\cT^{(4d)}_{\Sigma_\g,\fn}] \,.
}
The latter is closely related  to the 4d supersymmetric index  as  \cite{Assel:2014paa} 
\be \label{indpartrel}Z_{S^3_b \times S^1_\beta} (p,q,\mu) = e^{-\beta E_{\mathit{Casimir}}}   {\cal I}_{4d}(p,q,\mu)\,, \ee
where the 4d index may be defined as a trace in radial quantization \cite{Kinney:2005ej,Romelsberger:2005eg},
\be {\cal I}_{4d}(p,q,\mu) = \text{Tr} (-1)^F e^{-\beta' \{\cQ,\cQ^{\dagger} \} } p^{j_1+j_2-\frac{R}{2}} q^{j_1-j_2-\frac{R}{2}} \prod_i {\mu_i}^{F_i}\,, \ee	
where $j_1$ and $j_2$ are Cartan generators for the $SO(4) \cong SU(2)_1 \times SU(2)_2$ rotation symmetry, and $F_i$ run over a basis of the flavor symmetries. The quantity $E_{\mathit{Casimir}}$, to which we will return to more detail in Section \ref{sec:Casimir4d}, is the Casimir energy of the 4d theory quantized on $S^3_b$ \cite{Assel:2014paa}. 

The identification of parameters to those appearing naturally in the $S^3_b \times S^1_\beta$ partition function is 
\cite{Imamura:2011wg,Aharony:2013dha}
\be \label{indrel1} p = e^{- 2 \pi b Q^{-1} \beta}\,,\qquad   q = e^{- 2\pi b^{-1} Q^{-1} \beta}\,,\qquad   \mu_i = e^{2 \pi Q^{-1} \beta (\nu_i + iQ) } \,,\ee
where $\nu_i$ are the effective real mass parameters on $S^3_b$. Using \eqref{betag5gen} we find these are related to the parameters of the $S^3_b \times \Sigma_\g$ partition function by
\be \label{indrel2} p = e^{2 \pi b \lambda^{-1} \gamma^{-1}} \,,\qquad   q = e^{2 \pi b^{-1}\lambda^{-1} \gamma^{-1}} \,,\qquad   \mu = e^{2 \pi i \lambda^{-1} \gamma^{-1} (\nu_i+i Q)}\,, \ee
where $\gamma=-\frac{2 \pi Q}{{g_5}^2}$.
To summarize, this chain of reasoning leads us to identify
\be \label{indpfrel1} Z_{S^3_b \times \Sigma_\g}(\nu,\gamma)_\n[\cT^{(5d)}] = Z_{S^3_b \times S^1_β}(p,q,\mu)[\cT^{(4d)}_{\Sigma_\g,\fn}] = e^{-\beta E_{\mathit{Casimir}}}   {\cal I}_{4d}(p,q,\mu)[\cT^{(4d)}_{\Sigma_\g,\fn}]\,.\ee

In this section, we study this relation in more detail.  Although in principle this is an exact result, valid for any choice of parameters appearing in \eqref{indpfrel1}, in practice we are only able to compute the LHS in certain simplifying limits, where the contributions from instantons are under control.  We first consider in detail the case of the maximal, $\cN=2$ SYM theory.  Due to the extra supersymmetry, this admits a limit where the instanton contribution is very simple, and the partition function can be computed exactly.  Then we study the partition function in the ``Casimir limit,'' $\beta \sim {g_5}^2 \rightarrow \infty$.  Although na\"ively the instantons are important in this limit, we find their contribution can be explicitly characterized.

\subsection{\texorpdfstring{The 5d $\cN=2$ Yang-Mills theory}{The 5d N=2 Yang-Mills theory}}\label{sec:N=2SYM}

Let us start by considering the $S^3_b \times \Sigma_\g$ partition function for the maximally supersymmetric 5d $\cN=2$ Yang-Mills theory with simply laced gauge group $G$.\footnote{As mentioned in section \ref{sec:5dN=1curved}, this computation was also considered in \cite{Kawano:2012up,Fukuda:2012jr,Kawano:2015ssa}, where they also observed a relation to the Schur limit of the $4d$ index.  Below we clarify the precise relation of this observable to the 4d index of certain 4d $\cN=1$ SCFTs. }  We will often specialize to the case $G=SU(N)$ for concreteness, but will keep the discussion general when possible.  This theory contains an adjoint hypermultiplet, acted on by an $SU(2)_F$ flavor symmetry,\footnote{Specifically, we have $SU(2)_R \times SU(2)_F \subset USp(4)$, where the latter is the full R-symmetry group for this theory, and $SU(2)_R$ is the 5d $\cN=1$ R-symmetry.} for which we include a mass $\nu$ and flux $\n$.  Then the $S^3_b \times \Sigma_\g$ partition function of this theory can be written as a sum over Bethe vacua as
\be \label{n2bv}  Z_{S^3_b \times \Sigma_\g}(\nu,\gamma)_{\n} = \sum_{\hat{u} \in \cS_{BE}} \cH(\hat{u},\nu,\gamma)^{\g-1} \Pi_\nu(\hat{u},\nu,\gamma)^{\n}  \,.\ee
These operators can be constructed out of the twisted superpotential
\be \cW^{\cN=2}_{S^3_b \times \R^2}(\tu,\tnu,\tgamma) = \frac{1}{Q b}\cW^{\cN=2}_{NS}(\tu,\tnu,\tgamma;-b^2) +\frac{1}{Q b^{-1}}\cW^{\cN=2}_{NS}(\tu,\tnu,\tgamma;-b^{-2}) \,,\ee
in the notation of Section~\ref{sec:5d4d}, and similarly for the effective dilaton.  We may alternatively decompose this into a perturbative and instanton contribution,
\be \cW^{\cN=2}_{S^3_b \times \R^2}(\tu,\tnu,\tgamma) = \cW^{\cN=2,pert}_{S^3_b \times \R^2}(\tu,\tnu,\tgamma) + \cW^{\cN=2,inst}_{S^3_b \times \R^2}(\tu,\tnu,\tgamma) \,,\ee
and we may write an explicit expression for the first term, as in Section~\ref{sec:amodreduc}:
\be  \label{n2pertW} \cW^{\cN=2,pert}_{S^3_b \times \R^2}(\tu,\tnu,\tgamma)  = \frac{1}{2} \tgamma K^{ab} \tu_a \tu_b + \sum_{\alpha \in Ad(G)} g_b(\alpha(\tu) + \tnu) - \sum_{\alpha \in Ad(G)'} g_b(\alpha(\tu)+1) \, ,\ee
where $\tgamma=-\frac{2 \pi i}{{g_5}^2}$.  In general it is difficult to evaluate the non-perturbative contribution analytically.  However, we will describe some simplifying limits below.

To see what the $S^3_b \times \Sigma_\g$ partition function of this theory corresponds to in 4d, we recall that the 6d UV completion of this 5d theory is the $\cN=(2,0)$ theory compactified on a circle of radius
\be \beta = \frac{{g_5}^2}{2 \pi} \,.\ee
In other words, \eqref{betag5gen} holds with  $\lambda=1$ in this case.  The compactification of the 6d SCFT on a Riemann surface with a topological twist gives rise, in general, to a class of 4d $\cN=1$ SCFTs, which were described as the $A_{N-1}$ case in \cite{Bah:2012dg}.  The specific theory depends on the choice of R-symmetry used to perform the twist.  If we mix the $U(1)_R \subset SU(2)_R$ symmetry used to perform the topological twist on $\Sigma_\g$ with the maximal torus of the $SU(2)_F$ flavor symmetry with a coefficient $\hat{\n}$, this is equivalent to inserting a flux,\footnote{In \cite{Bah:2012dg} the parameter we call $\hat{\n}$ was denoted by $z$.}
\be \n = \hat{\n}(\g-1) \,,\ee
on $\Sigma_\g$.  In the special case
\be \hat{\n} = \pm 1\,, \ee
the twist preserves $\cN=2$ supersymmetry in 4d, and gives rise to the theories of class $\cS$ \cite{Gaiotto:2009we}.  Let us denote the theory  for general choices of flux, $\n$, by $\cT^{(4d)}_{\Sigma_\g,\n}$.

As a special case of the 4d-2d correspondence, we may relate the index of this 4d theory to the partition function of a suitable 2d TQFT \cite{Gadde:2009kb,Rastelli:2014jja}.  Then it is clear from the logic above that this TQFT is precisely the A-twist of the effective 5d $\cN=2$ theory compactified on $S^3_b$.  In other words, we have
\be \label{n2s3bind}Z_{S^3_b \times S^1_\beta}[\cT^{(4d)}_{\Sigma_\g,\n}] = Z_{S^3_b \times \Sigma_{\fg}}(\nu)_\n [\cT^{(5d)}] \,,\ee
where the RHS may be interpreted as a TQFT living on $\Sigma_\g$.  In particular, in writing the RHS as a sum over Bethe vacua, as in \eqref{n2bv}, we exhibit this TQFT structure explicitly.

While we expect the relation \eqref{n2s3bind} to hold for general parameters, it is in general difficult to evaluate the $S^3_b \times \Sigma_\g$ partition function explicitly due to the nontrivial instanton contributions.  Thus we first consider a special limit, where their contribution simplifies significantly.  We will see this limit turns out to be related to the ``Schur limit'' of the 4d $\cN=2$ superconformal index \cite{Gadde:2011ik}.  We then briefly comment on the general case.

\paragraph{The Schur limit.}

Let us first consider the perturbative contribution to the (ungauged) partition function in more detail.  Using \eqref{n2pertW}, and working in terms of the untilded variables, we have\footnote{Here we recall the product with the prime includes only the non-zero roots, while the unprimed product includes all roots.}
\eqst{Z^{\cN=2,pert}_{S^3_b \times \Sigma_\g}(u,\nu,\gamma)_{\m,\n} = e^{2 \pi i \gamma K^{ab} u_a \m_b} \prod_{\alpha \in Ad(G)} s_b\(\alpha(u)+\nu\)^{\alpha(\m) + \n} \\
\times \prod_{\alpha \in Ad(G)'} s_b\(\alpha(u) -i Q\)^{-\alpha(\m)+1-\g}.
} 
Now let us consider the limit\footnote{In \cite{Razamat:2014pta} this limit of the $S^3_b$ partition function was shown to correspond to the dimensional reduction of the Schur limit of the supersymmetric index.}
\be \label{nulim} \nu = \tfrac{i}{2}(b-b^{-1}) \,.\ee
Then, using the following identities of the double sine function,
\be \label{sbid1} s_b(x) = s_b(-x)^{-1}\,, \ee
\be \label{sbid2} s_b\(x+ \tfrac{i}{2} b^{\pm}\) = \frac{1}{2 \cosh (\pi b^{\pm} x)} s_b\(x- \tfrac{i}{2} b^{\pm}\)\,,  \ee
we find the perturbative contribution simplifies to
\eqst{Z^{\cN=2,pert}_{S^3_b \times \Sigma_\g}(u,\nu,\gamma)_{\m,\n} \\
	=e^{2 \pi i \gamma K^{ab} u_a \m_b} b^{-\n r_G} \prod_{\alpha>0} \left[2 \sinh\(\pi b \alpha(u)\)\right]^{1-\g-\n}\left[2 \sinh\(\pi b^{-1} \alpha(u)\)\right]^{1-\g+\n}\,.
} 
where we simplified the Cartan contribution of the adjoint hypermultiplet using $s_b(\nu) = s_b(\frac{i}{2}(b-b^{-1})) = b^{-1}$.  In particular, we see the dependence on the gauge flux $\m$, and hence also the perturbative Bethe equations, are very simple.  

Next, we consider the instanton contribution, which can be written as
\be Z^{\cN=2,inst}_{S^3_b \times \Sigma_\g}(u,\nu,\gamma)_{\m,\n} = \prod_\ell Z^{\cN=2,\mathit{inst}}_{\R^2_{\q^{(\ell)}_1} \times \R^2_{\q^{(\ell)}_2} \times S^1}\(x^{(\ell)},y^{(\ell)},z^{(\ell)}\) \,.\ee
Here the parameter, $y^{(\ell)}$, for the flavor symmetry is given by, using \eqref{ueldef5d2},
\be y^{(\ell)} = \left\{ \begin{array}{cc} e^{2 \pi b \nu} {\q_2^{(\ell)}}^{\n/2} = - {\q^{(\ell)}_1}^{1/2} {\q_2^{(\ell)}}^{\n/2} &\quad \ell=nn \; \text{or} \; ns \\ 
	e^{2 \pi b^{-1} \nu} {\q_2^{(\ell)}}^{\n/2} = - {\q^{(\ell)}_1}^{-1/2} {\q_2^{(\ell)}}^{\n/2} & \quad\ell=sn \; \text{or} \; ss \end{array}\right. \ee
Let us now specialize to the case $G=SU(N)$.  Then we note the following simplification of the instanton contribution to the Nekrasov partition function of the $\cN=2$ theory:
\eqs{
	Z^{\cN=2,\mathit{inst}}_{\R^2_{\q_1} \times \R^2_{\q_2} \times S^1}\(x,y=-(\q_1 \q_2)^{\pm 1/2},z\)=&\, \frac{1}{\eta(z)^{N-1}}\,, \\
	Z^{\cN=2,\mathit{inst}}_{\R^2_{\q_1} \times \R^2_{\q_2} \times S^1}\(x,y=-(\q_1 \q_2^{-1})^{\pm 1/2},z\)=&\,1\,. 
}
This is derived in Appendix \ref{sec:maxapp}, and was also noted in the context of the $S^5$ \cite{Kim:2012qf} and $\CP^2 \times S^1$ \cite{Kim:2013nva}, partition functions.  Then we see that if we set $\n=1$ we have
\eqs{Z^{\cN=2,\mathit{inst}}_{S^3_b \times S^2}&\(u,\nu=\tfrac{i}{2}(b-b^{-1}),\gamma\)_{\m,\n=1} \nn
&= \prod_{\ell=nn,ns} Z^{\cN=2,\mathit{inst}}_{\R^2_{\q^{(\ell)}_1} \times \R^2_{\q^{(\ell)}_2} \times S^1}\(x^{(\ell)},y^{(\ell)}=-{\q^{(\ell)}_1}^{1/2}{\q^{(\ell)}_2}^{1/2},z^{(\ell)}\) \nn
&\quad \times \prod_{\ell=sn,ss} Z^{\cN=2,\mathit{inst}}_{\R^2_{\q^{(\ell)}_1} \times \R^2_{\q^{(\ell)}_2} \times S^1}\(x^{(\ell)},y^{(\ell)}=-{\q^{(\ell)}_1}^{-1/2}{\q^{(\ell)}_2}^{1/2},z^{(\ell)}\) \nn
&= \frac{1}{\eta(z)^{2(N-1)}} \,,
}
where we recall from \eqref{yzdef} that
\be z^{(nn,ns)} = z = e^{-2 \pi b \gamma}\,,\qquad z^{(sn,ss)}=\bar{z} = e^{-2 \pi b^{-1} \gamma}\,. \ee
Similarly, for $\n=-1$, we have
\be Z^{\cN=2,\mathit{inst}}_{S^3_b \times S^2}\(u,\nu=\tfrac{i}{2}(b-b^{-1}),\gamma\)_{\m,\n=-1} =\frac{1}{\eta(\bar{z})^{2(N-1)}}\,. \ee 
On the other hand, we expect the instanton contribution to the integrand of the $S^3_b \times S^2$ partition function to be equal to
\be e^{-2 \pi i\Omega_{\mathit{inst}}} \({\Pi^{inst}_\nu}\)^{\n} \({\Pi^{inst}_a}\)^{\m_a} \,.\ee
Equating these expressions for $\n=\pm 1$, we deduce that (up to a sign)
\be e^{2 \pi i \Omega_{\mathit{inst}}} = \eta(z)^{N-1} \eta(\bar{z})^{N-1} \,, \qquad \Pi^{inst}_\nu = \frac{\eta(\bar{z})^{N-1}}{\eta(z)^{N-1}} \,,\qquad \Pi^{inst}_a = 1 \,. \ee
The last relation also implies that there is no instanton contribution to the Hessian determinant appearing in the handle-gluing operator.

It will be convenient to rewrite the $\eta$ functions appearing in the handle-gluing and flux operators to relate them to the index parameters in \eqref{indrel2},
\be \label{indrel3} p=e^{2 \pi \gamma^{-1} b} \,, \qquad q= e^{2 \pi \gamma^{-1} b^{-1}}\,. \ee 
Namely, using the modular properties of the $\eta$ functions, we have
\eqs{ \label{etamod} \eta(z) =&\, \eta(e^{-2 \pi b \gamma}) =(b \gamma)^{-1/2}  \eta(e^{2 \pi \gamma^{-1} b^{-1}})= (b \gamma)^{-1/2}   \eta(q) \,, \nonumber \\
	\eta(\bar{z}) =&\, \eta(e^{-2 \pi  b^{-1}\gamma}) =(b^{-1} \gamma)^{-1/2}  \eta(e^{2 \pi \gamma^{-1} b}) = (b^{-1} \gamma)^{-1/2}   \eta(p)\,.
}
We may now write the full, non-perturbatively complete operators for the $S^3_b \times \Sigma_\g$ partition function in the limit \eqref{nulim}.  We have
\eqs{ 
	\Pi_a =&\, e^{2 \pi i \gamma K^{ab} u_b}\,,\qquad  \Pi_\nu = \frac{\eta(p)^{N-1}}{\eta(q)^{N-1}}\prod_{\alpha>0} \frac{2 \sinh\(\pi b^{-1} \alpha(u)\)}{2 \sinh\(\pi b \alpha(u)\)}  \,, \\
	\cH =&\, \eta(p)^{N-1} \eta(q)^{N-1} \prod_{\alpha>0} \left[2 \sinh\(\pi b \alpha(u)\) 2 \sinh\(\pi b^{-1} \alpha(u)\)\right]^{-1}\,.
}
Note that the Hessian determinant contributes a factor of $\gamma^{N-1}$, which precisely cancels against the factors of $\gamma$ from \eqref{etamod}, and similarly the factor of $b^{-(N-1)}$ from the Cartan component of the adjoint hypermultiplet precisely cancels the factors of $b$ from \eqref{etamod}.

It is now trivial to write the solutions to the Bethe equations, $\Pi_a=1$, which are given by
\be \hat{u}_a =  \gamma^{-1} K^{-1}_{ab} n^b \equiv \gamma^{-1} \lambda_a \,,\qquad  n^b \in \Z \;\; \Leftrightarrow \;\; \lambda \in \Lambda_{cr} \;,\ee
where $\lambda_a \equiv K^{-1}_{ab} n^b$ runs over the coroot lattice of $G$ as we vary over $n^b \in \Z$.  Without loss of generality, we may restrict to coroots in the interior of the fundamental Weyl chamber,\footnote{Those on the boundary of the Weyl chamber will lead to Bethe vacua with enhanced Weyl symmetry, which we are instructed to discard.} or equivalently, those which can be written as $\delta+\lambda$, for $\lambda$ a dominant coroot, and $\delta$ the Weyl vector of $G$.  Then the set of Bethe vacua is
\be \cS_{BE}= \left\{ \ \hat{u}^{(\lambda)}_a \equiv \gamma^{-1} (\lambda_a + \delta_a) \;\; \big| \;\;  \lambda \in \Lambda_{cr}^+ \right\}\,. \ee

It remains to compute the partition function by evaluating the flux and handle-gluing operators at the above Bethe vacua.  First we observe
\be \prod_{\alpha>0} 2 \sinh\(\pi  b \alpha(\hat{u}^{(\lambda)})\) = \prod_{\alpha>0} (p^{\alpha(\lambda+\delta)/2} - p^{-\alpha(\lambda+\delta)/2}) = V(p) \dim_p R_\lambda  \,,\ee
and similarly
\be \prod_{\alpha>0} 2 \sinh\(\pi  b^{-1} \alpha(\hat{u}^{(\lambda)})\)  = V(q) \dim_q R_\lambda  \,,\ee
where $R_\lambda$ is the representation with highest weight $\lambda$, where we have identified the coroot and weight lattices using the Killing metric, and $\dim_p R$ is the ``quantum dimension'' of a representation $R$ of $G$.  Here we have made the identification to the parameters of the 4d index in \eqref{indrel3}.  Finally, $V(p)$ is the polynomial (here we specialize again to $SU(N)$, but the generalization is straightforward)
\be V(p) \equiv \det_{1 \leq a,b \leq N} p^{\(\frac{N+1}{2}-a\)(N-b)} = (-1)^{N(N-1)} p^{-\frac{1}{12}N(N^2-1)} \prod_{i=1}^{N-1} (1-p^i)^{N-i} \,.\ee

Putting this together, we see that the $S^3_b \times \Sigma_\g$ partition function for the $\cN=2$ $SU(N)$ theory, in the limit \eqref{nulim}, is given by
\eqst{Z_{S^3_b \times \Sigma_\g}^{\cN=2}\(\nu=\tfrac{i}{2}(b-b^{-1})\)_\n = \sum_{\hat{u} \in \cS_{BE}} \Pi_\nu(\hat{u})^{\n} \cH(\hat{u})^{\g-1} \\
=  \sum_{\lambda \in \Lambda_{cr}^+} \bigg(\frac{V(p) \dim_p (R_\lambda)}{\eta(p)^{N-1}} \bigg)^{-\n+1-\g} \bigg(\frac{V(q) \dim_q (R_\lambda)}{\eta(q)^{N-1}} \bigg)^{\n+1-\g} \,.
\label{n2ans}}

\begin{flushleft} 
	{\it 4d interpretation}
\end{flushleft} 

As described above, we expect this observable to compute the 4d superconformal index of the theory $\cT^{(4d)}_{\Sigma_\g,\n}$.  Specifically, using the map \eqref{indrel2}, we can see that the limit \eqref{nulim} corresponds to setting the fugacity $\mu$ corresponding to this flavor symmetry in 4d as
\be \mu = p \,.\ee
In the case of $\cN=2$ class $\cS$ theories, $\mu$ is usually referred to as $t$, and is the fugacity for the $U(1)_r$ symmetry in the 4d $\cN=2$ algebra.  This limit of the index was first studied for the $\cN=2$ class $\cS$ theories in \cite{Gadde:2011ik}, where it was referred to as the ``Schur limit'' of the index.  It was later generalized to the 4d $\cN=1$ theories corresponding to a more general choice of flux, $\n$, in \cite{Beem:2012yn}, where it was referred to as the ``mixed Schur index.''  The general result they found for the 4d index can be written in our notation as
\be {\cal I}_{4d}(p,q)[\cT^{(4d)}_{\Sigma_\g,\n}] \!=\! \bigg(\!\frac{p^{\frac{1}{12}N(N^2-1)} V(p)}{(p;p)^{N-1}}\!\bigg)^{-\ell_1}\!\bigg(\!\frac{q^{\frac{1}{12}N(N^2-1)} V(q)}{(q;q)^{N-1}}\!\bigg)^{-\ell_2}\!\! \sum_{\lambda \in \Lambda^+_{w}} \text{dim}_p(R_\lambda)^{-\ell_1} \text{dim}_q(R_\lambda)^{-\ell_2},\ee
where the sum is over representations, $R_\lambda$, of $SU(N)$, labeled by weights $\lambda$, $p$ and $q$ are the parameters of the index, and
\be \ell_1 = \g-1+\n \,,\qquad  \ell_2 = \g-1-\n \,.\ee
Comparing to \eqref{n2ans}, we see these agree for general choices of genus $\g$ and flux $\n$, up to an overall factor of \footnote{The terms proportional to $\frac{N-1}{24}$ in the exponents arise due to the relation between the Dedekind eta function and the $q$-Pochhammer symbol, \ie, $\eta(q) = q^{\frac{1}{24}}(q;q).$ }
\be \label{schurfact} p^{-\ell_1( \frac{N(N^2-1)}{12}+\frac{N-1}{24})} q^{-\ell_2( \frac{N(N^2-1)}{12}+\frac{N-1}{24})} \;  \,.\ee
Using \eqref{indrel1}, we may rewrite the powers of $p$ and $q$ appearing here as 
\eqsc{p^{-\ell_1 (\frac{N(N^2-1)}{12}+\frac{N-1}{24})} q^{-\ell_2( \frac{ N(N^2-1)}{12}+\frac{N-1}{24})} = e^{-\beta  \hat E_{\mathit{Casimir}} }\,;\nn
\hat E_{\mathit{Casimir}} = 2 \pi Q^{-1}\(\tfrac{N(N^2-1)}{12} + \tfrac{N-1}{24} \)(\ell_1 b+\ell_2 b^{-1} )\,.
\label{EhatCasimirSchur}}
As we show in Appendix~\ref{App:CasimirS3Sigma} this prefactor precisely matches the Casimir energy of the 4d theory  (see \eqref{CasimirMaxInstSchur}) and thus we precisely recover the expected relation  \eqref{indpartrel} between the $S^3_b \times \Sigma_\g$ partition function and the 4d index of these 4d $\cN=1$ theories.  This serves as a strong consistency check of our calculation. We discuss the Casimir energy  for general parameters and more general theories in Section~\ref{sec:Casimir4d}.

\paragraph{General parameters.}

Above we considered the partition function in a very special limit of parameters, where it drastically simplified, leading to an explicit evaluation formula.  A natural and (as far as the authors are aware) open question is to compute the 4d index of class $\cS$ theories for more general choices of the parameters, and in particular, find the dual TQFT that describes it.  

In principle the results derived above imply that the answer to this question is controlled by the Nekrasov-Shatashvili limit of the 5d instanton partition function of the maximal $\cN=2$ SYM theory.  That is, the TQFT in question is governed by the twisted superpotential
\be \cW^{\cN=2}_{S^3_b \times \R^2}(\tu,\tnu,\tgamma) = \cW^{\cN=2,pert}_{S^3_b \times \R^2}(\tu,\tnu,\tgamma) + \cW^{\cN=2,inst}_{S^3_b \times \R^2}(\tu,\tnu,\tgamma)\,. \ee
Then the equations determining the supersymmetric vacua of the effective 2d theory, and hence the states of the 2d TQFT, are given by solving the Bethe equations,
\be e^{2 \pi i \partial_{\tu_a} \cW^{\cN=2}_{S^3_b \times \R^2}(\tu,\tnu,\tgamma) } = 1\,, \qquad a= 1,...,r_G\,. \ee
By the gauge-Bethe correspondence, it is known \cite{Nekrasov:2009rc} that the vacuum equations associated to the NS-limit of the instanton partition function are equivalent to the Bethe equations of a certain integrable system.  In the case of the 5d $\cN=2$ theory, this system is a quantization of the {\it relativistic Calogero-Moser system}, also known as the {\it  Ruijsenaars-Schneider} (RS) model \cite{ruijsenaars2004elliptic}.  The relation between the 4d index and this integrable system was already pointed out in \cite{Gaiotto:2012xa}, where they observed that the indices of class $\cS$ theories are naturally eigenfunctions of certain difference operators associated to the RS model.  We expect this relation can be naturally understood in the above framework.  Moreover, it would be interesting to explore whether the perspective above allows one to practically compute the 4d index of non-Lagrangian class $\cS$ theories, which is a long-standing open problem.\footnote{There has been some partial progress via various indirect methods, such as constructing Lagrangians with reduced supersymmetry which flow to these non-Lagrangian class $\cS$ fixed points; see, \eg, \cite{Gadde:2015xta,Agarwal:2018ejn}.}  We hope to return to this in future work.

\subsection{\texorpdfstring{5d $\cN=1$ theories}{5d N=1 theories}}\label{sec:5dN=1theories}

We now consider the case of $\cN=1$ theories with a 6d $\cN=(1,0)$ UV completion.  We consider two known examples in detail. The first example is a set of 5d circular quiver gauge theories whose UV completion is a $\mathbb Z_{k}$ orbifold of the 6d $A_{N-1}$ theory, and which give rise, upon compactification of the 6d theory, to the 4d theories of class $\cS_{k}$ \cite{Gaiotto:2015usa}.  The second example is the Seiberg theory in the special case $N_{f}=8$, whose UV completion is the so-called E-string theory \cite{Witten:1995gx,Seiberg:1996bd}. The twisted compactifications of these two classes of theories were considered in \cite{Bah:2017gph} and \cite{Kim:2017toz}, respectively.  Then, as for the maximal theory above, we expect that the $S^3_b \times \Sigma_\g$ partition function of these 5d theories computes the 4d index of the corresponding 4d theories.

We also expect the $S^3_b \times \Sigma_\g$ partition function of these 5d theories, and therefore the supersymmetric indices of the corresponding 4d theories, to be related to an appropriate integrable system, through the gauge-Bethe correspondence for the Nekrasov-Shatashvili limit of the 5d instanton partition function \cite{Nekrasov:2009rc}.  Indeed, a connection of the corresponding 4d models to integrable systems was observed in these two examples, namely, to a generalization of the RS model discussed in \cite{Gaiotto:2015usa,Maruyoshi:2016caf,Ito:2016fpl,Yagi:2017hmj} for the class $\cS_k$ case, and to the {\it  van Diejen model} in the E-string case \cite{Nazzal:2018brc}.  It would be interesting to use the computation above to shed more light on the connection between these gauge theories and integrable systems.  As in the $\cN=2$ case discussed above, this requires a detailed understanding of the instanton corrections to the partition function.

In the present work, we will limit ourselves to a simplifying limit where we may evaluate the $S^3_b \times \Sigma_\g$ partition function analytically.  Specifically, we will consider the partition function in the strong coupling limit, $g_{5} \rightarrow \infty$. Although na\"ively in this limit we expect the contributions from instantons to be large, we will see that in fact their contribution in this limit can be simply characterized, and so we may compute the leading behavior of the partition function analytically. 

We first study the general behavior of 5d theories with 6d completions in the strong coupling limit in  Section~\ref{sec:Casimir4d}.  We then describe the instanton corrections arising in this limit, and in the process conjecture a general condition for 5d theories to admit such 6d completions in Section~\ref{sec:casimirinstantons}.  Finally, we compute the $S^3_b \times \Sigma_\g$ partition function this limit for the $\cN=2$ theory, as well as two examples mentioned above, and compare to the corresponding 4d computations, in Section~\ref{sec:Examples}.

\subsubsection{Casimir energy}\label{sec:Casimir4d}

We begin by showing that for generic theories with a 6d UV completion, the strong-coupling limit $g_{5}\to \infty$   of the partition function  exhibits the behavior
\equ{\label{Casimir5d}
	\lim_{g_{5}\to \infty} Z_{S^{3}_{b}\times \Sigma_{\fg}}\approx e^{-\beta \hat E_{\mathit{Casimir}}}\,,
}
where $\beta$ is the radius of the 6d circle, related to $g_{5}$ via \eqref{betag5gen}. The quantity $\hat E_{\mathit{Casimir}}$ appearing here has the following interpretation. In theories with a 6d UV completion, the limit $g_{5}\to \infty$ corresponds to the  $\beta\to \infty$ limit of the 6d partition function on $S^3_b \times \Sigma_\g \times S^1_\beta$. Since in this limit the partition function is dominated by the vacuum state, $\hat E_{\mathit{Casimir}}$ denotes the vacuum energy of the 6d theory quantized on $S^3_b \times \Sigma_\g$.  Equivalently, by shrinking $\Sigma_{\fg}$ to zero size, it also corresponds to the Casimir energy of the 4d theory thus obtained, quantized on $S^{3}_{b}$. 

On the other hand, this Casimir energy can be independently computed from the anomaly polynomial, $I_{8}$, of the parent 6d theory (see  Appendix~\ref{App:CasimirS3Sigma} for details) by performing the topological twist, integrating over the Riemann surface, and applying the method of equivariant integration of \cite{Bobev:2015kza}, \ie,  
\equ{\label{EcGenPrestext}
	E_{\mathit{Casimir}}=\int_{\varepsilon} \int_{\Sigma_{\fg}}\, I_{8}^{\mathit{twisted}}\,.
}	
For consistency, the two quantities  obtained from \eqref{Casimir5d} and \eqref{EcGenPrestext} should  coincide: $\hat E_{\mathit{Casimir}}=E_{\mathit{Casimir}}$. We have already seen an example of this in \eqref{EhatCasimirSchur} for the Schur limit. 

In the case of $\cN=1$, however, to make the computation feasible we limit ourselves to the perturbative partition function in \eqref{Casimir5d}. Thus, we expect the Casimir energy extracted from there to differ from the exact result \eqref{EcGenPrestext}. Nonetheless, the difference can be easily characterized. Namely, decomposing $I_{8}=I_{8, \mathit{ free}}+I_{8, \mathit{ int}}$, 
where $I_{8, \mathit{ free}}$ is the contribution from all the free multiplets in the tensor branch of the theory, and $I_{8, \mathit{ int}}$ is the remaining part, each term gives a corresponding contribution to \eqref{EcGenPrestext}.  

Then, we will find that the perturbative result matches exactly the latter and thus,  for consistency, the instanton contribution must be given by the former:
\equ{\label{caspertinst}
	\hat E_{\mathit{Casimir}}^{\mathit{pert}}=\int_{\varepsilon} \int_{\Sigma_{\fg}}\, I_{8, \mathit{ int}}^{\mathit{twisted}}\,,\qquad \qquad 
	\hat E_{\mathit{Casimir}}^{\mathit{inst}}=\int_{\varepsilon} \int_{\Sigma_{\fg}}\, I_{8, \mathit{ free}}^{\mathit{twisted}}\,.
} 
The analogous observation  for the case of $S^{5}$ was already made in \cite{Bobev:2015kza}, suggesting a deeper understanding of this fact. We now proceed with the explicit computation. 

\paragraph{Computation of the $S^3_b \times \Sigma_\g$ partition function in the $g_5 \rightarrow \infty$ limit.}

To study this limit, it will be useful to go to the Bethe sum formulation of the partition function.  Let us consider a general 5d theory, with gauge group $G$ and hypermultiplets in a representation $R = \bigoplus_i R_i$ of $G$.\footnote{Here we do not include a bare 5d Chern-Simons term, as one does not generally appear in 5d theories with 6d UV completions.}  Then recall the Bethe equations are given by
\bea \label{CasLimBE} \Pi_a = e^{2 \pi i \gamma K^{ab} u_b} \prod_{\rho \in R_i} s_b(\rho(u) + \nu_i)^{\rho_a} \prod_{\alpha \in Ad(G)'}s_b(\alpha(u) -iQ)^{-\alpha_a} \Pi_a^{inst}(u,\nu,\gamma)=1\,.
\eea
where $\Pi_a^{inst}$ is the contribution from instantons, and we recall
\equ{\label{defgammasec4}
	\gamma= -\frac{2 \pi Q}{g_5^2} \,,
}
Then the limit \eqref{Casimir5d} corresponds to taking $\gamma \rightarrow 0$, so we consider the solutions to \eqref{CasLimBE} in this limit.  More precisely, we expect the partition function to be divergent in this limit, so we are interested in the solutions which contribute to the leading divergence.

Let us first ignore the contribution of the instantons.  Then we see that, for any solutions at finite $u$, the exponent in the first factor of \eqref{CasLimBE} becomes negligible as $\gamma \rightarrow 0$.  Then we expect these solutions to have a finite contribution to the partition function as $\gamma \rightarrow 0$.  Since we are interested in extracting the leading divergence in the limit \eqref{Casimir5d}, we instead look for solutions at large $u$, scaling with $\gamma^{-1}$.  These can arise due to competition between this exponential factor and the double sine functions in \eqref{CasLimBE}.

Since we are working at large $u$, it will be useful to recall the expansion of the double sine function, as in \eqref{sbexp},
\be \label{CasLimsblim} s_b(x) \rightarrow e^{ \pm \frac{\pi i}{2}\(  x^2  + \frac{1}{3} Q^2 - \frac{1}{6} \) }\qquad \text{for $\text{Re}(x) \rightarrow \pm \infty$} \,,\ee
where $Q= \frac{1}{2}(b+b^{-1})$.  
At this point we must determine the sign of $\rho(u)$ for each weight $\rho$.  Without loss of generality we may take $u$ to lie in the fundamental Weyl chamber.  For certain representations, such as the adjoint, this fixes the sign of $\rho(u)$ for all weights $\rho$.  In this case we will call a weight ``positive,'' and write $\rho>0$, if $\rho(u)>0$ for $u$ in the chosen fundamental Weyl chamber, generalizing the notation for weights of the adjoint representation.  For general representations, we may need to say more about the region where $u$ lies before making this split into positive and negative weights, but we will not consider such representations here.\footnote{For example, for the fundamental representation of $SU(N)$, specifying we are in the fundamental Weyl chamber, with $u_1>...>u_N$ does not specify the sign of all $u_i$; there are still $N-1$ regions in this chamber where the separation into positive and negative weights are distinct.}  We also restrict to representations that are self conjugate, and then we may group the weights into pairs, $(\rho,-\rho)$, where $\rho>0$.   Then let us define, for a general such representation $R$ of $G$,
\be \cC_R^{abc} = \frac{1}{2} \sum_{\rho>0} \rho^a \rho^b \rho^c\,,\qquad \cC_R^a = \frac{1}{2} \sum_{\rho>0} \rho^a\,. \ee
For example, for $R=Ad$, the adjoint representation, $\cC^a_{Ad} = \delta^a$, where $\delta = \frac{1}{2}\sum_{\alpha>0} \alpha$ is the Weyl vector of the group.

With this background, and using \eqref{CasLimsblim}, we may approximate the LHS of the Bethe equations by
\eqst{ \label{CasLimBEexp} \Pi_a \underset{u \rightarrow \infty}{\longrightarrow} \exp \bigg\{2 \pi i \bigg( \gamma K^{ab} u_b + \sum_i \big( \cC^{abc}_{R_i} u_b u_c + \cC^a_{R_i} ({\nu_i}^2 + \tfrac{1}{3} Q^2 -\tfrac{1}{6}) \big) \\
-\cC^{abc}_{Ad} u_b u_c -\cC^a_{Ad} \( -\tfrac{2}{3} Q^2 -\tfrac{1}{6}\) \bigg) \bigg\}\,. 
}
Generically, the term quadratic in $u$ will dominate the first term at large $u$, and then the solutions will not scale with $\gamma^{-1}$ as $\gamma \rightarrow 0$.  However, suppose we consider a theory with field content such that\footnote{We note the same relation can be derived from the conditions \eqref{consNF} and \eqref{consNf} discussed in Section \ref{sec:largeN} as the condition to obtain $N^3$ scaling of the free energy.}
\be \label{CasLimC3vanish}  \cC^{abc}_{Ad}-\sum_i \cC^{abc}_{R_i}  =0\,,  \ee
then \eqref{CasLimBEexp} simplifies to (after using $ν=-\i Q\tnu$)
\be \label{CasLimBEexpnq} \Pi_a \underset{u \rightarrow \infty}{\longrightarrow} \exp \bigg\{2 \pi i \bigg( \gamma K^{ab} u_b + \sum_i \cC^a_{R_i} \(Q^2(1 -{\tnu_i}^2) -\tfrac{1+4Q^2}{6}\) +\cC^a_{Ad}\(\tfrac{1+4Q^2}{6}\) \bigg) \bigg\}\,. \ee
The solutions in this limit are then found by setting the exponent equal to $-2 \pi i n^a$, $n^a \in \Z$, and  we have
\be \label{CasLimsigmanj} \hat{u}^{(n)}_a \approx - \gamma^{-1} K_{ab} \bigg( \sum_i  \cC^b_{R_i} \(Q^2(1 -{\tnu_i}^2) -\tfrac{1+4Q^2}{6}\) +\cC^b_{Ad} \(\tfrac{1+4Q^2}{6}\) + n^a \bigg),
\ee
and we indeed find the expected scaling with $\gamma^{-1}$.  Here we must choose the $n^a$ such that $u$ lies in the interior of the fundamental Weyl chamber, as in the Schur limit discussed  above.

So far we have ignored the contribution of instantons to the Bethe equations.  As we will argue in Section \ref{sec:casimirinstantons} below, we expect these to have a subleading contribution at large $u$, and so we assume now that we may ignore their contribution in the Bethe equations above.

It now remains to evaluate
\be \label{casbs} Z_{S^3_b \times \Sigma_\g} = \sum_{\hat{u} \in \cS_{BE}} \cH(\hat{u},
\nu,\gamma)^{\g-1} \Pi_i(\hat{u},\nu,\gamma)^{\n_i} \ee
at the Bethe vacua \eqref{CasLimsigmanj}.  Since we are interested in extracting the leading divergence as $\gamma \rightarrow 0$, we will look for the vacuum which has the dominant contribution in this limit. As before, we first consider the perturbative contribution
\be \cH^{pert}(\hat{u},
\nu,\gamma)^{\g-1} \Pi^{pert}_i(\hat{u},\nu,\gamma)^{\n_i} = (H^{pert})^{{\g}-1} \prod_i \prod_{\rho \in R_i} s_b(\rho(\hat{u}) + \nu_i)^{\n_i} \prod_{\mathclap{\alpha \in \mathit{Ad(G)'}}} s_b(\alpha(\hat{u}) -iQ)^{1-\g},
\ee
where
\be H^{pert} = \det_{a,b} \frac{1}{2 \pi i} \frac{\partial \log \Pi^{pert}_a}{\partial u_b}\,. \ee
Expanding these at large $u$ using \eqref{CasLimsblim}, we find (defining $\hat{\n}_i = \frac{\n_i}{\g-1}$ as in the previous subsection)
\be \cH^{pert}(u,\nu)^{{\g}-1} \Pi^{pert}_i(u,\nu)^{\n_i} \approx (\gamma^{r_G} K)^{\g-1} \exp \bigg\{4 \pi i (\g-1)\bigg(-\sum_i \i Q\cC^a_{R_i} \hat{\n}_i \tnu_i + i Q \cC^a_{Ad} \bigg) u_a \bigg\}  \,,\ee
where we have used
\be  H^{pert} \approx  \det_{a,b} \gamma K^{ab} = K \gamma^{r_g}\,, \qquad K \equiv \det K^{ab} \,.\ee
For $\gamma \rightarrow 0$, the contribution from the Hessian factor is subleading, and we may approximate the contribution from the Bethe vacuum, \eqref{CasLimsigmanj}, as
\eqst{\label{CasLimHPiOS}
	\cH^{pert}(\hat{u}^{(n)},\nu,\gamma)^{{\g}-1} \Pi^{pert}_i(\hat{u}^{(n)},\nu,\gamma)^{\n_i}  \approx  \exp \bigg\{-4 \pi i (\g-1)\gamma^{-1} K_{ab} \\
	\bigg(\!\!-\sum_i \i Q\cC^a_{R_i} \hat{\n}_i \tnu_i + i Q \cC^a_{Ad} \bigg)\bigg( \sum_i  \cC^b_{R_i} \(Q^2(1 -{\tnu_i}^2) -\tfrac{1+4Q^2}{6}\) +\cC^b_{Ad} \(\tfrac{1+4Q^2}{6}\) + n^b \bigg) \bigg\}\,.
}
The final step is to determine the vacuum, $\hat{u}^{(n)}$, which has the leading contribution to the Bethe sum.  Recall that we assume $u$ lies in a fundamental Weyl chamber, and so we may choose the $n^b$ to take $u$ arbitrarily large within this chamber.  For concreteness, let us take the basis, $e_a$, of the Cartan to be dual to the fundamental weights, so that $c^a e_a$ spans the interior of the fundamental Weyl chamber as we take $c^a >0$.  Then we see that, in this basis we must impose
\be \text{Im} \bigg[ (\g-1) \gamma^{-1} K_{ab}  \bigg(-\sum_i \i Q\cC^b_{R_i} \hat{\n}_i \tnu_i + i Q \cC^b_{Ad} \bigg) \bigg]>0\,,\qquad  a=1,...,r_G \,;\ee
otherwise, by taking an appropriate $n^a$ large we can arrange the RHS of \eqref{CasLimHPiOS} to be arbitrarily large, and we do not find a well-defined $\gamma \rightarrow 0$ limit.

With this assumption, we can see the dominant contribution comes from taking the $n^a$ as small as possible, while keeping $u$ in the fundamental Weyl chamber.  In the basis above, this is achieved by setting $n^a=0$, and so we finally arrive at the leading behavior of the partition function,
\eqst{\label{CasLimZg0}
	Z^{pert}_{\Sigma_\g \times S^3_b} \underset{\gamma \rightarrow 0}{\approx} \exp \bigg\{4 \pi (\g-1)\gamma^{-1} K_{ab} \bigg(-\sum_i Q\cC^a_{R_i} \hat{\n}_i \tnu_i + Q \cC^a_{Ad} \bigg) \\
	×\bigg( \sum_i  \cC^b_{R_i} \(Q^2(1 -{\tnu_i}^2) -\tfrac{1+4Q^2}{6}\) +\cC^b_{Ad} \(\tfrac{1+4Q^2}{6}\) \bigg) \bigg\}\,.
}
Given the relation \eqref{defgammasec4} and identification \eqref{betag5gen} we see this has the expected behavior  \eqref{Casimir5d}. Specifically, we find
\eqst{\hat E^{pert}_{Casimir} =	4 \pi \lambda(\g-1) K_{ab} \bigg(-\sum_i \cC^a_{R_i} \hat{\n}_i \tnu_i + \cC^a_{Ad} \bigg) \\
	×\bigg( \sum_i  \cC^b_{R_i} \(Q^2(1 -{\tnu_i}^2) -\tfrac{1+4Q^2}{6}\) +\cC^b_{Ad} \(\tfrac{1+4Q^2}{6}\)  \bigg)\,,
	\label{ECasimirGen}}
where the label ``pert'' denotes that this has been computed using the perturbative approximation to the $S^3_b \times \Sigma_\g$ partition function.  We turn to consider the instanton contributions next.

\subsubsection{Instantons, the 5d prepotential, and a 6d uplift condition}  \label{sec:casimirinstantons}

So far we have ignored the contribution of instantons in the above analysis, and only used the perturbative approximation to the partition function.  We argue that these have a subleading contribution to the twisted superpotential in the regions of large $u$.  Therefore we expect they will not modify the analysis of the Bethe solutions obtained using the perturbative twisted superpotential above.

To motivate this claim, note that taking large $u$ is equivalent to exploring the asymptotic region of the Coulomb branch, where we expect the theory to be weakly coupled, and dominated by the perturbative calculation.  In fact, we claim that the large $u$ behavior of the twisted superpotential, which controls the flux operator $\Pi_a$ appearing in the Bethe equations, is related to the effective 5d prepotential, $\cF_{eff}$, which is known to be perturbatively exact.  Namely, using \eqref{CasLimsblim} we may write

\be \label{gblimcas} g_b(\tu) \;\; \underset{\text{Re}(\tu) \rightarrow \pm \infty}{\longrightarrow} \;\; \mp \frac{1}{12} Q^2 \tu^3 + \mathcal O(\tu) \,.\ee
This means the behavior of the perturbative twisted superpotential of the theory at large $\tu$ is given by\footnote{Here we assume $\text{Im}(\tu) \ll \text{Re}(\tu)$, so that we may approximate $\pm \tu$ in \eqref{gblimcas} as $|\tu|$.}
\eqst{\cW_{S^3_b \times \R^2}^{pert}(\tu,\tnu,\tgamma) =-\frac{1}{2} Q^2 \tgamma \tu^2 + \sum_i \sum_{\rho \in R_i} g_b(\rho(\tu) + \tnu_i) - \sum_{\alpha \in Ad(G)'} g_b(\alpha(\tu)+1) \\
	\underset{|\tu| \rightarrow \infty}{\longrightarrow} Q^2 \bigg(  \frac{\pi i}{{g_5}^2}  \tu^2 -\frac{1}{12} \sum_i \sum_{\rho \in R_i} |\rho(\tu) + \tnu_i|^3 +\frac{1}{12} \sum_{\alpha \in Ad(G)}|\alpha(\tu)|^3 \bigg) + \mathcal O(|\tu|)\,.
}
But the quantity in parentheses is precisely the effective prepotential of the 5d theory \cite{Seiberg:1996bd}, as in \eqref{Feffdef}.  A similar relation was noted for the integrand of the $S^5$ partition function in \cite{Jafferis:2012iv}.  Given that the prepotential of a 5d theory does not receive non-perturbative corrections, it is then natural to make a similar assumption for the large $|\tu|$ behavior of the twisted superpotential, and hence for the Bethe equations.

In light of the relation above, the condition \eqref{CasLimC3vanish} can be equivalently phrased by saying that, in the notation of \eqref{Feffdef},
\be \label{fcond} c_{abc}= 0\,,  \ee
\ie, the effective CS term vanishes for every direction in the Coulomb branch of the 5d theory.  As we saw above, when this condition is satisfied, we find the partition function has a leading divergence with $\log Z \sim \gamma^{-1}$ as $\gamma \rightarrow 0$, which is the expected Casimir behavior of a theory with an emergent circle of radius $\beta \sim \gamma^{-1}$.  Thus we may conjecture that the condition \eqref{CasLimC3vanish} (or equivalently \eqref{fcond}) is the relevant condition for the 5d theory to admit a 6d UV completion.  We will see in examples below that this is indeed satisfied in several examples where a 6d UV completion is expected to exist.  It  would be interesting to explore this conjecture further.

In addition to the Bethe equations, we did not include instanton contributions when substituting the Bethe solutions into the sum over vacua in \eqref{casbs}.  In fact, we will see below that these do have a nontrivial contribution, but they can be explicitly characterized in terms of free fields of the 6d theory as already mentioned above \eqref{caspertinst}.

\subsubsection{Examples}\label{sec:Examples}

Let us now consider some examples of the Casimir energy computation in 5d theories that are believed to admit a 6d UV completion.

\paragraph{Maximal theory.}

We begin with maximal 5d $\cN=2$ super Yang-Mills with gauge group $G$, consisting of an $\cN=1$ vector and an adjoint hypermultiplet, with mass $\tnu$. In this case \eqref{CasLimZg0} gives
\be \label{CasLimZg0max} Z_{\Sigma_\g \times S^3_b} \underset{\gamma \rightarrow 0}{\approx} \exp{\bigg\{\frac{\pi}{3} (\g-1)\gamma^{-1} h_G d_G( 1-\hat{\n}\tnu)  (1-\tnu^{2})Q^{3} \bigg\}}\,,\ee
where we used
\be K_{ab} \cC^a_{Ad} \cC^b_{Ad} = K_{ab} \delta^a \delta^b = \frac{1}{12} h_G d_G\,, \ee
where $d_G$ and $h_G$ are the dimension and and dual Coxeter number of $G$, respectively.  For example, for $G=SU(N)$, we have
\be K_{ab} \delta^a \delta^b = \frac{1}{12} N(N^2-1)\,. \ee 
Using \eqref{defgammasec4} and \eqref{Qdef} we thus have
\equ{
	\log Z_{\Sigma_\g \times S^3_b}\approx -\frac{g_{5}^{2}}{2\pi}\,\frac{π}{3}(\fg-1)N(N^{2}-1)(\tnu^{2}-1)(\hat \n\tnu-1)Q^{2}\,.
}
Now, with the identification  $\beta=\frac{g_{5}^{2}}{2\pi}$  (\ie, $\lambda=1$)  we see that the partition function has the expected behavior  \eqref{Casimir5d}, with 
\equ{\label{Casimirhatmax}
\hat{E}^{pert}_{Casimir}=\frac{π}{3}(\fg-1)N(N^2 -1)(\tnu^2-1)(\hat{\n}\tnu-1)Q^{2}\,.
}
This may be compared to the exact Casimir energy, computed from the anomaly polynomial of the 6d theory in Appendix~\ref{Sec:Class S theories}. As discussed there, separating the contribution from free fields to the anomaly polynomial and the remaining part one can write $E_{\mathit{Casimir}}=E^{\mathit{pert}}+E^{\mathit{non-pert}}$, with 
\eqs{  \label{CasimirMaxPtext}
	E^{\mathit{pert}}\equiv &\int_{\varepsilon} \int_{\Sigma_{\fg}}\, I_{8, \mathit{ int}}^{\mathit{twisted}}=\,\frac{\pi}{3} (\fg-1)  N(N^{2}-1)\left(\epsilon ^2-1\right)(\hat \n \epsilon -1) Q^{2}\,,\\ \label{CasimirMaxInstItext}
	E^{\mathit{non-pert}}\equiv &\int_{\varepsilon} \int_{\Sigma_{\fg}}\, I_{8, \mathit{ free}}^{\mathit{twisted}}=\,\frac{\pi}{6}(\fg-1)(N-1)\(1+\hat \n\epsilon + 2\hat \n\epsilon\,(\epsilon^{2}-1)Q^{2}\)\,,
}
where $\epsilon$ is the parameter controlling the mixing of R-symmetry with the flavor symmetry acting on the (adjoint) hypermultiplet; see \eqref{twistGen}. We see that identifying $\tnu=ε$, the perturbative localization calculation \eqref{Casimirhatmax}   coincides with \eqref{CasimirMaxPtext}. Thus, for consistency, the instanton contribution to the partition function should match the remaining free field contribution \eqref{CasimirMaxInstItext}, as claimed at the beginning of Section~\ref{sec:Casimir4d}. Next, we provide more examples of the same phenomenon. 

\paragraph{Class $\bs{\S_k}$ Theories.}

These theories have $k$ $SU(N)$ vector multiplets and $k$ bifundamental hypermultiplets,  forming a circular quiver \cite{Gaiotto:2015usa}. 
Let us assume that all the $u_a$'s have the same solution.  In this case each bifundamental contribution to the BAE is essentially that of an adjoint in which case we simply obtain $k$ times the contribution of maximal theory in the Casimir limit,
\equ{\label{CasEnSk}
	\log Z_{\Sigma_\g \times S^3_b}\approx -\frac{g_{5}^{2}}{2\pi k }\,\frac{πk^{2}}{3}(\fg-1)N(N^{2}-1)(1-\tnu^{2})(1-\hat\n\tnu)Q^{2}\,.
}
Again, identifying $\tnu=ε$ and the 6d radius with $\lambda=k$, \ie,
\equ{
	\beta = \frac{g_{5}^{2}}{2\pi k}\,,
}
we see that the perturbative Casimir energy reads
\equ{\hat{E}^{pert}_{Casimir}=\frac{πk^{2}}{3}(\fg-1)N(N^{2}-1)(1-ε^2)(1-\hat\nε)Q^{2}\,,
}
which precisely matches \eqref{CasimirSk}. Thus, we see that the missing instanton part \eqref{CasimirSkInst} is again identified with the contribution of free fields to the anomaly polynomial \eqref{CasimirSkFFs}.

\paragraph{E-string.}

Next, we consider the 5d E-string theory, a 5d $\cN=1$ $USp(2N)$ theory with $N_f=8$ fundamental hypermultiplets and an antisymmetric hyper.  First we will need to compute the quantities $\cC_R^{abc}$ and $\cC^{a}_R$ for the various representations.  We find, working with $u=(u_a)$, $a=1,...,N$, in the fundamental representation
{\allowdisplaybreaks
\eqs{\cC_R^{abc} u_a u_b u_c &= \begin{cases}
		\hfil\displaystyle	\frac{1}{2} \sum_{a=1}^{N} (u_a)^3, & R=\mathit{F} \\
		\hfil\displaystyle	\frac{1}{2} \sum_{a<b} \big( (u_a+u_b)^3+(u_a-u_b)^3 \big), & R=\mathit{AS} \\
		\displaystyle	\frac{1}{2} \sum_{a<b} \big( (u_a+u_b)^3+(u_a-u_b)^3 \big)+	\frac{1}{2} \sum_{a=1}^{N} {(2u_a)}^3, & R=\mathit{Ad}
	\end{cases} \\
	\cC_R^{a} u_a &= \begin{cases}
		\hfil\displaystyle	\frac{1}{2} \sum_{a=1}^{N} {u_a}, & R=\mathit{F} \\
		\displaystyle	\frac{1}{2} \sum_{a<b} \big( (u_a+u_b) + (u_a-u_b) \big) = \sum_{a=1}^{N} (N-a) u_a, & R=\mathit{AS} \\
		\hfil\displaystyle	\sum_{a=1}^{N} (N-a+1) u_a, & R=\mathit{Ad}
	\end{cases}
}}

\noindent Then note that
\be \cC_{Ad}^{abc} = \cC_{AS}^{abc} + 8\;  \cC_{F}^{abc}\,, \ee
and thus \eqref{CasLimC3vanish} is satisfied precisely for $N_f=8$ fundamental hypers.  In this case, plugging these into \eqref{CasLimZg0}, we have
\eqst{\label{CasLimZgE8}
	Z_{\Sigma_\g \times S^3_b} \underset{\gamma \rightarrow 0}{\approx} \exp \bigg\{4 \pi (\g-1)\gamma^{-1} K_{ab} \bigg(\!\!-\sum_{i=1}^8 \cC^a_{F} \hat{\n}_i \tnu_i -\cC^a_{AS} \hat{\n} \tnu +\cC^a_{Ad} \bigg)Q \\
	\times\!\bigg(\sum_{i=1}^8 \cC^b_{F} \(Q^2(1 -{\tnu_i}^2) -\tfrac{1+4Q^2}{6}\)\! +\cC^b_{AS} \(Q^2(1 -{\tnu}^2) -\tfrac{1+4Q^2}{6}\)\! +\cC^b_{Ad} \(\tfrac{1+4Q^2}{6}\)\!\!\bigg).
}
We may simplify this  using
\be \cC^a_{Ad} = \cC^a_{AS} +2 \cC^a_F\, ,\ee
which gives
\eqst{\label{CasLimZgE82}
	Z_{\Sigma_\g \times S^3_b} \underset{\gamma \rightarrow 0}{\approx} \exp \bigg\{4 \pi  (\g-1)\gamma^{-1} K_{ab} \bigg(\sum_{i=1}^8 \cC^a_{F}\(\tfrac{1}{4} -\hat{\n}_i \tnu_i \)+ \cC^a_{AS}\(1 -\hat{\n} \tnu\) \bigg)Q \\
	\times \bigg( \sum_{i=1}^8  \cC^b_{F} \(Q^2(1-{\tnu_i}^2) -\tfrac{1}{8}(1+4Q^2)\) + \cC^b_{AS}Q^2(1 -\tnu^2) \bigg) \bigg\}\,.
}
We may further expand this, using
\be  K_{ab} \cC^a_F \cC^b_F = \tfrac{1}{8} N\,,\quad   K_{ab} \cC^a_{AS} \cC^b_{AS} = \tfrac{1}{12}N(N-1)(2N-1)
\,,\quad  K_{ab} \cC^a_F \cC^b_{AS} = \tfrac{1}{8}N(N-1) \,,\ee
which yields
\bea \label{CasLimZgE8f} Z_{\Sigma_\g \times S^3_b}& \underset{\gamma \rightarrow 0}{\approx} \exp \bigg\{4 \pi (\g-1)\gamma^{-1} Q \bigg(\tfrac{1}{8} N \sum_{i,j=1}^8 \(\tfrac{1}{4}-\hat{\n}_i \tnu_i\)\(Q^2\(\tfrac{1}{2}- {\tnu_j}^2\) -\tfrac{1}{8}\) \\
&+ \tfrac{1}{8} N(N-1) \sum_{i=1}^8  \(\(\tfrac{1}{4} -\hat{\n}_i \tnu_i\)Q^2\(1 -\tnu^2\)+ \(1 -\hat{\n}\tnu\)\(Q^2(\tfrac{1}{2}- {\tnu_j}^2) -\tfrac{1}{8}\) \) \\
&+ \tfrac{1}{12} N(N-1)(2N-1) \(1 -\hat{\n} \tnu\)Q^2\(1 -\tnu^2\)\bigg) \bigg\} \,.\eea
The result of the anomaly polynomial calculation for the case of two flavor fluxes is given in \eqref{EcasimirE8Anom} and \eqref{EcasimirE8AnomInst}. To compare with that calculation, we set $\tnu=ε$, $\tnu_1=ε_E$, $\hat{\fn}_1=\hat{\fn}_E$ with all other $\hat{\fn}_{i=2,⋯,8}=\tnu_{i=2,⋯,8}=0$ in \eqref{CasLimZgE8f} to get
\eqst{\hat{E}^{pert}_{Casimir} =\pi(\g-1)λ\Big[\tfrac{1}{2} N(N-1)Q^2\(1 -ε^2\)\(\tfrac{2}{3}(2N-1)\(1 -\hat{\n}ε\) +\(2 -\hat{\n}_E ε_E\)\) \\
+\tfrac{1}{2} N\(Q^2\(4- ε_E^2\) -1\)\big((N-1)\(1 -\hat{\n}ε\) + \(2-\hat{\n}_E ε_E\)\!\big) \Big]\,,
}
which matches \eqref{EcasimirE8Anom} provided $\lambda=2$. This means the 6d radius is identified, as in \cite{Bobev:2015kza}, to be
\equ{\beta = \frac{g_{5}^{2}}{4\pi}\,.
}
Thus, once again, the instanton part \eqref{EcasimirE8AnomInst} missing from the above calculation is identified with the contribution of free fields to the anomaly polynomial \eqref{EcasimirE8AnomFree}.

\paragraph{Summary.} 

We have argued that computing the $S^{3}_{b}\times \Sigma_{\fg}$ partition function for theories with a UV completion as 6d $\cN=(1,0)$ theories is equivalent to computing the $S^{3}_{b}\times S^{1}$ partition function (or index, up to the Casimir energy), of the corresponding 4d $\cN=1$ theories obtained by compactification of the 6d theory on $\Sigma_{\fg}$.  We have checked this explicitly for the case of 5d maximal SYM and the corresponding 4d $\cN=1$ and $\cN=2$ class $\cS$ theories of \cite{Gaiotto:2009we,Benini:2009mz,Bah:2012dg} in the mixed Schur limit, finding perfect agreement with \cite{Beem:2012yn}.  We then considered examples of 5d $\cN=1$ theories. In this case we do not have an explicit handle on instanton contributions and we limited ourselves to computing the partition function asymptotically on the Coulomb branch, where instanton contributions are suppressed and the partition function is expected to be dominated by the Casimir energy, which we extracted. We find an exact match with an independent calculation of the Casimir energy from anomaly polynomials, up to the expected instanton corrections which, for consistency, should be identified with the contribution of free fields to the corresponding 6d anomaly polynomials, as noted in the case of $S^{5}$ in \cite{Bobev:2015kza}.  


\section{\texorpdfstring{Twisted partition function on $S^{2}\times S^{1} \times \Sigma_\g$ and AdS$_{6}$ black holes}{Twisted partition function on S²×S¹×Σg and AdS₆ black holes}}\label{sec:AdS6}

Although the main focus of this paper has so far been the partition function on $S^{3}_{b}\times \Sigma_{\fg}$, we observed in Section \ref{sec:5d4d} that there is a natural generalization of the above computation to $\cM_3 \times \Sigma_\g$ for $\cM_3$ a general lens space, $L(p,q)$, using the factorization of the twisted superpotential into ``holomorphic block''-like contributions, as in \eqref{wfact}.  In this section we briefly consider the simplest example, that of the space $\cM_3 = S^2 \times S^1$, corresponding to the topological index in 3d \cite{Benini:2015noa}.

After writing the result for the partition function on $S^2 \times S^1 \times \Sigma_\g$ according to the above framework in Section \ref{sec:S2S2S1}, we conjecture a natural generalization for the partition function on $\Sigma_{\fg_{1}}\times \Sigma_{\fg_{2}}\times S^{1}$, which we claim is valid perturbatively, and so also in the large $N$ limit. Based on this conjecture, we reproduce in Section~\ref{sec:AdS6Conjecture} the microscopic entropy of an infinite class of black holes in AdS$_{6}$. 

\subsection{\texorpdfstring{Partition function on $S^2 \times S^{1} \times \Sigma_\g$}{Partition function on S²×S¹×Σg}}\label{sec:S2S2S1}

Let us first consider the case where $\cM_3$ corresponds to the ``refined'' topological index, $S^2_{\q_1} \times S^1$, with a fugacity $\q_1$ for the angular momentum.\footnote{The subscript ``1'' denotes this $S^2$ sits as the first factor in the manifold, to avoid confusion below.}  Then we may build the integrand of the full $S^2_{\q_1} \times S^2 \times S^1$ partition function by including the contribution of the 5d Nekrasov partition function at the four fixed points, as in \eqref{s3s2inst}, namely,
\be \label{s2s2s1inst} \lim \limits_{\epsilon \rightarrow 0} \prod_\ell Z_{\R^2_{\q^{(\ell)}_1} \times \R^2_{\q^{(\ell)}_2} \times S^1}(x^{(\ell)},y^{(\ell)},z^{(\ell)}) \,,\ee
where now the parameters can be identified by a straightforward modification of the argument leading to \eqref{ueldef5d2}, giving
\bea \label{ueldef5d3} 	 & x^{(\ell)} = e^{2\pi i \tu} {\q^{(\ell)}_1}^{\m_1/2} {\q^{(\ell)}_2}^{\m_2/2}\,, \\
&\q_1^{(\ell)} = \left\{ \begin{array}{cc} \q_1 \,,&\quad \ell=nn \; \text{or} \; ns \\ {\q_1}^{-1}\,,&\quad \ell=sn \; \text{or} \; ss \end{array}\right.\;\;\;\;\, \\
&\q_2^{(\ell)} =
\left\{ \begin{array}{cc} 
	e^{2 \pi i \epsilon}\,, & \quad  \ell=nn \; \text{or} \; sn  \\
	e^{-2 \pi i  \epsilon} \,,& \quad \ell=ns \; \text{or} \; ss  
\end{array}\right. 
\eea
Then, using \eqref{5dNSlimit} and arguing as before, we find that the twisted superpotential controlling the partition function on $S^2_{\q_1} \times S^1 \times \Sigma_\g$  is given by
\eqss{ \label{wfact2} \cW_{S^2_{\q_1} \times S^1 \times \R^2}(\tu,\tnu,\tgamma)_{\m_1,\n_1} =&\;  \cW^{(5d)}_{\mathit{NS}}\(\tu+\tfrac{1}{2} \m_1 \epsilon_1,\tnu+\tfrac{1}{2} \n_1 \epsilon_1, \tgamma;\epsilon_1\)\\
	&+ \cW^{(5d)}_{\mathit{NS}}\(\tu-\tfrac{1}{2} \m_1 \epsilon_1,\tnu-\tfrac{1}{2} \n_1 \epsilon_1, \tgamma;-\epsilon_1\) \,,
}
and similarly for the effective dilaton,  $\Omega_{S^2_{\q_1} \times S^1 \times \R^2}(\tu,\tnu,\tgamma)_{\m_1,\n_1}$.  An important difference from the $S^3_b$ case is that the dependence on the fluxes, $\m_1,\n_1$, on $S^2_{\q_1}$ cannot be absorbed into a shift of $\tu$ and $\tnu$.  As a result, we obtain a direct sum of 2d theories indexed by the gauge flux, $\m_1 \in \Lambda_G$.  Then we may write the full partition function as\footnote{Here we denote the fluxes on $\Sigma_\g$ with a subscript ``$2$'' to distinguish them from those on $S^2_{\q_1}$.}
\be \label{s2s2s1bs} Z_{S^2_{\q_1} \times  S^{1} \times \Sigma_\g}(\tnu,\tgamma)_{\n_1,\n_2} = \sum_{\m_1 \in \Lambda_G} \sum_{ \hat{\tu} \in \mathcal {S_{\mathit{BE}}}_{\m_1,\n_1}} \Pi_i(\hat{\tu},\tnu,\tgamma)_{\m_1,\n_1}^{\n_{2,i}} \cH(\hat{\tu},\tnu,\tgamma)_{\m_1,\n_1}^{\g-1}\,, \ee
where we must sum over the 2d theories labeled by the gauge fluxes, $\m_1$ (as well as the flavor fluxes, $\n_1$), whose vacua are given by
\be {\cS_{BE}}_{\m_1,\n_1} = \{ \hat{\tu} \; \big| \; \Pi_a(\hat{\tu},\tnu,\tgamma)_{\m_1,\n_1} = 1 \}/ W_G \,, \ee
and, \eg,
\be \Pi_a(\tu,\tnu,\tgamma)_{\m_1,\n_1} = e^{2 \pi i \partial_{\tu_a} \cW_{S^2_{\q_1} \times S^1 \times \R^2}(\tu,\tnu,\tgamma)_{\m_1,\n_1}} \,, \ee
and similarly for the other operators.

Let us now focus on the unrefined limit, $\q_1 \rightarrow 1$, corresponding to the ordinary A-twist background on $S^2$.  We first discuss the perturbative contribution to the partition function in this limit.  For a single hypermultiplet, we have, by taking the $\epsilon_{1,2} \rightarrow 0$ limit of \eqref{zs2s2s1pert} using \eqref{epszero},
\be \label{zs2s2s1pert2} Z^{pert,U(1),hyp}_{S^2 \times S^2 \times S^1}(\tu)_{\m_1,\m_2} = \lim_{\epsilon_{1,2} \rightarrow 0} \(x;\q_1,\q_2\)_{\m_1,\m_2}^{-1} =  e^{\m_1 \m_2 s(\tu)}, \ee
where we introduced a function
\be s(\tu) \equiv \pi i \tu +\text{Li}_{1}(e^{2 \pi i \tu}) +\frac{i \pi}{2} \qquad \Rightarrow \qquad e^{s(\tu)} = \frac{ie^{\pi i \tu}}{1-e^{2\pi i \tu}} \,.\ee
The perturbative contribution for a general hyper and vector multiplet can be obtained similarly.  Equivalently, these can be obtained as in Section \ref{sec:amodreduc} by reduction on $\Sigma_\g$ to a 3d theory on $S^2 \times S^1$.  In either case, we find the perturbative approximation to the integrand of the $S^2 \times S^1 \times \Sigma_\g $ partition function is given by
\eqss{
	\label{pertS2S2S1}
	Z^{pert}_{S^2 \times S^1 \times \Sigma_\g }(\tu,\tnu,\tgamma)_{\fm_1,\fn_1;\fm_2,\fn_2}= z^{\tr(\m_1 \m_2)}& \prod_{\alpha\in \mathit{Ad}(G)'}\(e^{-s(\alpha(\tu))}\)^{[\alpha(\fm_{1})+1-\g][\alpha( \fm_{2})+1]}\\
	\times& \prod_{I}\prod_{\rho\in R_{I}}\(e^{s(\rho(\tu)+\tnu)}\)^{[\rho(\fm_{1})+\fn_{1}] [\rho( \fm_{2})+\fn_{2}]}\,,
}
where the index $I$ runs over the hypermultiplets  in the theory, in gauge representations $R_{I}$. Then the perturbative partition function is given by the integral formula
\be  Z^{pert}_{S^2 \times S^1 \times \Sigma_\g }(\tnu,\tgamma)_{\fn_1;\fn_2} = \sum_{\m_1,\m_2 \in \Lambda_G} \frac{1}{|W_G|} \oint_{\cC_{JK}} d\tu  \; Z^{pert}_{S^2\times\Sigma_\g \times S^{1}}(\tu,\tnu,\tgamma)_{\fm_1,\fn_1;\fm_2,\fn_2} \,.\ee
Equivalently, we may write this as a Bethe sum associated to the 2d theory obtained by compactification on $S^2 \times S^1$ as in \eqref{s2s2s1bs}, where now, to perturbative accuracy,
\eqs{ \nonumber
	&\Pi^{pert}_a(\tu,\tnu,\tgamma)_{\m_1,\n_1}= z^{{\m_1}^a} \prod_{\alpha\in \mathit{Ad}(G)'}\(e^{-s(\alpha(\tu))}\)^{[\alpha(\fm_{1})]\alpha^a} \prod_{I}\prod_{\rho\in R_{I}}\(e^{s(\rho(\tu)+\tnu)}\)^{[\rho(\fm_{1})+\fn_{1}] \rho^a}\,,\\ \label{gfs2s2s1} 
	& \Pi^{pert}_I(\tu,\tnu,\tgamma)_{\m_1,\n_1} = \prod_{\rho\in R_{I}}\(e^{s(\rho(\tu)+\tnu)}\)^{[\rho(\fm_{1})+\fn_{1}]}\,,\\ \nonumber
	&\cH(\tu,\tnu,\tgamma) = \prod_{\alpha\in \mathit{Ad}(G)'}\(e^{-s(\alpha(\tu))}\)^{-\alpha( \fm_{2})} H^{pert}(\tu,\tnu,\tgamma)_{\m_1,\n_1}\,,
}
where $H^{pert}_{\m_1,\n_1}=\det_{a,b} \frac{1}{2 \pi i} \frac{\partial}{\partial \tu_b} \log \Pi^{\mathit{pert}}_a(\tu,\tnu,\tgamma)_{\m_1,\n_1}$ is the Hessian factor.

\paragraph{5d prepotential and the Bethe equations.}

We can gain another perspective on this calculation, and characterize the instanton contributions, by observing the leading behavior of the Nekrasov-Shatashvili limit of the twisted superpotential in the $\epsilon \rightarrow 0$ limit \cite{Nekrasov:2009rc},
\be  \cW^{(5d)}_{\mathit{NS}}(\tu,\tnu,\tgamma;\epsilon) \underset{\epsilon \rightarrow 0}{\cong} \frac{1}{\epsilon} \cF_{\R^4 \times S^1}(\tu,\tnu,\tgamma) + \cdots ,\ee
where $\cF_{\R^4 \times S^1}(\tu,\tnu,\tgamma)$ is the effective prepotential of the 5d theory compactified on a circle \cite{Nekrasov:1996cz}.  Then we have, from \eqref{wfact2}, that the flux dependence of the twisted superpotential on $S^2 \times S^1 \times \R^2$ is governed by the 5d prepotential, via
\bea \label{wfact3}\cW_{S^2 \times S^1 \times \R^2}(\tu,\tnu,\tgamma)_{\m_1,\n_1}  &=  \lim_{\epsilon_1 \rightarrow 0} \cW_{S^2_{\q_1} \times S^1 \times \R^2}(\tu,\tnu,\tgamma)_{\m_1,\n_1} \\
&=  \m_{1,a} \frac{\partial \cF_{\R^4 \times S^1}(\tu,\tnu,\tgamma)}{\partial \tu_a} + \n_{1,I} \frac{\partial \cF_{\R^4 \times S^1}(\tu,\tnu,\tgamma)}{\partial \tnu_I} + \cdots \eea

We can see this dependence on the fluxes explicitly at the perturbative level.  For example, for a hypermultiplet, we have
\be \label{Zw}
Z^{pert,U(1),hyp}_{S^2 \times S^2 \times S^1}(\tu)_{\m_1,\m_2} = e^{2 \pi i \m_1 \m_2 \partial_\tu^2 w(\tu)} , \;\;\; w(\tu) = \frac{1}{(2 \pi i)^3} \text{Li}_3(e^{2 \pi i \tu})+ \frac{1}{24} \tu(\tu+1)(2\tu+1) \,,
\ee
where $w(\tu)$ is the perturbative contribution to the effective prepotential of a 5d $\cN=1$ hypermultiplet.  Then we indeed observe that the dependence on fluxes for the full perturbative contribution is governed by the perturbative prepotential
\be \label{pertFR4S1}
\cF^{pert}_{\R^4 \times S^1}(\tu,\tnu,\tgamma) = \tgamma \tu^2  + \sum_I \sum_{\rho \in R_I} w(\rho(\tu) + \tnu_I) - \sum_{\alpha \in Ad(G)'} w(\alpha(\tu)) \,.
\ee
Namely, one can check that the gauge flux operator in \eqref{gfs2s2s1} is given by
\be \label{pifromf} \Pi^{pert}_a(\tu,\tnu,\tgamma)_{\m_1,\n_1}=\exp \bigg\{2 \pi i \bigg(\m_{1,b} \partial_{\tu_a} \partial_{\tu_b} \cF^{pert}_{\R^4 \times S^1}(\tu,\tnu,\tgamma)+\n_{1,I} \partial_{\tnu_I} \partial_{\tu_b} \cF^{pert}_{\R^4 \times S^1}(\tu,\tnu,\tgamma) \bigg) \bigg\} \,,\ee
and similarly for the flavor flux operator.  

We expect that the instanton corrections to the $S^2 \times S^1 \times \Sigma_\g$ partition function are given by replacing the perturbative expression for the prepotential above by the full, non-perturbative result, as discussed, \eg, in \cite{Nekrasov:1996cz}.   However, we will not require these instanton corrections when we consider the large $N$ limit below, and leave a detailed discussion of them to future work.

Before moving on to discuss the large $N$ limit, let us conjecture a generalization to the perturbative result above where $S^2$ is replaced by $\Sigma_{\g_1}$ (and correspondingly we rename $\Sigma_\g \rightarrow \Sigma_{\g_2}$).  This can be motivated as in Section \ref{sec:amodreduc}, by reducing the theory on $\Sigma_{\g_2}$ and considering the $\Sigma_{\g_1} \times S^1$ partition function of the resulting theory.  This leads us to propose
\eqst{\label{pertS2S2S12}
	Z^{pert}_{\Sigma_{\g_1} \times  \Sigma_{\g_2} \times S^1}(\tu,\tnu,\tgamma)_{\fm_1,\fn_1;\fm_2,\fn_2}= z^{\tr(\m_1 \m_2)} \prod_{\alpha\in \mathit{Ad}(G)'}\(e^{-s(\alpha(\tu))}\)^{[\alpha(\fm_{1})+1-\g_1][\alpha( \fm_{2})+1-\g_2]}\\
	\times \prod_{I}\prod_{\rho\in R_{I}}\(e^{s(\rho(\tu)+\tnu)}\)^{[\rho(\fm_{1})+\fn_{1}] [\rho( \fm_{2})+\fn_{2}]}\,.
}
Since the reduction to 3d gave us the correct perturbative contribution to the $1$-loop determinant in the genus zero case, it is natural to conjecture the same holds here.  However, we do not have a direct localization derivation of this claim, nor do we make any claims about the non-perturbative contribution.

\subsection{\texorpdfstring{A large $N$ conjecture and AdS$_{6}$ black holes}{A large N conjecture and AdS₆ black holes}}\label{sec:AdS6Conjecture} 

In this final section, we propose a conjecture on the large $N$ behavior of the partition function  \eqref{pertS2S2S12}, motivated by holography and the entropy of black holes in AdS$_{6}$. 

\paragraph{AdS$_{6}$ black holes.} 

On general grounds, we expect the gravity dual of a 5d $\cN=1$  SCFT on $M_{4}\times S^{1}$  with a topological twist on $M_{4}$ to be given by a supersymmetric solution interpolating between asymptotically locally AdS$_{6}$ (with an $M_{4}\times S^{1}$ boundary) and an AdS$_{2}\times M_{4}$ geometry for small values of the radial coordinate, \ie, an extremal black hole in locally AdS$_{6}$. Such a background was considered  in 6d $F(4)$ minimal gauged supergravity in \cite{Naka:2002jz}, with metric of the form 
\eqss{\label{solUnivBH}
	ds^{2}_{\text{BH}}=&\,e^{2f(r)} (-dt^{2}+dr^{2})+e^{2g(r)}ds^{2}(M_{4})\,,
} 
and a nonzero flux for the graviphoton through 2-cycles in $M_{4}$, which is K\"ahler. Since the graviphoton is the only nontrivial 1-form potential, and is dual to the R-symmetry of the field theory, it was argued in \cite{Bobev:2017uzs} that such a background should describe the IR behavior of generic 5d $\cN=1$ theories with AdS$_{6}$ gravity duals, with a universal topological twist on $M_{4}$. Taking $M_{4}=\Sigma_{\fg_{1}}\times \Sigma_{ \fg_{2}}$, a product of two negatively curved Riemann surfaces, \ie, $\fg_{1}>1$, $\fg_{2}>1$, the Bekenstein-Hawking entropy of the black hole reported in  \cite{Naka:2002jz} can be written as\footnote{However, see the comment at the end of this section, which was added in v2 of this paper.} 
\equ{\label{SBHFS5}
	S_{\text{BH}}=-\frac{4}{9}(\fg_{1}-1)( \fg_{2}-1)\,F_{S^5}\,,
}
where $F_{S^5}$ is the free energy on $S^{5}$ of the UV 5d SCFT.

\paragraph{A large $N$ conjecture.}

In light of the recent success in accounting for the microscopic entropy of AdS$_{4}$ black holes from 3d twisted partition functions, initiated in \cite{Benini:2015eyy} and followed up in \cite{Benini:2016rke,Hosseini:2016tor,Hosseini:2016ume,Cabo-Bizet:2017jsl,Azzurli:2017kxo,Hosseini:2017fjo,Benini:2017oxt}, it is natural to ask whether the entropy of AdS$_{6}$ black holes is  similarly captured by the 5d partition function, \ie, $ S_{\text{BH}}=\log Z_{M_{4}\times S^{1}}$, which can be addressed with the results  presented here for the case $M_{4}=\Sigma_{\fg_{1}}\times \Sigma_{\fg_{2}}$.  

The basic observation we make is the following. Recall that in Section~\ref{sec:largeN}, we argued the dominant vacuum contributing to the  $S^{3}_{b}\times \Sigma_{\fg}$ partition function was determined by extremizing the Bethe potential, $\cW_{S^3_b \times \R^2}$.  In the present case (and specializing first to $\g_1=0$), we can also attempt to extremize $\cW_{S^2 \times S^1 \times \R^2}$, however, in this case we have seen the twisted superpotential and vacuum equations depend on the flux, $\m_1$, through $S^2$.  Then we must also extremize over this choice of flux.  Given the form of \eqref{pifromf}, which implies this flux dependence is controlled by the prepotential, we will conjecture that at large $N$ the dominant vacuum is found by extremizing the prepotential, $\cF_{\R^4 \times S^1}$, itself.  As we discuss below, this leads to the expected behavior of the partition function at large $N$. 
To fully justify this would require a detailed analysis of the matrix model \eqref{pertS2S2S12} and the corresponding vacua arising from performing the double sum over magnetic fluxes $\m_{1},\m_{2}$ and the contour integral over the Coulomb branch. This is an interesting problem, which lies beyond the scope of this paper.

Let us evaluate the prepotential in the large $N$ limit, where it is dominated by its perturbative contribution.  Then we simply need the expansion
\equ{
	w(iz+\tnu)\approx -\frac{1}{8}\(1+2\tnu\)z|z|+ i\(-\frac{1}{12}|z|^{3}+\frac{1}{4}C_{\tnu}|z|\) \qquad \text{for $\text{Re}|z|\gg1$}\,,
}
where we defined $C_{\tnu}\equiv\frac{1}{6}+\tnu(1+\tnu)$. 

Let us consider the Seiberg theory (the extension to the orbifold theories is straightforward). The manipulations are very similar to those of Section~\ref{sec:largeN} and thus here we are more concise. Adding the contributions to the prepotential \eqref{pertFR4S1}  from the vector, antisymmetric and fundamental hypers (with corresponding real masses, which we denote by $\tnu_{I}=-\Delta_{I}$), and using the Ansatz $\tu_{i}\to i N^{\alpha}x$ for the eigenvalue distribution we see that $\alpha=\frac{1}{2}$ is required to have a  nontrivial extremum, hence an $N^{5/2}$ scaling of the prepotential, and the classical term in  \eqref{pertFR4S1} is subleading. After a convenient rescaling of the coordinate,  $x\to \tfrac23 \sqrt{\Delta_{\mathit{AS}}(1-\Delta_{\mathit{AS}})}x$, and corresponding inverse rescaling of the density $\rho$, we find
\eqst{\label{OmegaFS5}
	\cF^{pert}_{\R^4 \times S^1}= \frac{4i}{27 \pi}N^{5/2}\(\Delta_{\mathit{AS}}(1-\Delta_{\mathit{AS}})\)^{3/2}\bigg[\frac{\pi(8-N_{f})}{3}\int dx \,\rho(x) \,|x|^{3}  \\
	-\frac{9 \pi}{8} \int dx \,dy \, \rho(x)\rho(y)\( |x+y|+|x-y|\)\bigg]\,,
}
where all integrals run from $0$ to $x^{\ast}$. Note that to this order in $N$ this depends only on the real mass  $Δ_{\mathit{AS}}$  for the antisymmetric field, but not on the one for the fundamental fields and that the quantity inside the brackets is precisely the free energy functional on $S^{5}$ for the theory, with saddle point configuration \eqref{density}.  An analogous calculation can be done for the $(A,B,C)$ orbifolds of the Seiberg theory, with the same conclusion. Given the similarity of the discussion in Section~\ref{sec:largeN} for $S^3_b \times \Sigma_\g$ this leads us to conjecture, as mentioned above, that at large $N$ the partition function is dominated by the eigenvalue distribution that extremizes the 5d prepotential.

Next, to determine the value of the partition function on the eigenvalue distribution above, one should determine the dominant contribution in different gauge flux sectors,  $\fm_{1},\fm_{2}$. Let us now observe that the zero gauge sector, $\fm_{1}=\fm_{2}=0$, reproduces the RHS of \eqref{SBHFS5}. Indeed, setting  $\fn_{1}=\fn_{2}=0$ in  \eqref{pertS2S2S12} for the universal twist, performing the rescaling of the coordinate mentioned above \eqref{OmegaFS5},  and evaluating the zero gauge flux partition function on the configuration \eqref{density},  we obtain\footnote{Here we used the expansion $s(z)\approx -\pi|z|$ for $z$ real and large,  $|z|\gg1$.  Additionally, as in the case of the $S^3_b \times \Sigma_\g$ partition function, we expect the contribution from the Hessian determinant to be subleading at large $N$, which we assume. }
\eqs{\nonumber
	\log Z^{\fm=0}_{\Sigma_{\fg_{1}}\times \Sigma_{ \fg_2}\times S^{1} }\Big|_{\partial \cF=0}
	= &\,\frac{\pi}{3}(\fg_{1}-1)(\fg_2-1) \sqrt{\Delta_{\mathit{AS}}(1-\Delta_{\mathit{AS}})} \\ \nonumber
	&\times N^{5/2}\int dx dy\, \rho(x)\rho(y)\(|x+y|+|x-y|\) \\
	=&\, -\frac{8}{9}(\fg_{1}-1)(\fg_2-1)\sqrt{\Delta_{\mathit{AS}}(1-\Delta_{\mathit{AS}})}\, F_{S^{5}}\,.
} 
Then, extremizing this expression with respect to the real mass parameter (or fugacity, $\Delta_{\mathit{AS}}$), as in \cite{Benini:2015eyy}, sets $\Delta_{\mathit{AS}}=\tfrac{1}{2}$, with value at the extremum 
\equ{\label{Z=SBH}
	\log Z^{\fm=0}_{\Sigma_{\fg_{1}}\times \Sigma_{ \fg_2}\times S^{1} }\Big|_{\partial \cF=0}= -\frac{4}{9}(\fg_{1}-1)(\fg_2-1)\,F_{S^{5}}\,,
}
which coincides with the RHS of \eqref{SBHFS5}.

\paragraph{Comment added in version 2:} 

In light of the new paper \cite{Minwoo}, noting that the black hole reported in  \cite{Naka:2002jz} needs to be corrected,\footnote{We  thank Minwoo Suh for sharing this result with us prior to  publication.} and, correspondingly, its entropy \eqref{SBHFS5}, we discuss how the results above relate to this observation. To summarize, the main points emphasized in v1 (and which are unchanged in v2)  are as follows:
\begin{enumerate}[a)]
	\item \label{parta}  We have proposed that the eigenvalue configuration dominating the large $N$ behavior of the partition function is obtained by extremizing the 5d prepotential, and showed (after appropriate rescalings) that this coincides with the configuration extremizing the free energy of the same theory on $S^{5}$.
	\item \label{partb} Evaluating the  partition function on this extremum, and keeping only the zero gauge flux sector, reproduces the entropy of the AdS$_{6}$ black hole described in \cite{Naka:2002jz}.  
\end{enumerate}
Due to the exact  match of \eqref{Z=SBH} with \eqref{SBHFS5}, this suggests that the zero gauge flux sector completely accounts for the entropy of the black hole described in \cite{Naka:2002jz},  as  noted in v1 of this paper on the arXiv. However, as mentioned above it was recently pointed out in \cite{Minwoo} that the background of \cite{Naka:2002jz} may not be valid, as it assumed the vanishing of the two-form gauge field, $B_{\mu\nu}=0$, which is inconsistent with the equations of motion.  Instead, it was shown that by solving for $B_{\mu\nu}$ from the equations of motion, that the corrected supersymmetric solution has twice the entropy \cite{Minwoo}:
\equ{\label{EntropyMinwoo}
	S_{\text{BH}}'=-\frac{8}{9}(\fg_{1}-1)( \fg_{2}-1)\,F_{S^5}\,.
} 
Indeed, we note that this new result is  consistent with first reducing from 5d to 3d via the universal relation $F_{S^{3}\times \Sigma_{\fg_{1}}}=-\frac{8}{9} (\fg_{1}-1) F_{S^{5}}$ of Section~\ref{sec:largeN} and then reducing from 3d to 1d via the universal relation $F_{\Sigma_{\fg_{2}}\times S^{1}}=-(\fg_{2}-1)F_{S^{3}}$ discussed in \cite{Azzurli:2017kxo}. Then, composing these two relations leads to $F_{\Sigma_{\fg_{1}}\times \Sigma_{ \fg_2}\times S^{1}}=\frac{8}{9}(\fg_{1}-1)( \fg_{2}-1)\,F_{S^5}$, consistent with  \eqref{EntropyMinwoo} and the expected identification $S_{\text{BH}}=\log Z_{\Sigma_{\fg_{1}}\times \Sigma_{ \fg_2}\times S^{1}}$.

Assuming this new solution to be correct, the zero gauge flux sector  \eqref{Z=SBH} then accounts for only half the entropy of the corrected solution. Thus, it seems that to account for the full entropy one must also include contributions from non-zero gauge flux sectors. An interesting proposal on how to do so is discussed in \cite{Hosseini:2018uzp} and,  indeed, following the procedure described there one obtains the remaining half of the entropy, matching \eqref{EntropyMinwoo} and clarifying an initial numerical mismatch. It would be interesting to have a first-principles understanding of this proposal purely in field theory. 


\section{Outlook}

In this work, we have considered five dimensional $\cN=1$ gauge theories and computed their exact partition function on various manifolds with partial topological twists using localization. As discussed, this is a fruitful vantage point from which we can also access the physics of field theories of various dimensions, from 1d up to 6d, arising at either the IR or UV end of RG flow.  There are a number of possible directions for future work.

\

As discussed in Section~\ref{sec:derivation}, one can consider partition functions on $\cM_{3}\times \Sigma_{\fg}$ with more general $\cM_{3}$.  Our formalism already suggests the result for $\cM_3$ a lens space, and we discussed the example of $S^2 \times S^1$ in Section \ref{sec:AdS6}, but in principle one can consider arbitrary Seifert manifolds \cite{Closset:2018ghr}.  It would also be interesting to better understand the relation of our computation to partition functions on more general five-manifolds considered in the literature, such as $S^5$, $\CP^2 \times S^1$, $S^4 \times S^1$, and $Y^{p,q}$, as well as studying new geometries.  For example, $Y^{p,q}$ is topologically an $S^1$ fibration over $S^2 \times S^2$, and may be related to our results by introducing a suitable ``fibering operator,'' as in \cite{Closset:2017zgf,Closset:2017bse,Closset:2018ghr}.

\

For a given $\cM_{3}$, the partition function of the 3d theory obtained by reduction on $\Sigma_{\fg}$ is computed by an appropriate 2d TQFT, in what may be called a ``3d-2d correspondence.''  We described these 3d theories as a direct sum of ordinary 3d theories in Section \ref{sec:amodreduc}, however the full, non-perturbative computation suggests they are something more exotic.  The results of Section~\ref{sec:largeN}, in particular \eqref{FExtQvrs}, suggest that a large class of novel 3d $\cN=2$ SCFTs exist, arising as the IR fixed point of the orbifolded Seiberg theories in 5d, compactified on a Riemann surface. These are labeled by the discrete flavor fluxes $\fn$ and their free energy scales as $N^{5/2}$. It would be worth investigating whether some of these theories can be understood purely in terms of simpler building block three-dimensional theories, analogous to $T_{N}$ theories for four-dimensional class $\mathcal S$.   

\

The results of Section~\ref{sec:5d6d4d} provide a new method for computing the superconformal indices of non-Lagrangian theories in 4d, by Lagrangian methods in 5d. Our results coincide with those which have been previously computed and in principle provide a powerful tool to compute superconformal indices of these theories, for which very few methods are currently available.  In practice, the feasibility of this calculation will depend on whether the  instanton contribution is explicitly computable.  Conversely, in cases where the 4d theory is Lagrangian,  the exact 4d index is computable, and so gives a prediction for the $S^3_b \times \Sigma_\g$ partition function, which we may use to help characterize the instanton contributions.  We may similarly compute the generalized indices of these 4d theories \cite{Nishioka:2014zpa,Closset:2017bse} by studying the $\cM_3 \times \Sigma_\g$ partition function for more general $\cM_3$.  Another obvious generalization is to include punctures on the Riemann surface, which correspond to insertions of local defect operators in the TQFT, and give access to more general 4d theories.  Finally, this computation gives a new entry in the gauge-Bethe correspondence dictionary \cite{Nekrasov:2009rc}, and it would be very interesting to understand in more detail and generality the relation between 4d compactifications of 6d theories and integrable systems, as discussed in some examples in Section \ref{sec:5d6d4d}.

\

The results of Section~\ref{sec:AdS6} and their relation to entropy of black holes in AdS$_{6}$  deserve a better understanding. On the field theory side, one should rigorously extend the exact computation of the $S^{2}\times \Sigma_{\fg}\times S^{1}$ partition function to the case  $\Sigma_{\fg_{1}}\times \Sigma_{\fg_{2}}\times S^{1}$, including instanton corrections. It would also be interesting to have a first-principles understanding on the proposed conjecture that the dominant eigenvalue distribution at large $N$ is given by that extremizing the 5d prepotential.  On the supergravity side, it would be interesting to construct extremal black hole solutions with an AdS$_{2}\times \Sigma_{\fg_{1}}\times \Sigma_{\fg_{2}}$ near-horizon geometry, for generic $\fg_{1,2}$ and flavor fluxes. In particular, nonzero flavor fluxes may allow for solutions with $\fg_{2}=0$, in which case we have provided an expression for the full non-perturbative partition function, determined implicitly by studying the Nekrasov-Shatashvili limit of the twisted superpotential. This could allow for precision tests of the ideas presented here, including subleading corrections to the black hole entropy. 

\

We plan to return to some of these questions in future work.


\acknowledgments{We would like to thank Sujay K. Ashok, Nikolay Bobev, Cyril Closset, Michele Del Zotto, Kazuo Hosomichi, Heeyeon Kim, Guli Lockhart,  Shlomo Razamat, and Minwoo Suh for insightful discussions. We would also like to thank the authors of \cite{Hosseini:2018uzp} for agreeing to coordinate the submissions of our papers.  BW was supported in part by the National Science Foundation under Grant No. NSF PHY11-25915. PMC is supported by Nederlandse Organisatie voor Wetenschappelijk Onderzoek (NWO) via a Vidi grant and is also part of the Delta ITP consortium, a program of the NWO that is funded by the Dutch Ministry of Education, Culture and Science (OCW). DJ thanks the hospitality of ICTS, Bangalore during the visits to two programs: Kavli Asian Winter School 2018 (\texttt{ICTS/Kaws2018/01}) and Quantum Fields, Geometry and Representation Theory (\texttt{ICTS/qftgrt/2018/07}), where part of this work was done. DJ also thanks the hospitality of IMSc, Chennai where a part of this work was done and presented.}


\appendix
\section{\texorpdfstring{Casimir Energy on $S^{3}_{b}\times \Sigma_{\fg} \times S^{1}_{\beta}$}{Casimir Energy on S³b×Σg×S¹β}}\label{App:CasimirS3Sigma}

Here we provide some general formulas for the Casimir energy for the 4d theories obtained by twisted compactification of 6d $\cN=(1,0)$ theories. The expression is obtained by twisting the 6d anomaly polynomial, $I_{8}$, integrating over the Riemann surface to obtain the anomaly polynomial of the 4d theory, and then using the results of  \cite{Assel:2015nca,Bobev:2015kza}, \ie, 
\equ{\label{EcGenPres}
	E_{\mathit{Casimir}}=\int_{\varepsilon} \int_{\Sigma_{\fg}}\, I_{8}^{\mathit{twisted}}\,,
}
where $\int_{\Sigma_{\fg}}$ is a regular integral over the Riemann surface, $\int_{\varepsilon} $ is an equivariant integral. We give an explicit expression below (namely \eqref{CasimirN=1Sigma}) for a rather general 6d $\cN=(1,0)$ theories and consider in detail the examples of the 6d $\cN=(2,0)$ theory, its orbifolds, and the E-string theory. As discussed in Section~\ref{sec:Casimir4d} this quantity matches with the one extracted from the corresponding $S^{3}_{b}\times \Sigma_{\fg}$ partition function. 

We begin by assuming the untwisted 6d anomaly polynomial is of the form\footnote{This is a rather generic 6d anomaly polynomial which contains all the cases we consider below. If any of the flavor symmetries is Abelian, as will be the case for $\mathcal S_{k}$ theories, one may use  $C_{2}(F)=-C_{1}(F)^{2}$.}
\equ{\label{AnPGen}
	I_{8}=\frac{1}{2}k^{AABB}C_{2}(A)C_{2}(B)+k^{AA}C_{2}(A)p_{1}(TM)+k_{1}p_{1}(TM)^{2}+k_{2}p_{2}(TM)\,.
}
Here  $A,B$ run over all global symmetries of the theory (both R-symmetry and flavor symmetries), $C_{2}$ are their corresponding second Chern classes, and $k^{AABB}=k^{BBAA}$ the anomaly coefficients.   $p_{1,2}(TM)$ are  Pontryagin classes for the tangent bundle $TM$, and $ k_{1,2}$ are the corresponding gravitational anomaly coefficients.

Performing a topological twist on $\Sigma_{\fg}$ by $U(1)_{R}\subset SU(2)_{R}$ and a generic $U(1)_{F}$ subgroup of the flavor symmetry amounts to the replacements
\eqss{\label{twistGen}
	C_{2}(R)&\to -\(C_{1}(R)-\frac{\kappa}{2}\, t_{\fg}\)^{2}\,,\\
	C_{2}(F)&\to-\(  \frac{\hat{\mathfrak n}_{F}}{2} \, t_{\fg}+ \epsilon_{F} \, C_{1}(R)+ C_{1}(F)\)^2+\sum_{j}C_{2}(G_{j})\,,
}
where the sum over $j$ is over all simple commutants of  $U(1)_{F}$ in the full flavor group, and $\epsilon_{F}$ controls the amount of mixing of the R-symmetry with the flavor symmetry along the flow to the IR. We now implement the twist \eqref{twistGen} in $I_{8}$ and integrate over the Riemann surface, normalized as $\int_{\Sigma_{\fg}}t_{\fg}=\frac{1}{2\pi}\text{Vol}(\Sigma_{\fg})=\eta_{\Sigma}$, with $\eta_{\Sigma}=2|\fg-1|$ for $\fg\neq 1$ and $\eta_{\Sigma}=1$ for $\fg=1$, to obtain a six-form anomaly polynomial. Comparing this to the anomaly polynomial for a 4d theory,
\equ{
I_{6}=\frac{k_{RRR}}{6}\,C_{1}(R)^{2}-\frac{k_{R}}{24}\, C_{1}(R)p_{1}(TM)+\cdots\,,
}
where ellipses denote contributions involving the flavor symmetries, we read off the 4d R-symmetry anomaly coefficients:
\equ{ \label{anomcoeff4d}
	k_{RRR}=\,12(\fg-1)\, k^{AABB}\hat{ \mathfrak n}_{A}\epsilon_{A}\epsilon_{B}^{2}\,, \qquad k_{R}=\, 48 (\fg-1)\, k^{AA}\hat{\mathfrak n}_{A}\epsilon_{A}  \,,
} 
where we have defined $\epsilon_{R}=\hat{\mathfrak{n}}_{R}=1$ to write the expressions in a compact form. 

We may now use these to compute the Casimir energy of the 4d theory on $S^{3}_{b}$. Recall this is generally given in terms of the R-symmetry anomaly coefficients by  \cite{Assel:2015nca,Ardehali:2015hya,Bobev:2015kza} 
\equ{\label{EC4dgen}
	E_{\mathit{Casimir}}=\frac{1}{24}k_{R}\,(\omega_{1}+\omega_{2})+\frac{1}{48}(k_{RRR}-k_{R})\,\frac{\(\omega_{1}+\omega_{2}\)^{3}}{\omega_{1}\omega_{2}}\,,
}
where $\omega_{1,2}$ are the squashing parameters of the $S^{3}_{b}$, which in our conventions are given by
\equ{\label{identw1w2}
	\omega_{1}=\frac{ 2\pi b}{Q} \,, \qquad \omega_{2}=\frac{2\pi b^{-1}}{Q}\,;\qquad Q=\frac{1}{2} (b+b^{-1})\,.
}
Plugging the anomaly coefficients \eqref{anomcoeff4d} into \eqref{EC4dgen} then gives the desired expression
\equ{\label{CasimirN=1Sigma}
	E_{\mathit{Casimir}}=\,8\pi (\fg-1)\, k^{AA}\hat{\mathfrak n}_{A}\epsilon_{A} +16\pi(\fg-1)\(\frac{1}{4}\, k^{AABB}\hat{ \mathfrak n}_{A}\epsilon_{A}\epsilon_{B}^{2} -\, k^{AA}\hat{\mathfrak n}_{A}\epsilon_{A}\)Q^{2}\,.
}
This represents the full Casimir energy for the theories at hand. It can be viewed as the Casimir energy of the 6d $\cN=(1,0)$ theory, quantized on $S^{3}_{b}\times \Sigma_{\fg}$ or as the Casimir energy of the 4d theory obtained by reduction on $\Sigma_{\fg}$, quantized on $S^{3}_{b}$.

As we discuss in the main text, this should coincide exactly with the expression derived from the non-perturbative partition function on $S^{3}_{b}\times \Sigma_{\fg}$.  However, since in practice we can often only evaluate the perturbative part explicitly, it is convenient to decompose this as
\eqss{\label{caspertnonpert}
	E_{\mathit{Casimir}}=\, &E^{\mathit{pert}}+ E^{\mathit{non-pert}}\,,
}
where $E^{\mathit{pert}}$ is, by definition, the quantity extracted from the perturbative $S^{3}_{b} \times \Sigma_{\fg}$ partition function and $E^{\mathit{non-pert}}$ is the remaining part. As we shall show below, in all the examples we study the remaining part is due to the contribution to the 6d anomaly polynomial from free multiplets. That is, decomposing $I_{8}=I_{8, \mathit{ int}}+I_{8, \mathit{ free}}$ 
where $I_{8, \mathit{ free}}$ is the contribution from all the free multiplets in the tensor branch of the theory and $I_{8, \mathit{ int}}$ is the remaining part, we find  
\equ{
	E^{\mathit{pert}}=\int_{\varepsilon} \int_{\Sigma_{\fg}}\, I_{8, \mathit{ int}}^{\mathit{twisted}}\,,\qquad E^{\mathit{non-pert}}=\int_{\varepsilon} \int_{\Sigma_{\fg}}\, I_{8, \mathit{ free}}^{\mathit{twisted}}\,.
}
The analogous observation  for the case of $S^{5}$ was already made in \cite{Bobev:2015kza}.\footnote{To avoid possible confusions, we emphasize that the computations in \cite{Bobev:2015kza}  lead to the Casimir energy of the 6d theory on $S^{1}\times S^{5}$ instead of the 4d Casimir energy computed here. }  The new examples we provide suggest  there should be a deeper understanding of this fact, which we do not address here. 

\subsection[\texorpdfstring{5d maximal theory and class ${\mathcal S}$}{5d maximal theory and class S}]{5d maximal theory and class ${\mathcal S}$}\label{Sec:Class S theories}

The simplest example is 5d $\cN=2$ SYM theory, consisting of an  $\cN=1$ vector and a hypermultiplet in the adjoint representation of the gauge group $G$. This theory is believed \cite{Douglas:2010iu} to correspond to the 6d $\cN=(2,0)$ theory on a circle $S^{1}_{\beta}$, with radius 
\equ{\label{radiusR6}
	\beta = \frac{g^{2}_{5}}{2\pi}\,,
}
with $g_{5}$ the gauge coupling constant of the 5d gauge theory.  The anomaly polynomial of the 6d $(2,0)$ theory is given by
\equ{\label{AnP6dmax}
	I_{8}=r_{G}A_{8}(1)+d_{G}h_{G}\, \frac{p_{2}(NM)}{24}\,,
}
where  $A_{8}(1)$ is the anomaly polynomial of one free tensor multiplet,
\equ{\label{An(0,2)tensor}
	A_{8}(1)=\frac{1}{48}\(p_{2}(NM)-p_{2}(TM)+\frac14\(p_{1}(NM)-p_{1}(TM)\)^{2}\)\,,
}
and $d_{G}$, $r_{G}$, and $h_G$ are the dimension, rank, and dual Coxeter number of the group $G$, respectively, and $TM$ and $NM$ refer to the tangent and normal $SO(5)_{R}$ bundles, respectively. To evalute the full Casimir energy we first write \eqref{AnP6dmax} in $(1,0)$ language we take $SO(5)_{R}\supset SU(2)_{R}\times SU(2)_{L}$ and use the relations
\equ{\label{p1p210}
	p_{1}(NM) =\, -2(C_{2}(L)+C_{2}(R))\,, \quad p_{2}(NM) = (C_{2}(L)-C_{2}(R))^{2}\,.
}
Then, comparing the anomaly polynomial to \eqref{AnPGen} we read off the corresponding anomaly coefficients, and using these in \eqref{CasimirN=1Sigma} gives 
\eqss{
	E_{\mathit{Casimir}}=&\,\frac{\pi}{6}(\fg-1)(N-1)(1+\hat \n\epsilon)\\
	&+\frac{\pi}{3} (\fg-1) \left(\epsilon ^2-1\right) (N-1)[N(N+1)(\hat \n \epsilon -1)+\hat \n\epsilon]Q^{2}\,,
}
where we have specified $G=SU(N)$ and used $d_{G}=N^{2}-1\,,r_{G}=N-1\,,h_{G}=N$. We wish to compare this to the quantity extracted from the 5d localization computation. With this in mind, let us split the quantity above into the two pieces:
\eqs{\label{CasimirMaxP}
	E^{\mathit{pert}}= &\,\frac{\pi}{3} (\fg-1) N(N^{2}-1)\left(\epsilon ^2-1\right) (\hat \n \epsilon -1) Q^{2}\,,\\ \label{CasimirMaxInstI}
	E^{\mathit{non-pert}}= &\,\frac{\pi}{6}(\fg-1)(N-1)\(1+\hat \n\epsilon + 2\hat \n\epsilon\,(\epsilon^{2}-1)Q^{2}\)\,.
}
As discussed in the main text one can  easily check that the perturbative Casimir energy matches the contribution from the last term in \eqref{AnP6dmax}, while the remaining non-perturbative part is entirely due to the free part,
\equ{\label{EcInstMax}
	E^{\mathit{non-pert}}=(N-1)\int_{\varepsilon}  \int_{\Sigma_{\fg}}\, A_{8}^{\mathit{twisted}}(1)\,.
}
In general, we expect instantons in the 5d computation to contribute $r_{G}\int_{\varepsilon} \int_{\Sigma_{\fg}} A_{8}^{\mathit{twisted}}(1)$. 

\paragraph{Mixed Schur limit.}

As we discuss in the main text, a special value for the fugacity is given by the mixed Schur limit:
\equ{
	\epsilon=\frac{b-b^{-1}}{b+b^{-1}}\,.
}
In this limit, the perturbative and non-perturbative contributions to the Casimir energy \eqref{CasimirMaxP} and \eqref{CasimirMaxInstI} simplify to
\eqss{\label{CasimirMaxInstSchur}
	E^{\mathit{pert}}= &\,\frac{ \pi}{Q}\frac{N(N^{2}-1)}{6}\((\fg-1- \n)b+(\fg-1+ \n)b^{-1}\)\,,\\
	E^{\mathit{non-pert}}= &\,\frac{ \pi}{Q}\frac{(N-1)}{12}\((\fg-1-\n)b+(\fg-1+\n)b^{-1}\)\,,
}
where we have used the definition $\fn=\hat \n(\fg-1)$. Note the values $\fn=\pm(\fg-1)$, for which 4d supersymmetry is enhanced to $\cN=2$, are special. In this case,  as we discuss below, the full Casimir energy is proportional to the central charge of the associated 2d chiral algebra, in the sense of \cite{Beem:2013sza}.

\paragraph{Chiral algebra (or Schur) limit.} 

We consider for concreteness $ \fn=\fg-1$.  Recall for a generic  4d $\cN=2$ theory the Casimir energy is given by (see (4.36) in \cite{Bobev:2015kza}):
\equ{
	E_{\mathit{Casimir}} = \frac12 (c_{4d}-2a_{4d})\frac{\sigma (\sigma+\omega_{1}+\omega_{2})^{2}}{\omega_{1}\omega_{2}}+(c_{4d}-a_{4d})\frac{\sigma(\sigma^{2}-\omega_{1}^{2}-\omega_{2}^{2})}{\omega_{1}\omega_{2}}\,,
} 
where $\sigma\equiv\gamma-\omega_{1}-\omega_{2}$, with $\gamma, \sigma$ fugacities for $SU(2)_{R}\times U(1)_{r}$, respectively, and we have set all flavor fugacities to zero. The ``chiral algebra limit'' corresponds to setting $\gamma=\omega_{2}$ and hence $\sigma=-\omega_{1}$, which gives the simple expression
\equ{
	E_{\mathit{Casimir}} =\frac{\omega_{2}}{2}\,c_{4d} = -\frac{\omega_{2}}{24}\, c_{2d}\,,
}
where in the second equality we used the chiral algebra relation $c_{2d}\equiv-12c_{4d}$ \cite{Beem:2013sza}.  Thus, in this limit the Casimir energy of the 4d $\cN=2$ theory is  proportional to the central charge of its corresponding 2d chiral algebra.\footnote{We note a similar relation  was found in Eq. (3.21) in \cite{Bobev:2015kza} for the Casimir energy of the 6d theory on the squashed $S^{5}$.} To write this explicitly for the maximal theory on $\Sigma_{\fg}$ with no punctures, recall\footnote{See, \eg, \cite{Bah:2012dg}, where their parameter $z$ is identified with $\hat \fn$ here.}
\eqss{
	c_{4d}=\,&\frac{1}{6}(\fg-1)r_{G}(1+2h_{G}(1+h_{G}))\,,\\
	a_{4d}=\,&\frac{1}{24}(\fg-1)r_{G}(5+8h_{G}(1+h_{G}))\,,
} 
and thus
\equ{\label{InstCAlimit}
	E^{\mathit{pert}}= \frac{\pi}{3Q}(\fg-1)r_{G}h_{G}(1+h_{G})b^{-1}\,, \qquad E^{\mathit{non-pert}}=\frac{\pi}{6Q}(\fg-1)r_{G}b^{-1}\,.
}
For $G=SU(N)$ these are a special case of \eqref{CasimirMaxInstSchur} with $\fn=\fg-1$. For $\fn=1-\fg$ the roles of $b$ and $b^{-1}$ are exchanged.

\subsection[\texorpdfstring{Class $\mathcal S_{k}$ theories}{Class Sk theories}]{Class ${\mathcal S_{k}}$ theories}

These are a class of 4d $\cN=1$ theories which arise from the twisted compactification of $\mathbb Z_{k}$ orbifolds of the maximal 6d  $A_{N-1}$  theory \cite{Bah:2017gph}. For general values of $k$ and $N$ the theory has  an $SU(k)_{b} \times  SU(k)_{c} \times u(1)_{s}$ flavor symmetry. Before twisting the anomaly polynomial of the 6d theory is given by\footnote{Note that compared to Eq. (2.1) in \cite{Bah:2017gph} we have a sign difference in the $C_{2}(R)C_{1}(s)^{2}$ term. Also, we have subtracted the contribution from $k^{2}$ hypermultiplets so that for $k=1$ this matches the expression used in \eqref{AnP6dmax} for the maximal theory. }
\eqss{\label{ISk}
	I_{8}=&\,\frac{k^{2}N^{3}}{24}C_{2}(R)^{2}-\frac{N(k^{2}-1)}{48}C_{2}(R)\(4C_{2}(R)+p_{1}(TM)\)\\
	&+N\( \frac{C_{2}(R)p_{1}(TM)}{48}-\frac{p_{2}(TM)}{48}+\frac{p_{1}(TM)^{2}}{192}\)\\
	&+(k^{2}-1)\( \frac{1}{24}C_{2}(R)^{2}+\frac{1}{48} C_{2}(R)p_{1}(TM)+\frac{7p_{1}(TM)^{2}-4p_{2}(TM)}{5760}\)\\
	&-\(\frac{1}{24}C_{2}(R)^{2}+\frac{1}{48} C_{2}(R)p_{1}(TM)+\frac{23}{5760}p_{1}(TM)^{2}-\frac{29}{1440}p_{2}(TM)\)\\
	&-\frac{Nk^{2}}{48}\,p_{1}(TM)C_{1}(s)^{2}+\frac{k^{2}N(N^{2}-1)}{12}C_{2}(R)C_{1}(s)^{2}+\frac{k^{2}N^{3}}{24}C_{1}(s)^{4}\\
	&-k^{2}I_{{(1,0) \mathit{ hyper}}}+I_{8}(F_{b,c})\,,
}
where $I_{8}(F_{b,c})$ denotes contributions involving at least one of the $SU(k)_{b}$ or $ SU(k)_{c}$ symmetries. Since we do not turn background fields for these symmetries, we may ignore these terms in what follows.  To identify the parts encoding the perturbative and non-perturbative terms in the Casimir energy it is convenient to recall the anomaly polynomial for a $(1,0)$ tensor multiplet  (see, \eg, Eq. (2.3) in \cite{Bah:2017gph}): 
\equ{
	I_{(1,0)\text{ tensor}}=\frac{1}{24}C_{2}(R)^{2}+\frac{1}{48} C_{2}(R)p_{1}(TM)+\frac{23}{5760}p_{1}(TM)^{2}-\frac{29}{1440}p_{2}(TM)\,,
}
and for a free half-hypermultiplet charged under $U(1)_{s}$ as\footnote{Here we have taken the anomaly polynomial  for a free half-hypermultiplet in the doublet of $SU(2)_{L}$ (see, \eg, (1.5) in \cite{Ohmori:2014pca})  and set $C_{2}(L)=-C_{1}(s)^{2}$. }
\equ{
	I_{(1,0) \mathit{ hyper}}=\frac{1}{24}C_{1}(s)^{4}-\frac{1}{48} C_{1}(s)^{2}p_{1}(TM)+\frac{7p_{1}(TM)^{2}-4p_{2}(TM)}{5760}\,,
}
and finally
\equ{
	I_{(1,0) \mathit{ vector}} = -\frac{1}{24}C_{2}(R)^{2}-\frac{1}{48} C_{2}(R)p_{1}(TM)- \frac{7p_{1}(TM)^{2}-4p_{2}(TM)}{5760}\,.
}
In terms of these combinations \eqref{ISk} takes the following form 
\eqss{\label{ISkNice}
	I_{8}=&\,(N-1)\left[I_{(1,0)\mathit{ tensor}}+(k^{2}-1)I_{(1,0)\mathit{ vector}}+k^{2}I_{(1,0) \mathit{ hyper}}\right]\\
	&+k^{2}N(N^{2}-1)\frac{\(C_{2}(R)+C_{1}(s)^{2}\)^{2}}{24}\,.
}
We note this has a very similar form to \eqref{AnP6dmax}.\footnote{Indeed, in the special case $k=1$, $U(1)_{s}$ is enhanced to $SU(2)_{L}$ with  $C_{1}(s)^{2}=-C_{2}(L)$ and the contribution from the $(1,0)$ tensor combines with that of the hyper into that of a $(2,0)$ tensor   to form the combination $A_{8}(1)=I_{(1,0)\mathit{ tensor}}+I_{(1,0)\mathit{ hyper}}$. 
}
Indeed, from the 5d localization calculation we find the perturbative part of the Casimir energy arises entirely from the terms in the second line of \eqref{ISkNice}, and is given by
\equ{\label{CasimirSk}
	E^{\mathit{pert}}= \frac{\pi k^{2}}{3} (\fg-1) \left(\epsilon ^2-1\right) [N(N^{2}-1)(\hat \n \epsilon -1)] Q^{2}\,,\\
}
while the remaining part, 
\equ{\label{CasimirSkInst}
	E^{\mathit{non-pert}}= \frac{\pi}{6}(\fg-1)(N-1)\(2-k^{2}+k^{2}\hat \n\epsilon +2k^{2}\hat \n\epsilon\,(\epsilon^{2}-1)Q^{2}\)\,,
}
arises from 
\equ{\label{CasimirSkFFs}
	E^{\mathit{non-pert}}=(N-1)\int_{\varepsilon}\int_{\Sigma_{\fg}} \,\(I_{(1,0)\mathit{ tensor}}^{\mathit{twisted}}+(k^{2}-1)I_{(1,0)\mathit{ vector}}^{\mathit{twisted}}+ k^{2}I_{(1,0) \mathit{ hyper}}^{\mathit{twisted}}\)\,.
}

We thus see once again that the instanton correction to the Casimir energy is encoded in the 6d anomaly polynomial of free fields.

\subsection{E-string theory}\label{Sec:E-string}

This is the 5d Seiberg theory in the special case $N_{f}=8$. The twisted compactification of this theory on a Riemann surface was considered in  \cite{Kim:2017toz}. The anomaly polynomial before twisting reads \cite{Ohmori:2014pca}\footnote{Here we follow the conventions in \cite{Bobev:2015kza} and we have included an additional contribution from a $(1,0)$ tensor in the $\mathcal O(1)$ part to complete $A_{8}(1)$. This is done so that for $k=1$ this anomaly polynomial matches \eqref{AnP6dmax} for the maximal theory.}
\eqs{
	I_{E} &=\frac{N^{3}}{6}p_{2}(NM)+\frac{N^{2}}{2}e(NM) A_{4}+N\(A_{4}^{2}-\frac{p_{2}(NM)}{24}\) +(N-1) A_{8}(1)\,,\nn
	I_{(1,0) \mathit{ hyper}} &=\frac{1}{24}C_{2}(L)^{2} +\frac{1}{48} C_{2}(L)p_{1}(TM) +\frac{7p_{1}(TM)^{2}-4p_{2}(TM)}{5760}\,, \nn
	A_{4} &=\frac{1}{4}\(p_{1}(NM)+p_{1}(TM)+\frac{1}{15}C_{2}(E_{8})\)\,.
\label{IEs1}}
To write the anomaly polynomial in the form \eqref{AnPGen} we use the relations \eqref{p1p210} and $e(NM)= C_{2}(L)-C_{2}(R)$, which gives
\eqs{
	I_{E}&=\frac{N(4N^{2}+6N+3)}{24}\, C_{2}^{2}(R)+\frac{(N-1)(4N^{2}-2N+1)}{24}\, C_{2}^{2}(L) \nn
	&\quad -\frac{N(N^{2}-1)}{3}\, C_{2}(R)C_{2}(L)+\frac{(N-1)(6N+1)}{48}C_{2}(L)p_{1}(TM) \nn
	&\quad -\frac{N(6N+5)}{48}\, C_{2}(R)p_{1}(TM)+\frac{N(N-1)}{120} C_{2}(L)C_{2}(E_{8})_{\mathbf{248}} \nn
	&\quad -\frac{N(N+1)}{120} C_{2}(R)C_{2}(E_{8})_{\mathbf{248}}+\frac{N}{240}\, p_{1}(TM) C_{2}(E_{8})_{\mathbf{248}}+\frac{N}{7200}\, C_{2}^{2}(E_{8})_{\mathbf{248}} \nn
	&\quad +(30 N-1)\frac{7p_{1}(TM)^{2}-4p_{2}(TM)}{5760}- I_{(1,0)\mathit{ tensor}}\,.
\label{IEs2}}
This anomaly polynomial coincides with the one in \cite{Kim:2017toz}, apart from the contribution $I_{(1,0)\mathit{tensor}}$ from a free tensor which we have included. We now perform the twisted reduction on the Riemann surface following this reference. In addition to the R-symmetry flux there are nine possible Abelian flavor fluxes; eight along the Cartan of $E_{8}$ and one for the Cartan of $SU(2)_{L}$. For simplicity we consider only two flavor fluxes, picking a generic $U(1)_{F}\subset E_{8}$ and one along $U(1)_{L}\subset SU(2)_{L}$. Twisting amounts to 
\eqss{\label{twistEstring}
	C_{2}(R)&\to -\(C_{1}(R)-\frac{\kappa}{2}\, t_{\fg}\)^{2}\,,\\
	C_{2}(E_{8})_{\mathbf{248}}&\to -\xi' \( \frac{\tilde{\mathfrak n}_{E}}{2}\, t_{\fg}+ \tilde{\epsilon}_{E} \, C_{1}(R)+ C_{1}(F)\)^2+d_{E_{8}}\sum_{j}C_{2}(G_{j})\,,\\
	C_{2}(L)&\to-\(\frac{\hat{\mathfrak n}}{2}\, t_{\fg}+ \epsilon\, C_{1}(R)+ C_{1}(L)\)^2\,,
}
where $\xi'\equiv 2 \xi d_{E_{8}}$, with  $d_{E_{8}}=30$  the Dynkin index for the fundamental of $E_{8}$ and $\xi$ is numerical constant that depends on the choice of $U(1)_{F}$ inside $E_{8}$.  The sum over $j$ is over all simple commutants of  $U(1)_{F}$ in $E_{8}$. See \cite{Kim:2017toz} for details.

To write things in a compact form, let us write the anomaly polynomial in the form  
\eqs{
	I&=\frac{k_{RRRR}}{2}\, C_{2}^{2}(R)+\frac{k_{LLLL}}{2} \, C_{2}^{2}(L)+k_{RRLL}\, C_{2}(R)C_{2}(L)\nn
	&\quad+k_{LL}\, C_{2}(L)p_{1}(TM)+k_{RR}\, C_{2}(R)p_{1}(TM)+\frac{k_{LLEE}}{\xi'}\, C_{2}(L)C_{2}(E_{8})_{\mathbf{248}}\nn
	&\quad+ \frac{k_{RREE}}{\xi'}\,C_{2}(R)C_{2}(E_{8})_{\mathbf{248}}+\frac{k_{EE}}{\xi'}\, p_{1}(TM) C_{2}(E_{8})_{\mathbf{248}}+\frac{k_{EEEE}}{2\xi'^{2}}\, C_{2}^{2}(E_{8})_{\mathbf{248}}\nn
	&\quad+k_{1}\,p_{1}(TM)^{2}+k_{2}\,p_{2}(TM)\,.
\label{Igen6d}}
The coefficients for the E-string theory are easily read off by comparing this expression to \eqref{IEs2}. We subtract the contribution of the free fields from the full expression (continuing the pattern from above) to get
\eqst{\label{EcasimirE8Anom}
E^{\mathit{pert}} =\pi(\fg-1)\Big[N(N-1)Q^{2}(1-\epsilon^{2})\big(\tfrac{2}{3}(2N-1)(1-\hat{\mathfrak n}\epsilon) +(2-\hat{\mathfrak n}_{E}\epsilon_{E})\big) \\
+N\(Q^2(4-\epsilon_{E}^{2})-1\)\big((N-1)(1-\hat{\mathfrak n}\epsilon) +(2-\hat{\mathfrak n}_{E}\epsilon_{E})\big)\Big]\,,
}
where we defined $\tilde{\mathfrak n}_{E}=\frac{1}{\sqrt{2\xi}}\hat{\mathfrak n}_{E}$, $\tilde{\epsilon}_{E}=\frac{1}{\sqrt{2\xi}}\epsilon_{E}$. This exactly matches the perturbative result derived in the main text. The remaining piece,
\eqss{\label{EcasimirE8AnomInst}
	E^{\mathit{non-pert}} =\frac{\pi}{6}(\fg-1)(N-1)\(1+\hat{\mathfrak n}\epsilon +2\hat{\mathfrak n}\epsilon (\epsilon^{2}-1)Q^2\)\,,
}
is given by the free fields 
\equ{\label{EcasimirE8AnomFree}
	E^{\mathit{non-pert}}=(N-1)\int_{\varepsilon}\int_{\Sigma_{\fg}} \,\(I_{(1,0)\mathit{ tensor}}^{\mathit{twisted}}+I_{(1,0) \mathit{ hyper}}^{\mathit{twisted}}\)\,,
}
and must be generated entirely due to instanton corrections in the 5d partition function.

We have thus shown by explicit computation of the perturbative piece that the instanton corrections to the 4d Casimir energy arise from the free anomaly polynomial of the 6d theory in all the examples we have considered (for class $\mathcal S$, class $\mathcal S_{k}$, and the compactification of the E-string theory). It would be interesting to have a better understanding of whether this is always the case and, if so, the reason behind this.


\section{\texorpdfstring{Instanton partition function for $\cN=2$ $U(N)$ and $SU(N)$ SYM}{Instanton partition function for N=2 U(N) and SU(N) SYM}}\label{sec:maxapp}

Here we consider the instanton partition function for the maximal  SYM theory with gauge group $U(N)$ or $SU(N)$.  This consists of an adjoint vector multiplet (as always) and an adjoint hypermultiplet, which we assign a mass $\tnu$.  Then, taking into account the extra factors needed to pass between the Dolbeault, self-dual, and Dirac complexes, we may write the total equivariant index for the fields of this theory as, first in the 4d case,
\bea \text{ind}_{4d  \; \cN=4}(\tu,\tnu) &= \( \frac{1+ e^{i (\epsilon_1+\epsilon_2)}}{2} - e^{i \frac{\epsilon_1+\epsilon_2}{2}}\frac{e^{i \tnu}+e^{-i\tnu}}{2} \)  \bigg(\frac{\tr_{adj}(e^{i \tu})}{(e^{i \epsilon_1}-1)(e^{i \epsilon_2}-1)} \\
&\quad -e^{-i \frac{\epsilon_1+\epsilon_2}{2}}\( \tr_{N} (e^{i \tu}) \tr_{\bar{k}} (e^{i \phi}) + c.c.\) + (1-e^{-i \epsilon_1})(1-e^{-i \epsilon_2}) \tr_{adj}(e^{i \phi}) \bigg),
\eea
where $\epsilon_i$ are the equivariant parameters.

We are interested here in the 5d case, and for  this we need to incorporate the KK modes, $n$, as in Section \ref{sec:5d4d}.  One additional subtlety is that, for the Dirac complex determining the hypermultiplets, the quantization of the KK momenta are shifted by $n \rightarrow n+\frac{1}{2}$ \cite{Kim:2012qf}.  Then we find\footnote{In this appendix we find it convenient to keep the factors of $r$, the radius of the $S^1$ fiber, explicit.  To compare with the formulas in the main text one may set $r=1$.}
\bea \label{AppB5dind}\text{ind}_{5d \; \cN=2}(\tu,\tnu) &= \sum_{n \in \Z} e^{ \frac{i n}{r}}\( \frac{1+ e^{i (\epsilon_1+\epsilon_2)}}{2} - e^{\frac{i}{2r}} e^{i \frac{\epsilon_1+\epsilon_2}{2}}\frac{e^{i \tnu}+e^{-i \tnu}}{2} \) \bigg(\frac{\tr_{adj}(e^{i \tu})}{(e^{i \epsilon_1}-1)(e^{i \epsilon_2}-1)} \\
&\quad - e^{-i \frac{\epsilon_1+\epsilon_2}{2}}\( \tr_{N} (e^{i \tu}) \tr_{\bar{k}} (e^{i \phi}) + c.c.\) +(1-e^{-i \epsilon_1})(1-e^{-i \epsilon_2}) \tr_{adj}(e^{i \phi}) \bigg).
\eea
where we write the equivariant parameters as $\q_i = e^{2 \pi i \epsilon_i}$, as in Section \ref{sec:5d4d}.

Let us consider in more detail the instanton contribution, which comes from the second and third terms inside the second parantheses in \eqref{AppB5dind}.  Using the rule \eqref{indrep} to pass from the equivariant index to the $1$-loop determinant, and integrating over the equivariant parameters for the $U(k)$ instanton symmetry, we may write the contribution to the partition function from the $k$ instanton sector as
\bea \label{5dn2Zk} Z_k =&  \ds \frac{1}{k!} \bigg(\frac{\sin\(\pi r(\tnu+\frac{1}{2r}+\epsilon_-)\) \sin\(\pi r(\tnu+\frac{1}{2r}-\epsilon_-)\)}{\sin\( \pi r(\tnu+\frac{1}{2r}+\epsilon_+)\) \sin\( \pi r(\tnu+\frac{1}{2r}-\epsilon_+)\)}\frac{\sin\( 2\pi r\epsilon_+\)} {\sin\( \pi r \epsilon_1\) \sin\( \pi r \epsilon_2\)} \bigg)^k  \\
& \ds \times \int d^k \phi \prod_{i=1}^N \prod_{\alpha=1}^k \frac{\sin\(\pi r(\tu_i - \phi_\alpha +\tnu+\frac{1}{2r})\) \sin\( \pi r(\tu_i- \phi_\alpha -\tnu-\frac{1}{2r})\)}{\sin\( \pi r(\tu_i - \phi_\alpha + \epsilon_+)\)\sin\( \pi r(\tu_i- \phi_\alpha - \epsilon_+)\)}  \\
&  \ds\times \prod_{\alpha < \beta} \frac{\sin\( \pi r(\phi_\alpha - \phi_\beta+\tnu+\frac{1}{2r}+\epsilon_-)\) \sin\( \pi r(\phi_\alpha-\phi_\beta +\tnu+\frac{1}{2r}-\epsilon_-)\)}{\sin\( \pi r(\phi_\alpha - \phi_\beta+\tnu+\frac{1}{2r}+\epsilon_+)\) \sin\(\pi r(\phi_\alpha-\phi_\beta +\tnu+\frac{1}{2r}-\epsilon_+)\)} \\
&\times \frac{\sin\( \pi r(\phi_\alpha - \phi_\beta+2\epsilon_+)\)\sin\( \pi r(\phi_\alpha-\phi_\beta)\)} {\sin\( \pi r(\phi_\alpha - \phi_\beta+\epsilon_1)\) \sin\( \pi r(\phi_\alpha-\phi_\beta +\epsilon_2)\)} \,,
\eea
where we have defined $\epsilon_\pm=\frac{1}{2}(\epsilon_1 \pm \epsilon_2)$, and the prefactor is the contribution from the Cartan components of the adjoint representation of $U(k)$.  This is integrated over a suitable contour.  As mentioned in Section \ref{sec:5d4d}, the result can be expressed as a sum over $N$-colored Young diagrams, $Y_i$.  Namely, we have \cite{Nekrasov:2004vw,Kim:2012qf}
\be Z_k = \sum_{\vec Y\,,\;|\vec Y|=k} Z_{\vec Y} \,,\ee
with
\be \label{AppBZydef} Z_{\vec Y} = \prod_{i,j=1}^N \prod_{s \in Y_i} \frac{\sin\( \pi r (E_{ij} + \tnu + \frac{1}{2r} - \epsilon_+)\) \sin\( \pi r (E_{ij} - \tnu - \frac{1}{2r} - \epsilon_+)\)}{\sin\( \pi r E_{ij} \) \sin\( \pi r (E_{ij} - 2 \epsilon_+)\)} \,,\ee
where we have defined
\be E_{ij} = \tu_i - \tu_j - \epsilon_1 h_i(s) + \epsilon_2 (v_j(s)+1) \,,\ee
where $h_i(s)$ is the horizontal distance from the box to the right side of the Young diagram, and $v_i(s)$ is the vertical distance to the bottom.

Let us note some simplifying limits of this instanton partition function.  First, if we set
\be \tnu = \pm  \epsilon_+ - \tfrac{1}{2r} \, \;\;\;\; \Leftrightarrow \;\;\;\; y \equiv e^{2 \pi i r \tnu} = -(\q_1 \q_2)^{\pm 1/2}\;,\ee
one can check from \eqref{AppBZydef} that $Z_{\vec Y}=1$ for all $\vec Y$, and so the full instanton partition function is given by
\be Z^{inst}_{\R_{\q_1}^2 \times \R_{\q_2}^2 \times S^1}\(x,y=-(\q_1 \q_2)^{\pm 1/2},z\) \propto \sum_{k=0}^\infty z^k  \sum_{\vec Y\,,\;|\vec Y|=k} 1 \,. \ee
This is simply the generating function for the $N$-colored Young diagrams, which can be expressed in terms of the $q$-Pochhammer symbol, 
\be \label{qpo} Z^{inst}_{\R_{\q_1}^2 \times \R_{\q_2}^2 \times S^1}\(x,y=-(\q_1 \q_2)^{\pm 1/2},z\) \propto (z;z)^{-N}\,. \ee
The precise coefficient depends on the detailed normalization of the instanton measure, and it is not known how to fix this from first principles.  In \cite{Kim:2012qf,Kim:2013nva} it was found that the appropriate normalization is such that we obtain the Dedekind eta function, $\eta(z) = z^{1/24} (z;z)$, and we will make the same assumption here.  Then

\be \label{qpo2} Z^{inst}_{\R_{\q_1}^2 \times \R_{\q_2}^2 \times S^1}\(x,y=-(\q_1 \q_2)^{\pm 1/2},z\)= \eta(z)^{-N}.
\ee
Next, consider
\be \tnu = -\tfrac{1}{2r} \pm \epsilon_- \, \;\;\;\; \Leftrightarrow \;\;\;\; y = -(\q_1 \q_2^{-1})^{\pm 1/2}.\ee
Then we can see the prefactor in \eqref{5dn2Zk} vanishes, and so $Z_k=0$ for $k>0$, implying
\be \label{qpo3} Z^{inst}_{\R_{\q_1}^2 \times \R_{\q_2}^2 \times S^1}\(x,y=-(\q_1 \q_2^{-1})^{\pm 1/2},z\)= 1\,, \ee
Both of these identities remain true if we shift $\tnu \rightarrow \tnu + \frac{\ell}{r}$, $\ell \in \Z$, as this is a symmetry of the partition function.

In the $SU(N)$ case, we expect the above identities to remain true, however we need to strip off the overall $U(1)$ contribution from above, and find that $N \rightarrow N-1$ in \eqref{qpo2}. 


\bibliographystyle{utphys}
\bibliography{bib5d}

\end{document}